\documentclass[aps,prx,twocolumn,superscriptaddress,nofootinbib]{revtex4-2}
\pdfoutput=1
\usepackage{graphicx,amssymb,amsmath,amsfonts,empheq}
\usepackage[dvipsnames]{xcolor}
\usepackage{color}
\usepackage{braket}
\usepackage{latexsym}
\usepackage{upgreek}
\usepackage{dsfont}
\usepackage{bm}
\usepackage{makecell}
\usepackage{multirow}
\usepackage{enumitem}
\usepackage[normalem]{ulem}
\usepackage{url}
\usepackage{hyperref}
\hypersetup{
colorlinks = true,
linkcolor = blue, 
citecolor = blue,
urlcolor = [rgb]{0.25, 0.41, 0.88}}
\usepackage{bbm}
\usepackage[utf8]{inputenc}
\usepackage[english]{babel}
\usepackage[T1]{fontenc}
\usepackage{lipsum}
\usepackage{physics}
\usepackage{comment}
\usepackage{wasysym} %
\usepackage{slashed}

\usepackage{tikz}
\usepackage{tikz-cd}
\usepackage{tkz-base}
\usetikzlibrary{arrows}
\usetikzlibrary{intersections}
\usetikzlibrary{shapes.geometric}
\usetikzlibrary{decorations.pathmorphing, patterns,shapes}
\usetikzlibrary{decorations.markings}

\usepackage{diagbox}

\tikzset{
	partial ellipse/.style args={#1:#2:#3}{
		insert path={+ (#1:#3) arc (#1:#2:#3)}
	}
}

\tikzset{
	mid arrow/.style={postaction={decorate,decoration={
				markings,
				mark=at position .575 with {\arrow[#1]{stealth}}
	}}},
	near arrow/.style={postaction={decorate,decoration={
				markings,
				mark=at position .275 with {\arrow[#1]{stealth}}
	}}},
	far arrow/.style={postaction={decorate,decoration={
				markings,
				mark=at position .800 with {\arrow[#1]{stealth}}
	}}},
}

\pgfdeclarepatternformonly{south west lines}{\pgfqpoint{-0pt}{-0pt}}{\pgfqpoint{3pt}{3pt}}{\pgfqpoint{3pt}{3pt}}{
	\pgfsetlinewidth{0.4pt}
	\pgfpathmoveto{\pgfqpoint{0pt}{0pt}}
	\pgfpathlineto{\pgfqpoint{3pt}{3pt}}
	\pgfpathmoveto{\pgfqpoint{2.8pt}{-.2pt}}
	\pgfpathlineto{\pgfqpoint{3.2pt}{.2pt}}
	\pgfpathmoveto{\pgfqpoint{-.2pt}{2.8pt}}
	\pgfpathlineto{\pgfqpoint{.2pt}{3.2pt}}
	\pgfusepath{stroke}}

\usepackage[dvipsnames]{xcolor}

\newcommand{\rket}[1]{\mbox{$\left| #1 \right)$}}
\newcommand{\rbra}[1]{\mbox{$\left( #1 \right|$}}
\renewcommand{\bar}{\overline}
\renewcommand{\leq}{\leqslant}
\renewcommand{\geq}{\geqslant}
\renewcommand{\Re}{\operatorname{Re}}
\renewcommand{\Im}{\operatorname{Im}}
\newcommand{\rd}{\mathrm{d}}
\newcommand{\ri}{\mathrm{i}}

\newcommand{\sgn}{\operatorname{sgn}}
\newcommand{\nn}{\nonumber}
\newcommand{\Pf}{\operatorname{Pf}}

\newcommand{\U}{\operatorname{U}}
\renewcommand{\O}{\operatorname{O}}
\newcommand{\SU}{\operatorname{SU}}
\newcommand{\Sp}{\operatorname{Sp}}
\newcommand{\SO}{\operatorname{SO}}

\newcommand{\opt}{\mathrm{opt}}
\newcommand{\subopt}{\mathrm{sub}}

\newcommand{\bbC}{\mathbb{C}}

\newcommand{\bbZ}{\mathbb{Z}}

\newcommand{\calC}{\mathcal{C}}
\newcommand{\calD}{\mathcal{D}}

\newcommand{\calF}{\mathcal{F}}

\newcommand{\calL}{\mathcal{L}}

\newcommand{\calO}{\mathcal{O}}
\newcommand{\calP}{\mathcal{P}}
\newcommand{\calQ}{\mathcal{Q}}

\newcommand{\calS}{\mathcal{S}}

\newcommand{\calY}{\mathcal{Y}}
\newcommand{\calZ}{\mathcal{Z}}

\newcommand{\sfa}{\mathsf{a}}
\newcommand{\sfb}{\mathsf{b}}
\newcommand{\sfg}{\mathsf{g}}

\newcommand{\sfv}{\mathsf{v}}
\newcommand{\sfh}{\mathsf{h}}
\newcommand{\sfT}{\mathsf{T}}
\newcommand{\sfQ}{\mathsf{Q}}
\newcommand{\ML}{\mathrm{ML}}
\newcommand{\bfr}{\mathbf{r}}
\newcommand{\bfv}{\mathbf{v}}
\newcommand{\bfh}{\mathbf{h}}
\newcommand{\bfT}{\mathbf{T}}
\newcommand{\bfH}{\mathbf{H}}
\newcommand{\bfV}{\mathbf{V}}
\newcommand{\bfY}{\mathbf{Y}}
\newcommand{\bfZ}{\mathbf{Z}}

\newcommand{\eff}{\mathrm{eff}}
\newcommand{\Vol}{\mathrm{Vol}}

\begin{document}

\title{Non-linear Sigma Model for the Surface Code with Coherent Errors}

\author{Stephen W. Yan}\thanks{These authors contributed equally to this work.}
\affiliation{Department of Physics, University of California,
Santa Barbara, CA 93106, USA}
\author{Yimu Bao}\thanks{These authors contributed equally to this work.}
\affiliation{Kavli Institute for Theoretical Physics, University of California,
Santa Barbara, CA 93106, USA}
\author{Sagar Vijay}
\affiliation{Department of Physics, University of California,
Santa Barbara, CA 93106, USA}

\begin{abstract}
The surface code is a promising platform for a quantum memory, but its threshold under coherent errors remains incompletely understood. We study maximum-likelihood decoding of the square-lattice surface code in the presence of single-qubit unitary rotations that create electric anyon excitations. We microscopically derive a non-linear sigma model with target space $\mathrm{SO}(2n)/\mathrm{U}(n)$ as the effective long-distance theory of this decoding problem, with distinct replica limits: $n\to1$ for optimal decoding, which assumes knowledge of the coherent rotation angle, and $n\to0$ for suboptimal decoding with imperfect angle information. This exposes a sharp distinction between the two decoders. The suboptimal decoder supports a ``thermal-metal'' phase, a non-decodable regime that is qualitatively distinct from the conventional non-decodable phase of the surface code under incoherent Pauli errors. By contrast, the metal phase cannot arise in optimal decoding, since the metallic fixed-point becomes unstable in the $n\to 1$ replica limit. We argue that optimal decoding may be possible up to the maximally-coherent rotation angle. Within the sigma model description, we show that the decoding fidelity is related to twist defects of the order-parameter field, yielding quantitative predictions for its system-size dependence near the metallic fixed point for both decoders. We examine our analytic predictions for the decoding fidelity as well as other physical observables with extensive numerical simulations. We discuss how the symmetries and the target space for the sigma model rely on the lattice of the surface code, and how a stable thermal metal phase can arise in optimal decoding when the syndromes reside on a non-bipartite lattice.
\end{abstract}

\maketitle

\section{Introduction}

Topological quantum error-correcting codes are central to fault-tolerant quantum computation, leveraging local stabilizer measurements together with a macroscopic code distance to protect quantum information~\cite{kitaev2003fault,dennis2002topological,semeghini2021probing,bluvstein2022quantum,satzinger2021realizing,google2023suppressing,andersen2023observation,bluvstein2023logical,qec_below_threshold}. 
In this setting, decoding is naturally formulated as a statistical inference problem~\cite{dennis2002topological}. 
From this perspective, environmental noise can induce sharp transitions between quantum many-body phases distinguished by information encoding, with close ties to information-theoretic probes of intrinsic topological quantum order~\cite{fan2023diagnostics,bao2023mixed,lee2023quantum,wang2025intrinsic,sohal2025noisy,hauser2026information,tang2025phases,chen2024separability,chen2024symmetry,ellison2025toward}. 
For stochastic Pauli errors, this correspondence is well established: the surface code exhibits a decodable phase—where logical information is recoverable with high probability—separated from a non-decodable phase.  
In this simple case, the phases and phase transitions are understood through mappings to classical statistical mechanical models with quenched disorder~\cite{wang2003confinement,katzgraber2009error,bombin2010topological,bombin2012strong,kubica2018three,chubb2021statistical,song2022optimal,ferris2014tensor,bravyi2014efficient,Darmawan_2017,Darmawan_2018,tuckett2018ultrahigh,tuckett2019tailoring,chubb2021statistical,chubb2021general,darmawan2024optimaladaptationsurfacecodedecoders,tang2025phases,li2025perturbative,wan2025revisiting} and as intrinsic transitions in the decohered mixed state~\cite{fan2023diagnostics,bao2023mixed,lee2023quantum}.

Coherent error qualitatively enriches this picture. 
It can generate interference between error histories that are indistinguishable at the level of stabilizer syndromes. 
In this setting, the decodablity depends on the allowed recovery operations.
By allowing arbitrarily non-local or non-Pauli operations, it has been shown~\cite{bravyi_coherent,cheng2025emergent} that single-qubit unitary errors in stabilizer codes with odd code distance and even stabilizer weight can always be corrected in principle.  
However, the status of \emph{Pauli decoding}—a practically relevant decoding scheme, restricted to implementing a syndrome-conditioned transversal Pauli gate—remains to be fully investigated.

Recent work has revealed a richer landscape of decoding in topological codes under coherent errors~\cite{bravyi_coherent,venn2022coherent,cheng2025emergent,venn_entanglement_phases,behrends_general_x,phases_of_decodability_bao,lavasani2025stability}. 
In the two-dimensional surface code subject to single-qubit unitary rotations, Ref.~\cite{bravyi_coherent} first studied the decoding based on minimum-weight-perfect-matching, demonstrating an error threshold at a critical rotation angle.  
Subsequent studies of the same model by Refs.~\cite{venn2022coherent,venn_entanglement_phases} demonstrated connections between Pauli decoding and the physics of Anderson localization in two-dimensional fermionic systems with quenched disorder.
More recently, it has been shown that retaining quantum information in the post-measurement state need not imply practical decodability: under generic unitary errors, the surface code may lie in a regime where encoded information is preserved, yet efficient Pauli decoding is believed to be impossible~\cite{phases_of_decodability_bao}. 

In this work, we revisit the problem of Pauli decoding in the square-lattice surface code under single-qubit unitary rotations that only create electric anyons, which was originally studied in Refs.~\cite{bravyi_coherent,venn2022coherent,venn_entanglement_phases}.
We formulate a replica non-linear sigma model as the continuum field theory that governs the fidelity of Pauli decoding in the rotated surface code after syndrome measurements.
Our effective theory reveals universal features of error-correcting topological quantum matter in this setting.
We highlight key differences between optimal Pauli decoding, which uses the complete knowledge of the underlying coherent errors to determine the best Pauli recovery, and suboptimal decoding, a physically-relevant setting in which imperfect knowledge of the underlying error model further limits recovery. 
We show that the optimal and the suboptimal Pauli decoders are associated with distinct ``replica limits'' of the effective non-linear sigma model. 
Within this framework, we argue that on the square lattice, optimal decoding may remain possible up to the rotation angle that generates the maximum amount of coherence.
However, the approach to the decodable regime is controlled by an unusually slow renormalization group flow away from the fixed point at the maximally coherent rotation angle, obscuring the emergence of the decodable phase at accessible system sizes.
In contrast, a similar fixed point which arises in the replica limit associated with the suboptimal decoding is stable, giving rise to a non-correctable ``thermal metal'' phase for the suboptimal decoder and a phase transition when tuning the rotation angle.
We further develop quantitative predictions for the behavior of the decoding fidelity from the sigma model description. 

We note that closely related sigma models appear as effective theories of 1+1D monitored fermion dynamics~\cite{fava2023nonlinear,jian2022criticality,jian2023measurement,poboiko2023theory,fava2024monitored,poboiko2025measurement,guo2025field}, underscoring a broader connection between the universality of the decoding problem in two dimensions and that of monitored dynamics in 1+1D.

\subsection{Overview}
We begin with a detailed overview of the main results in this work and connections to previous results on the surface code subject to coherent errors~\cite{venn2022coherent,venn_entanglement_phases,behrends_general_x}. 
In this work, we investigate the square-lattice surface code subject to single-qubit unitary rotations $e^{\ri \theta_\ell Z_\ell}$ about the Pauli $Z$ axis, with rotation angles which are (i) spatially uniform or (ii) drawn from a Gaussian distribution independently for each qubit. 
We study the recovery of quantum information when decoding is restricted to transversal Pauli corrections. 

Decoding can be viewed as a statistical inference task: after a syndrome $s$ is measured with Born probability $\calQ_{s}$, the decoder selects a Pauli recovery operation, labeled by its homology class $\alpha$, intended to best restore the logical information. The decoder is thus characterized by its posterior belief $\calP_{\alpha|s}$, the probability it assigns to the underlying error strings belonging to class $\alpha$, given an observed syndrome $s$. If the Pauli recovery is sampled from this posterior distribution, then the decoding fidelity can be written as
\begin{align}\label{eq:intro_fidelity}
    \mathcal{F}=\sum_{s} \calQ_{s} \sum_{\alpha}\calP_{\alpha|s}\,\calQ_{\alpha|s} \,,
\end{align}
where $\calQ_{\alpha|s}$ is the true fidelity with the initial code state after applying the Pauli recovery associated with $(\alpha,s)$. This framework provides a clear distinction between optimal decoding—which corresponds to Bayesian consistency with the true error model $\calP_{\alpha|s}= \calQ_{\alpha |s} \equiv \calQ_{\alpha,s}/\sum_{\alpha}\calQ_{\alpha,s}$—and suboptimal decoding, which departs from this due to imperfect knowledge of the underlying error model, so that $\calP_{\alpha|s}\ne \calQ_{\alpha |s}$.

The decoding fidelity, the key figure of merit for error-correction, is a non-linear function of the quantum state after syndrome measurements.
As a result, an analytic treatment typically requires a replica construction—introducing $n$ identical copies of the post-measurement state—so that the fidelity can be re-cast as the expectation value of a linear observable in the replicated theory.
The physical fidelity $\calF$ in Eq.~\eqref{eq:intro_fidelity} is then obtained in an appropriate replica limit. 
This approach is familiar from studies of statistical mechanics with quenched disorder and is frequently used in the study of measurement-induced quantum many-body phenomena~\cite{bao2020theory,jian2020measurement,fisher2023random}.  

\begin{figure}
$\begin{array}{c}
\includegraphics[width=.43 \textwidth]{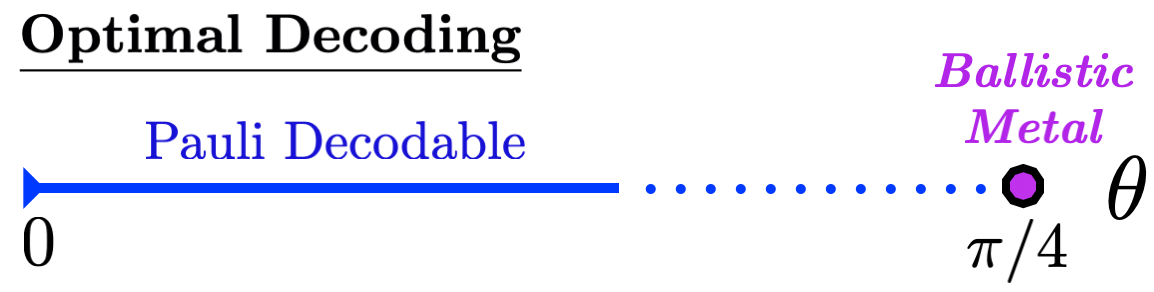}\\
\text{(a)}\\ \\ \\
\includegraphics[width=.425 \textwidth]{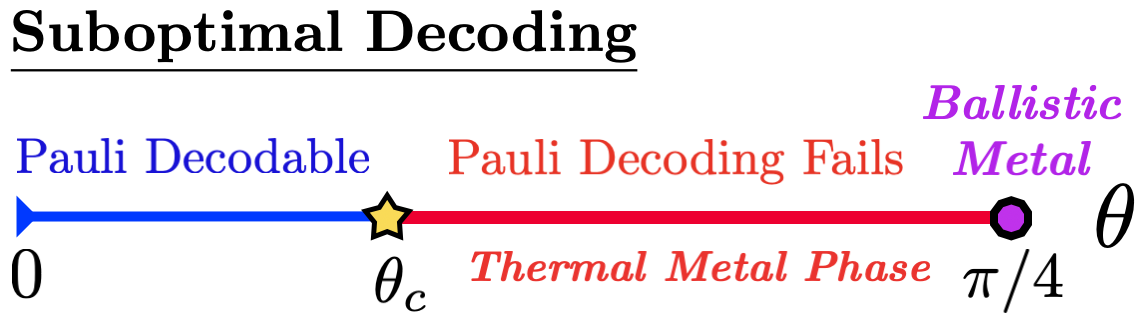}\\
\text{(b)}
\end{array}$
\caption{Proposed phase diagrams for optimal (a) and suboptimal (b) Pauli decoding in the square-lattice surface code subject to single-qubit unitary rotations with a uniform rotation angle $\theta$. For optimal decoding, the effective description near $\theta=\pi/4$ is close to an unstable metallic fixed point, resulting in a large crossover length-scale, after which we believe an asymptotic decodable phase is reached.}
\label{fig:phase_diagram}
\end{figure}

Our main finding is that for the square-lattice surface code in an appropriate regime of coherent error, the fidelity for both optimal and suboptimal Pauli decoding is described by correlations in a non-linear sigma model (NLsM).
In the continuum, the replica theory governing decoding is 
\begin{align}\label{eq:intro_nlsm}
\calS[Q]=-\frac{1}{2g_0}\int \rd x\, \rd t\;\mathrm{tr}\left(\nabla Q \cdot \nabla Q\right)
\end{align}
with the matrix-valued field $Q(x,t)\in \SO(2n)/\U(n)$.
We derive this NLsM microscopically, after identifying an $\O(2n)$ replica symmetry of the decoding problem.
A path-integral formulation yields saddle points which spontaneously break this $\O(2n)$ symmetry, leading to the effective action (\ref{eq:intro_nlsm})  describing fluctuations of the order-parameter field $Q$.
Optimal and suboptimal decoding are governed by the distinct replica limits $n\rightarrow 1$ and $n\rightarrow 0$, respectively.

A key consequence of the NLsM description of decoding is a sharp distinction between the phase structure of optimal and suboptimal Pauli decoding with coherent error.  The NLsM at weak coupling describes a stable fixed point of the renormalization group (RG) in the $n\rightarrow 0$ replica limit, but is known to be unstable in the $n\rightarrow 1$ limit~\cite{hikami_three_loop_beta,wegner_four_loop_beta}. Thus, the suboptimal decoder can, in principle, witness an additional ``phase'' of quantum error correction, with universal properties, which is absent in an optimal decoding scheme.

The stable phase of the NLsM in the $n\to 0$ replica limit can be understood by recalling that the same effective description (\ref{eq:intro_nlsm}) also arises in the study of disorder-induced localization in fermionic systems in two spatial dimensions—specifically, Bogoliubov quasiparticles in a spinless superconductor with broken time-reversal symmetry, i.e. symmetry class D in the Altland–Zirnbauer classification~\cite{AZ_10fold}. This connection becomes explicit through a mapping between the probability amplitude for syndrome measurements and the propagation amplitude of a Chalker–Coddington network model~\cite{chalker1988percolation}, as first noted in Ref.~\cite{venn2022coherent}. In symmetry class D, three phases generically arise: a trivial superconductor, a topological superconductor, and a ``thermal metal,'' named for its logarithmic growth of thermal conductance with system size $L$. In the replica limit $n\to0$, relevant both for localization and for understanding suboptimal decoding, this thermal metal is the stable phase described by the NLsM. %

The NLsM (\ref{eq:intro_nlsm}) thus describes suboptimal decoding within the thermal metal phase, and optimal decoding in the vicinity of the metallic fixed point: in the latter case, the system drifts away from the unstable fixed point at large scales.
In the simplest scenario, the $n \to 1$ NLsM flows to one of two stable insulating fixed points, which correspond to a decodable or a non-decodable phase.
We develop these predictions—and a quantitative account of Pauli decoding in and near the thermal metal—from the NLsM. 
For the error model with spatially uniform rotation angles, we propose a phase diagram for optimal and suboptimal decoding shown in Fig.~\ref{fig:phase_diagram} based on analytic arguments and large-scale numerical simulations.
In this case, we believe that optimal Pauli decoding is possible in the thermodynamic limit as long as $\theta < \pi/4$.  
The analytic argument for this is that the critical point separating a decodable from a non-decodable phase should be described by the NLsM (\ref{eq:intro_nlsm}) with a non-trivial topological $\Theta$-term~\cite{Bocquet_Class_D_Fermions_with_Random_Mass} ($\Theta = \pi$) in the $n\rightarrow 1$ replica limit.
This critical theory is known to have a statistical Kramers-Wannier duality symmetry~\cite{fava2023nonlinear,wang_self_dual_cho_fisher}, which is not present microscopically in our decoding problem for coherent rotation angle $\theta < \pi/4$. 

We now preview how the fidelity arises as an observable in the NLsM and its consequences for decoding in the surface code on a cylinder with circumference $L$, height $T$ (aspect ratio $\kappa = T/L$). 
For optimal decoding, the fidelity maps to expectation values of ``twist'' defects of the order-parameter field $Q$ inserted in the longitudinal direction of the cylinder; near the metallic fixed point, this expectation value is suppressed due to the large stiffness.
We also develop a dual formulation in which the decoding fidelity is related to the expectation values of twist defects inserted along the periodic direction of the cylinder.
The predictions in these two formulations allow us to quantitatively understand the optimal decoding fidelity in various parameter regimes presented in Sec.~\ref{sec:fidelity_predictions} [see Eq.~(\ref{eq:fid_kappa_greater_than_1/gR}), (\ref{eq:opt_decoding_kappa_large_ thermal_metal}), (\ref{eq:opt_fidelity_small_kappa_1}), (\ref{eq:opt_fidelity_kappa_smaller_1})]. 
The finite-size scaling of the optimal decoding fidelity is set by the renormalization group (RG) flow of the sigma-model coupling $g$, which is marginally relevant at the metallic fixed point. 
An important consequence of the marginal flow of the stiffness is a striking aspect ratio dependence of the optimal decoding fidelity near the metallic fixed point: the fidelity decreases with system size at a fixed $\kappa \ll 1$, but increases with size for $\kappa \gg 1$. 
This prediction is valid up to a scale at which the renormalized coupling becomes $\calO(1)$, and the system ultimately crosses over to a stable infrared phase.

For suboptimal decoding, by contrast, the metallic fixed point is stable and corresponds to a non-decodable phase. 
The suboptimal decoding fidelity can be formulated in terms of twist defects of $Q$, which also carry a ``domain wall'' that flips the sign of the Pfaffian of $Q$, forcing the field to interpolate between disconnected components of the target space.  
When the suboptimal decoder is ``close'' to decoding optimally, the decay of the fidelity with system size is controlled not only by the renormalized stiffness, but also by the scaling dimension of a relevant perturbation that breaks the enlarged replica symmetry for the optimal decoder to the reduced symmetry of the suboptimal decoder, which ultimately produces the sigma model description (\ref{eq:intro_nlsm}) with the replica limit $n\to 0$. 
When approaching the thermal metal fixed point, this analysis yields the asymptotic form for the fidelity in Eq.~\eqref{eq:suboptimal_fidelity}.

\begin{figure}[t]
\begin{tikzpicture}[baseline={([yshift=-.8ex]current bounding box.center)}, scale=0.7]
  \def\h{3}    %
  \def\r{1.5}  %
  \draw (0, \h) ellipse (1.5 and 0.5);
  \draw (0,0) [partial ellipse=180:360:1.5 and 0.5];
  \draw[dashed] (0,0) [partial ellipse=0:180:1.5 and 0.5];
  \draw (-\r,0) -- (-\r,\h);
  \draw (\r,0) -- (\r,\h);
  \draw[red,decorate, decoration={snake, amplitude=0.5mm, segment length=2mm}] (-0.474,-0.474) -- (-0.474,-0.474+\h);
  \draw[red,-Stealth] (0,\h/2) [partial ellipse=240:270:1.5 and 0.5];
  \node[red] at (-1.1,-0.474+\h/2+0.05) {$Q$};
  \node[red] at (0.6,-0.474+\h/2-0.05) {$\Lambda Q \Lambda$};
  \node at (-2.5,3) {(a)};
  \draw[-Stealth] (-1.8,0) -- (-1.8,2) node[left] {$t$};
  \draw[-Stealth] (0,0) [partial ellipse=210:330:1.9 and 0.7] node[right] {$x$};
\end{tikzpicture}
\quad
\begin{tikzpicture}[baseline={([yshift=-.8ex]current bounding box.center)}, scale=0.7]
  \def\h{3}    %
  \def\r{1.5}  %
  \draw (0, \h) ellipse (1.5 and 0.5);
  \draw (0,0) [partial ellipse=180:360:1.5 and 0.5];
  \draw[dashed] (0,0) [partial ellipse=0:180:1.5 and 0.5];
  \draw (-\r,0) -- (-\r,\h);
  \draw (\r,0) -- (\r,\h);
  \draw[red,decorate, decoration={snake, amplitude=0.5mm, segment length=2mm}] (0,2) [partial ellipse=180:360:1.5 and 0.5];
  \draw[dashed,red,decorate, decoration={snake, amplitude=0.5mm, segment length=2mm}] (0,2) [partial ellipse=0:180:1.5 and 0.5];
  \draw[-Stealth,red] (0,0.2+\h/2) node[above]{$Q$} -- (0,-0.4+\h/2) node[below]{$\Lambda Q \Lambda$};
  \draw[-Stealth] (-1.8,0) -- (-1.8,2) node[left] {$t$};
  \draw[-Stealth] (0,0) [partial ellipse=210:330:1.9 and 0.7] node[right] {$x$};
  \node at (-2.5,3) {(b)};
\end{tikzpicture}
\caption{The decoding fidelity is related to a twist of the order-parameter field $Q \rightarrow \Lambda Q\Lambda$.  In one picture, this twist is inserted along the open direction of the cylinder (a), while in a dual description, the twist is inserted along the compact direction of the cylinder (b).}
\label{fig:twist_intro}
\end{figure}
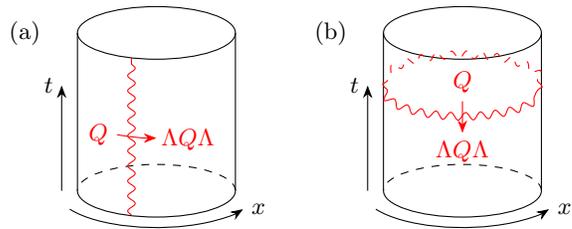

We corroborate these predictions using large-scale numerical simulations.
For the suboptimal decoder, we observe a non-decodable thermal metal phase. 
By simulating the associated Chalker-Coddington network model, we directly extract the logarithmic growth of the thermal conductance with a universal coefficient consistent with the known perturbative beta-function~\cite{evers2008anderson}. 
In contrast, our numerical study suggests that the thermal metal phase is absent in the network model associated with the optimal decoder.
Available numerical evidence further suggests that optimal Pauli decoding is likely possible in the thermodynamic limit across the parameter range $\theta < \pi/4$.
However, away from the metallic fixed point at $\theta = \pi/4$, the renormalization group flow is extremely slow, leading to an unusually large crossover length scale before the system ultimately flows to the fixed point describing a decodable phase.  
Right at $\theta = \pi/4$, the decoding problem has a mean free path that is strictly infinite in its network model description, leading to a fine-tuned ``ballistic metal'' that is not described by the non-linear sigma model.
We show that the decoding fidelity is an \emph{oscillating} function of system size and aspect ratio (see Sec.~\ref{sec:ballistic_metal}). 

The symmetry of the replicated partition function and the target space of the associated non-linear sigma model depend on the bipartiteness of the lattice that syndromes live on (see Sec.~\ref{sec:nlsm_opt_dual}).
In Sec.~\ref{sec:conclusion}, we discuss how the sigma model with target space $\SO(n)$ arises as the effective theory for decoding in the triangular lattice surface code with coherent rotation $e^{\ri \theta_\ell Z_\ell}$ creating anyon excitations that live on sites.  This sigma model also describes disordered fermions but in a distinct symmetry class DIII~\cite{evers2008anderson}.
In this case, the sigma model has a stable thermal metal phase in both replica limits $n \to 1$ and $n \to 0$~\cite{hikami_three_loop_beta} associated with the optimal and suboptimal decoders, respectively.

We will now explain the relationship of our results with that of Refs.~\cite{venn2022coherent,venn_entanglement_phases} which studied the decoding problem of the surface code under the same single-qubit unitary errors.
The decoding problem was first shown to be governed by a Chalker-Coddington network model with symmetry class D in Ref.~\cite{venn2022coherent}.
In that work, the authors argued for the existence of a metal-to-insulator transition based on the assumption that the decoding problem is governed by the same replica limit ($n \to 0$) as in the Anderson localization problem in class D.
In another paper by the same authors, they studied the decoding transition in a setup that belongs to what we call suboptimal decoding~\cite{venn_entanglement_phases}.
In their setup, the syndromes were drawn from the Born distribution associated with incoherent errors, while decoding was performed under the assumption that coherent errors had occurred.
A metal-to-insulator transition was observed with better numerical agreement with the analytic predictions compared to that in Ref.~\cite{venn2022coherent}.
According to our current understanding in this paper, a metal-to-insulator transition should only occur in the setup considered in Ref.~\cite{venn_entanglement_phases}, while what is observed in Ref.~\cite{venn2022coherent} may be attributed to pronounced finite-size effects in numerical simulations.

\subsection{Organization}
The rest of the paper is organized as follows.
First, in Sec.~\ref{sec:setup} we review the surface code with coherent errors and define the decoding problem of interest.
In Sec.~\ref{sec:stat_mech_of_decoding}, we review the statistical mechanics model for the decoding problem, as developed in Ref.~\cite{venn2022coherent,venn_entanglement_phases}.
In Sec.~\ref{sec:replica_decoding}, we formulate the decoding fidelity as the limit of a replica sequence, thereby laying the foundation for the analytic theory developed in the following sections.
Through an explicit microscopic derivation, we show that the replicated decoding fidelity is governed by a non-linear sigma model (NLsM) in Sec.~\ref{sec:nlsm}.
We present the resulting predictions for the decoding fidelity in Sec.~\ref{sec:fidelity_predictions} and other physical quantities in Sec.~\ref{sec:other_nlsm_predictions}.
We highlight the key differences between the optimal and suboptimal decoders, which we verify through numerical study.
For completeness, we discuss the decoding fidelity of the ballistic metal point $\theta = \pi/4$ in Sec.~\ref{sec:ballistic_metal}, which is not captured by the sigma model.
Finally, we conclude in Sec.~\ref{sec:conclusion}.

\tableofcontents

\section{Setup}\label{sec:setup}

\begin{figure}[t]
\centering
\includegraphics[width=.48 \textwidth]{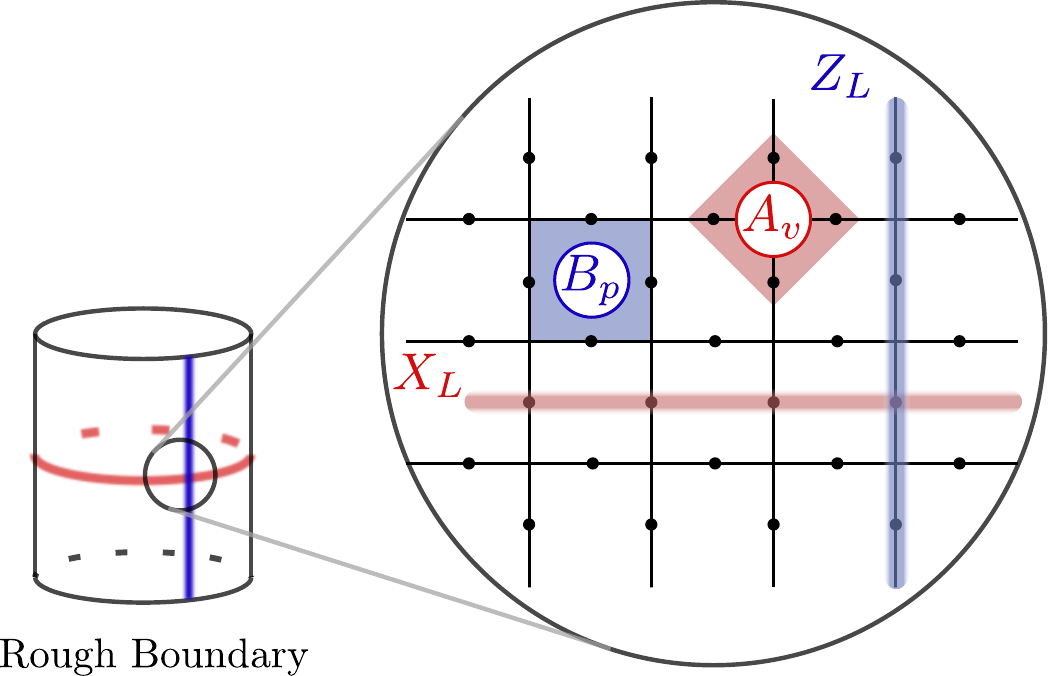}
\caption{
Square-lattice surface code on the cylinder of circumference $L$ and length $T$ terminated with rough boundaries at the top and bottom.
The code is defined with $X$-vertex stabilizers $A_v$ and $Z$-plaquette stabilizers $B_p$.
Logical $X_L$ and $Z_L$ operators traverse the horizontal and vertical directions, respectively, encoding one qubit of quantum information.
}
\label{fig:surface_code}
\end{figure}

We consider the two-dimensional surface code on the square-lattice of size $L \times T$, which involves qubits defined on links~\cite{kitaev2003fault,bravyi1998quantum}.
We focus on the cylindrical geometry with open (rough) boundary conditions in the vertical direction and periodic boundary conditions in the horizontal direction as illustrated Fig.~\ref{fig:surface_code}.
In the bulk of the cylinder, the surface code has stabilizers $A_v = \prod_{\ell \in v} X_\ell$ and $B_p = \prod_{\ell \in p} Z_\ell$ associated with each vertex $v$ and plaquette $p$, respectively.
At the top and bottom rough boundaries, we include the incomplete three-body operators associated with incomplete plaquettes as part of the stabilizer group.
The codespace, specified by the $+1$ eigenvalues of all the stabilizers, encodes one logical qubit of quantum information. 
The logical Pauli-Z (X) operator is a string operator $Z_L = \prod_{\ell \in \gamma} Z_\ell$ ($X_L =\prod_{\ell \in\tilde{\gamma}} X_\ell$) along a path $\gamma$ ($\tilde{\gamma}$) on the direct (dual) lattice spanning from one rough boundary to the other (wrapping around the cylinder).
The operators $X_L$ and $Z_L$ have minimal support $L$ and $T$, respectively.

The excitations in the surface code are specified by the stabilizers with $-1$ eigenvalues; we refer to the excitations with $A_v = -1$ and $B_p = -1$ as electric $e$ and magnetic $m$ anyons, respectively, which have bosonic self-statistics and semionic mutual statistics. 
In the decoding problem, the excitations detected by stabilizer measurements are called ``syndromes'' and provide information about the underlying errors that have occurred in the code.

\subsection{Coherent error model}
We consider the surface code subject to single-qubit Pauli-Z rotations~\cite{bravyi_coherent}, which create a coherent superposition of excited states with distinct configurations of $e$-anyons in the code state $|\psi_{0}\rangle$.
These errors can originate from imperfect state preparation, unintended unitary errors, or stabilizer measurements in a tilted basis. 
The resulting corrupted state is 
\begin{align}\label{eqn:rotated_surface_code_state}
\ket{\psi}=\prod_{\ell} e^{\ri\theta_\ell Z_\ell}\ket{\psi_0} \,,
\end{align}
where the rotation angle $\theta_\ell$ can have spatial dependence.

We consider two specific cases: 
\begin{itemize}
    \item[(1)]  $\theta_\ell = \theta$, which is spatially uniform in the system.
    \item[(2)] $\theta_\ell$ drawn from a Gaussian distribution 
    \begin{align}
        p(\theta_\ell) = \frac{1}{\sqrt{2 \pi g}}\exp\left[{-\frac{(\theta_\ell-\pi/4)^2}{2g}} \right]\,,\label{eq:angle_distribution}
    \end{align}
    with mean $\pi/4$ and variance $g$.
\end{itemize}
The second case is more amenable to analytic study, and allows us to provide a microscopic derivation of the effective non-linear sigma model (NLsM) of the decoding problem.
However, we do not believe that the universal predictions of the effective NLsM depend on how the coherent rotation angles are modeled, a claim we later verify through numerical simulation.

\subsection{Pauli decoding}
In this work, we focus on a decoding strategy that begins with measuring the stabilizers of the corrupted surface code state.
Based on the measurement outcomes, i.e. syndromes, we apply transversal Pauli operations to recover the encoded state.
We refer to this decoding strategy as \emph{Pauli decoding}.

This strategy is motivated by practical considerations. 
First, unitary rotations are always information-preserving and in principle can be undone by applying the inverse rotation.
However, the required operation is generically non-Clifford, which is difficult to benchmark and potentially costly to implement.  
Moreover, this approach also requires precise knowledge of the coherent rotation angle, while Pauli decoding does not.

We also note that there are specific settings in which the combination of coherent rotations followed by syndrome measurements always preserves the logical information. 
In the square-lattice surface code with odd code distance (i.e. logical operators have odd Pauli weights) subject to single-qubit unitary errors~\cite{bravyi_coherent}, the state after syndrome measurements is always related to the encoded state by a unitary rotation.
However, the recovery unitary is typically non-local, and its implementation requires a circuit of depth that scales linearly with the system size.
This process is complicated in practice and also undesirable for fault-tolerant information processing, as it may propagate  potential remaining errors through the state.

\subsection{Decoding algorithms}\label{sec:decoders}
When considering a recovery scheme based on Pauli operations,  maximum-likelihood (ML) decoding can produce an optimal decoding fidelity~\cite{niwa_coherent_information_css,hauser2026information}.
For an observed syndrome configuration $s$, the error strings $\calC$ compatible with the syndrome $s$, i.e. $\partial \calC = s$, can be divided into several homological classes.
The ML decoder evaluates the total probability $\calQ_{\alpha,s}$ of all the error strings in the same homological class $\alpha$ and chooses a recovery string belonging to the class with the highest probability, resulting in a decoding fidelity
\begin{align}\label{eqn:def_ML}
    \calF_{\text{ML}} = \sum_s \calQ_s \frac{\max_\alpha \calQ_{\alpha,s}}{\sum_\alpha \calQ_{\alpha,s}} = \sum_s \max_\alpha \calQ_{\alpha,s}
    \,,
\end{align}
where $\calQ_s = \sum_\alpha \calQ_{\alpha,s}$.
We will primarily consider another asymptotically optimal decoder—the ``probabilistic'' decoder—whose behavior is more amenable to analytic study. 
In this decoder, the homology class $\alpha$ of the applied recovery string is randomly selected according to the probability $\calQ_{\alpha | s} = \calQ_{\alpha, s}/\calQ_s$.
This results in a decoding fidelity
\begin{align}\label{eqn:prob_decoder}
    \calF_\mathrm{opt} = \sum_s \calQ_s \sum_\alpha \left( \frac{\calQ_{\alpha,s}}{\calQ_s}\right)^2 = \sum_{s,\alpha} \frac{\calQ_{\alpha,s}^2}{\calQ_s} \,.
\end{align}
We show in Appendix~\ref{app:performance_prob_decoder} that the probabilistic decoder has the same decoding threshold as that of the ML decoder.

So far, we assume that the decoder has perfect knowledge of the underlying error model; that is, the error model used to infer the probability for each homology class is the same as the one that generates the syndromes. Realistically, the coherent rotation angle that produces the syndromes may not be known, and in this case, the decoder may use a different distribution $\calP_{\alpha,s}\ne \calQ_{\alpha,s}$ to model its observations.
This results in a \emph{suboptimal decoder} which exhibits a decoding fidelity
\begin{align}\label{eq:subopt_prob_decoder}
    \calF_\mathrm{sub} = \sum_s \calQ_s \sum_\alpha \frac{\calQ_{\alpha,s} \calP_{\alpha,s}}{\calQ_s \calP_s} \,.
\end{align}
We will focus on a specific suboptimal decoder in which the decoder has an incorrect estimate of the rotation angle $\theta_\ell$.
In this case, the syndromes are generated from the code with coherent error $e^{\ri \theta_\ell Z_\ell}$, while the decoder assumes the error is given by $e^{\ri \theta'_\ell Z_\ell}$ with 
\begin{align}
\frac{\pi}{4} - \theta_\ell = (1+\epsilon)\left(\frac{\pi}{4} - \theta'_\ell\right) \,,\label{eq:model_suboptimal_decoder}
\end{align}
where $\epsilon$ is a small number assumed to be uniform in the system and parametrizes the estimation error.
This model is chosen for concreteness; we expect that the qualitative features of the suboptimal decoder are independent of the precise relation between the estimated rotation angle $\theta'$ and  $\theta$.

\section{Statistical mechanics of decoding}\label{sec:stat_mech_of_decoding}
In this section, we review the statistical mechanical description of the decoding fidelity in the surface code with coherent rotations developed in Ref.~\cite{venn2022coherent}.
Specifically, the fidelity is governed by the random-bond Ising model (RBIM) with complex couplings, which also has an equivalent formulation as a Chalker-Coddington network model~\cite{chalker1988percolation} in class D~\cite{venn2022coherent}.

Without loss of generality, we consider the surface code initialized in the logical-X eigenstate $\ket{+}_L$ and corrupted by coherent rotations,
\begin{align}
\ket{\psi} &= \prod_\ell e^{\ri\theta Z_\ell} \ket{\Psi_0} \sim \sum_{\calC_z} (\ri \tan\theta_\ell)^{\abs{\calC_z}} Z^{\calC_z} \ket{+}_L,\label{eq:error_conf_expansion}
\end{align}
where $\calC_z$ is a binary vector that labels a chain of $Z$ errors, $|\calC_z|$ its Hamming weight, and we have neglected the overall normalization of the state in the final expression. 

The central quantity of optimal decoding is the coset probability $\calQ_{\alpha,\,s}$, i.e., the total probability of error strings in homology class $\alpha$ and compatible with the observed syndrome $s$. 
For the surface code on the cylinder, the compatible error strings fall into two homologically inequivalent classes $\alpha = 0, 1$.
The probability of error chains in class $\alpha$ is given by $\calQ_{\alpha,\, s} = \abs{\braket{\psi_{\alpha,\,s}}{\psi}}^2$, where $\ket{\psi_{\alpha,\,s}} = Z^{\calC^\mathrm{ref}_{z,\,\alpha}}\ket{+}_L$ is a state with syndrome $s$ created by a fixed reference string $\calC^\mathrm{ref}_{z,\,\alpha}$ in class $\alpha$.
This probability can be expressed as an expansion in error configurations $\calC_z$,
\begin{align}\label{eqn:syndrome_probability}
    \calQ_{\alpha,\, s} = \abs{\calZ_{\alpha,\,s}}^2, \quad \calZ_{\alpha,\,s}
    := \sum_{\calC_z}\!{}' \prod_\ell (\ri \tan\theta)^{\calC_{z,\ell}},
\end{align}
where $\sum'$ represents a constrained summation over error configurations $\calC_z$ that are compatible with the syndromes and belong to class $\alpha$.

The error chains $\calC_z + \calC_{z,\alpha}^{\mathrm{ref}}$ consist of closed, topologically trivial loops, and are naturally thought of as the fluctuating domain walls in an Ising magnet~\cite{dennis2002topological}.  
To make this connection precise, we introduce Ising variables $\sigma_\bfr$ on the plaquette $\bfr$ and bond variables $\eta_{\bfr\bfr'}$ to represent the error $\calC_z$ and reference string $\calC^\mathrm{ref}_{z,\, \alpha}$ as
\begin{align}\label{eqn:low_T_ising_spin_def}
    (\calC_z)_{\bfr\bfr'} = \frac{1 - \eta_{\bfr\bfr'} \sigma_\bfr \sigma_{\bfr'}}{2} \,,\;\;
    (\calC^\mathrm{ref}_z)_{\bfr\bfr'} = \frac{1 - \eta_{\bfr\bfr'}}{2} \,. 
\end{align}
With these definitions $\calZ_{\alpha,\,s}$ becomes
\begin{equation}\label{eqn:single_replica_complex_rbim}
    \calZ_{\alpha,\, s} = 
    \sum_{\sigma}
    e^{\sum_{\langle \bfr,\, \bfr'\rangle}\left(J_{\bfr\bfr'} - \frac{\ri\pi}{4} \right) \eta_{\bfr\bfr'} \sigma_\bfr \sigma_{\bfr'}} \,,
\end{equation}
with $ J_{\bfr\bfr'} = (1/2)\ln(1/\tan\theta_{\bfr\bfr'})$.
We note that decoding in the surface code under incoherent error is also governed by the classical statistical mechanics of an RBIM~\cite{dennis2002topological}, however, the probability $\calQ_{\alpha,\,s}$ maps to one copy of the partition function and differs from the case of coherent errors.

The partition sum $\calZ_{\alpha,\, s}$ can also be rewritten using transfer matrices acting on Majorana fermions, after performing a Jordan-Wigner transformation. 
This approach provides a starting point for large-scale numerical studies of the decoding problem and for deriving the effective field theory in Sec.~\ref{sec:nlsm} which governs optimal and suboptimal decoding.  

To begin, we write the partition sum in terms of transfer matrices associated with each row
\begin{equation}\label{eqn:single_copy_rbim_prob_amplitude}
    \mathcal{Z}_{\alpha,\, s} = \rbra{+}^{\otimes L} \hat{\mathsf H}_{T}\hat{\mathsf T}_{T-1} \hat{\mathsf T}_{T-2} \cdots \hat{\mathsf T}_1 \rket{+}^{\otimes L} 
    \,,
\end{equation}
where $\hat{\mathsf T}_{j} = \hat{\mathsf V}_{j} \hat{\mathsf H}_{j}$ is the many-body transfer matrix of the $j$-th row with $\hat{\mathsf V}_{j} = \bigotimes_{i=1}^{L} \hat{\sfv}_{\bfr,\, \bfr+\hat{e}_t}$ and $\hat{\mathsf H}_{j} = \bigotimes_{i=1}^{L}\hat{\sfh}_{\bfr,\,\bfr+\hat{e}_x}$ with $\bfr = (i, j)$.
We use $|\cdot)$ to denote the states in the Hilbert space on which the transfer matrix acts, with the partition sum given by the transition amplitude between the initial and final state $|+)^{\otimes L}$.
We consider a system on a cylinder with circumference $L$ and length $T$.
The transfer matrices $\hat{\sfv}_{\bfr,\,\bfr+\hat{e}_t}$ and $\hat{\sfh}_{\bfr,\,\bfr+\hat{e}_x}$, associated with each link, take the form
\begin{align}\label{eqn:complex_coupling_RBIM_transfer_matrix_spin_representation}
    \hat{\mathsf v}_{\bfr,\,\bfr+\hat{e}_t} &= e^{\ri\theta_{\bfr,\,\bfr+\hat{e}_t}\hat\sigma^x_{i} - \ri \frac{\pi}{4}(1-\eta_{\bfr,\,\bfr+\hat{e}_t})(1-\hat\sigma^x_{i})} \,, \\
    \hat{\mathsf h}_{\bfr,\,\bfr+\hat{e}_x} &= e^{(J_{\bfr,\,\bfr+\hat{e}_x}-\frac{\ri\pi}{4})\eta_{\bfr,\,\bfr+\hat{e}_x}\hat\sigma^z_{i} \hat\sigma^z_{i+1}} \,.
\end{align}
After a Jordan-Wigner transformation, each transfer matrix describes the Gaussian (free) evolution of $2L$ Majorana fermions ($\sigma^{x}_{i}\rightarrow \ri\gamma_{2i-1}\gamma_{2i}$, $\sigma^{z}_{i}\sigma^{z}_{i+1}\rightarrow \ri\gamma_{2i}\gamma_{2i+1}$). 

\begin{figure}
\includegraphics[width=.5 \textwidth]{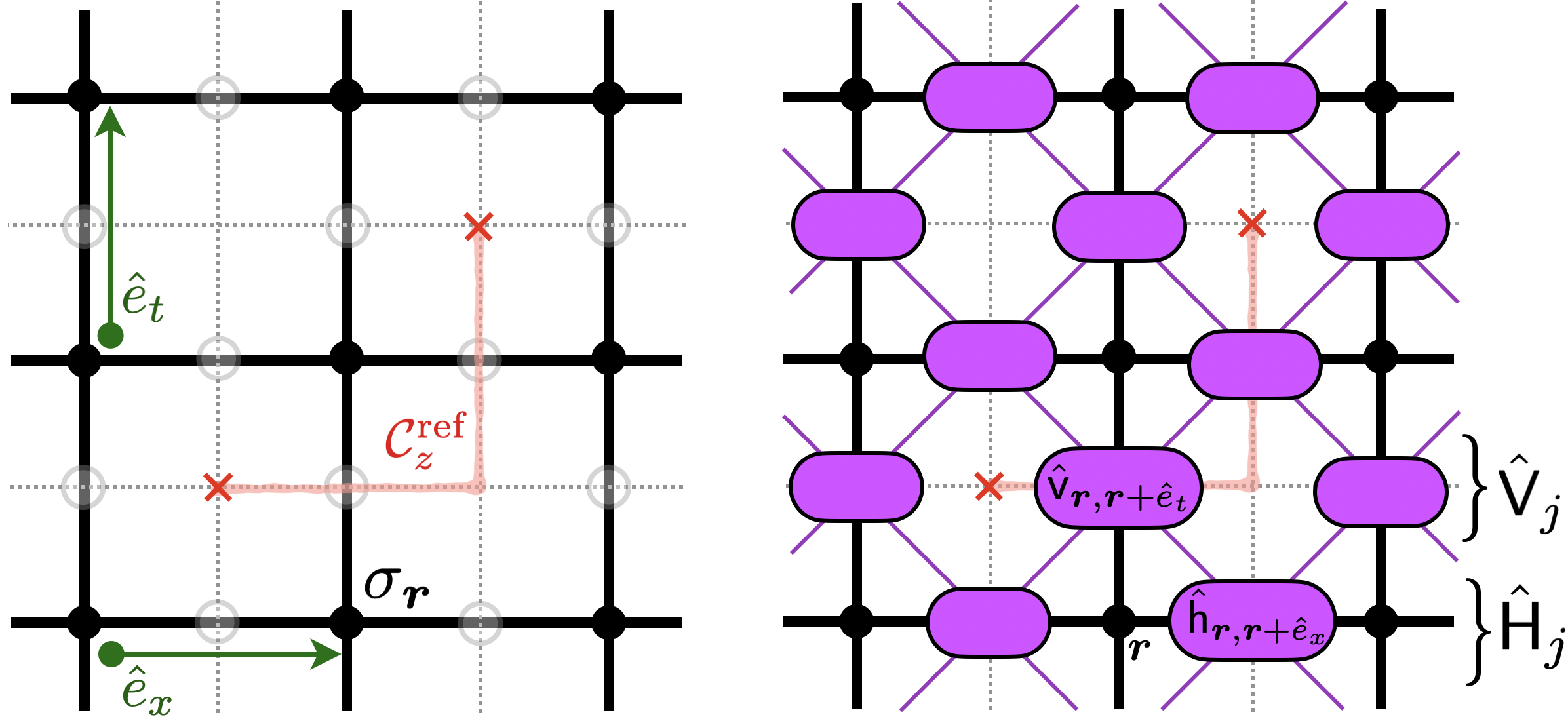} 
\caption{The partition sum $\calZ_{\alpha,s}$ (\ref{eqn:single_replica_complex_rbim}) is defined for Ising spins on the dual square lattice of the surface code.  A reference error string $\calC_{z}^{\mathrm{ref}}$ for a syndrome intersects bonds of this lattice, and fixes a configuration of the Ising interactions $\eta_{\boldsymbol{r}\boldsymbol{r}'}$.  On the right, this partition sum is represented as a product of transfer matrices, as in Eq. (\ref{eqn:single_copy_rbim_prob_amplitude}).
}
\label{fig:RBIM}
\end{figure}

It is convenient to understand the action of the transfer matrices as linear maps on the $2L$ Majorana modes, $\hat{\mathsf T} \gamma_\alpha \hat{\mathsf T}^{-1} = \mathsf{T}_{\alpha\beta}\,\gamma_\beta$.
In particular, the transfer matrix at each node takes the form
\begin{equation}\label{eqn:network_model_node_transfer_matrices_form}
\begin{aligned}
    \sfh_{\bfr,\,\bfr+\hat{e}_x} &= \frac{\ri}{\sin2\theta} \begin{pmatrix}
        -\cos2\theta & -\ri\eta \\
        \ri\eta & -\cos2\theta
    \end{pmatrix}\,,\\
    \sfv_{\bfr,\,\bfr+\hat{e}_t} &= \eta\begin{pmatrix}
        \cos2\theta & -\sin2\theta \\
        \sin2\theta & \cos2\theta
    \end{pmatrix}\,.
\end{aligned}
\end{equation}
Here, $\eta$ and $\theta$ are associated with the spacetime location, and we suppress the subscript for simplicity.
The row transfer matrices acting on single-particle modes is a $2 L \times 2 L$ matrix, and is given by $\mathsf{V}_{j} = \bigoplus_{i=1}^{L} \mathsf{v}_{\bfr,\,\bfr+\hat{e}_t}$ and $\mathsf{H}_{j} = \bigoplus_{i=1}^{L} \sfh_{\bfr,\,\bfr+\hat{e}_x}$.

Furthermore, the single-particle transfer matrices at each node can be written as unitary scattering matrices from incoming to outgoing modes.
Here, $\sfv^\dagger \sfv = \mathds{1}$ is unitary from bottom to top, and $\sfh^\dagger \sigma^z \sfh = - \sigma^z$ is unitary from left to right, as indicated in Fig.~1 of Ref.~\cite{venn2022coherent}.
The transfer matrix dynamics is thus equivalent to the propagation amplitude in a Chalker-Coddington network model.\footnote{We note that the mapping to the network model only works for the system with periodic boundary conditions and an even circumference $L$. However, we expect the universal features of the decoding problem in the thermodynamic limit to be independent of these details.}
The transfer matrix preserves only particle-hole symmetry, describing a network model in Altland-Zirnbauer symmetry class D~\cite{AZ_10fold}.

Finally, we comment that when the rotation angle $\theta < \pi / 4$, the vertical transfer matrix $\hat{\sfv}$ is purely unitary while $\hat{\sfh}$ involves imaginary-time evolution, violating the Kramers-Wannier symmetry that acts by $\gamma_{2i} \mapsto \gamma_{2i+1}$ in the fermion description.
On the other hand, we show in Appendix.~\ref{app:kw_duality} that exactly at $\theta = \pi / 4$, the coset probability $\calQ_{\alpha, \,s} = |\calZ_{\alpha,\, s}|^2$ that involves both the bra and ket transforms into $\calQ_{\alpha, \, s'}$ for a syndrome configuration $s'$ equally probable as $s$.
Thus, $\calQ_{\alpha,\,s}$ has the statistical Kramers-Wannier duality symmetry discussed in Ref.~\cite{wang_self_dual_cho_fisher}.

\section{Replica theory of decoding}\label{sec:replica_decoding}
An analysis of the averaged decoding fidelities in Eqs.~\eqref{eqn:prob_decoder} and~\eqref{eq:subopt_prob_decoder} is challenging because the fidelities are non-linear functions of the syndrome probability distributions.  
To this end, we develop a replica theory for the decoding fidelities of the optimal and suboptimal decoders, analogous to standard replica sequences used to study the physics of quenched disorder.
The replicated fidelity is then expressed in terms of the expectation values of symmetry defect insertions in the partition functions of the replicated complex RBIM (Sec.~\ref{sec:replica_decoding_RBIM}).
We also develop an alternative formulation of the replicated fidelity in the stat-mech model dual to the replicated complex RBIM (Sec.~\ref{sec:replica_decoding_dual}).

\subsection{Fidelity in the RBIM picture}\label{sec:replica_decoding_RBIM}
We first formulate the fidelities as the limits of replica sequences that will be amenable to analytic study in Sec.~\ref{sec:nlsm}. 
Specifically, we consider a replica sequence for the optimal decoding fidelity $\calF_\opt$ in Eq.~\eqref{eqn:prob_decoder} as
\begin{align}\label{eqn:n_rep_fidelity}
    \calF^{(n)}_\opt = \frac{\sum_{s} \left(\sum_\alpha \calQ_{\alpha,\,s}^2 \right)\calQ_s^{n-1}}{\sum_{s} \calQ_{s}^{n+1}} \,,
\end{align}
recovering the desired fidelity in the replica limit $n \to 0$, i.e., $\calF_\opt = \lim_{n \to 0} \calF_\opt^{(n)}$.

To express the replicated fidelity $\calF^{(n)}_\opt$ in an informative way, we first introduce the disordered averaged partition function of $2n+2$ copies of the RBIM
\begin{align}
\mathbf{Z}_{0} := \sum_{s,\alpha} \calQ_{\alpha,s}^{n+1} &= \sum_{s,\alpha} \calZ_{\alpha,s}^{n+1}\calZ_{\alpha,s}^{*n+1} \\ &\propto \sum_{\eta} \abs{\calZ(\eta)}^{2n+2} \,.\label{eq:replicated_partition_rbim}
\end{align}
In the last line, we have used the fact that $\calZ_{\alpha,\,s} = \calZ(\eta)$ depends only on the random bond configuration $\eta$ \eqref{eqn:low_T_ising_spin_def} in the complex RBIM, such that the summation over $s$ and $\alpha$ may be replaced by the summation over $\eta$ up to an overall prefactor.

We further introduce the partition function 
\begin{equation}
\begin{aligned}
    \bfZ_{2k} :=& \sum_s \calQ_{0,\,s}^{n+1-k} \calQ_{1,\,s}^k + \calQ_{1,\,s}^{n+1-k} \calQ_{0,\,s}^k \\
    \propto& \sum_{\eta} \abs{\calZ(\eta \, \zeta)}^{2k} \, \abs{\calZ(\eta)}^{2n+2-2k} \,,
\end{aligned}
\end{equation}
where $\zeta$ is a collection of Ising bond variables and takes the value $\zeta_{ij} = -1$ along a path in the longitudinal direction connecting two boundaries.
This partition function is related to $\bfZ_0$ by flipping the sign of random bond coupling $\eta$ along a longitudinal path in $2k$ copies of the RBIM, equivalent to imposing anti-periodic boundary conditions.

In this way, one can express the replicated fidelity in Eq.~\eqref{eqn:n_rep_fidelity} as
\begin{align}\label{eqn:rbim_replicated_opt_fidelity}
\calF_\opt^{(n)} = \frac{2\sum_{k=0}^{n-1}\binom{n-1}{k} \Phi_{2k}}{\sum_{k=0}^{n+1} \binom{n+1}{k} \Phi_{2k}}\,,\quad 
\Phi_{2k} := \frac{\bfZ_{2k}}{\bfZ_0}\,,
\end{align}
where $\Phi_{2k}$ is naturally thought of as the expectation value of an appropriate \emph{symmetry defect} in the replica theory.  We will argue for a coarse-grained description of this quantity in Sec.~\ref{sec:nlsm}. 

Similarly, we can formulate the fidelity of the suboptimal decoder as the replica limit $n \to 0$ of the following sequence
\begin{align}
\calF_\subopt^{(n)} = \frac{\sum_{s} \left(\sum_\alpha \calQ_{\alpha,s}\calP_{\alpha,s}  \right)\calP^{n-1}_s}{\sum_s \calQ_{s}\calP^n_{s}}\,.
\end{align}
Here, we again introduce the replicated partition functions in the presence of $2k$ symmetry defect insertions
\begin{align}
    \bfY_{2k} :=& \sum_s \calQ_{0,\,s} \calP^{n-k}_{0,\,s}\calP^k_{1,\,s} + \calQ_{1,\,s} \calP^{n-k}_{1,\,s}\calP^k_{0,\,s} \nn \\
    \propto&\sum_{\eta} \abs{\calZ(\eta)}^2 \abs{\calY(\eta)}^{2n-2k} \abs{\calY(\eta \,\zeta)}^{2k}\,,
\end{align}
which allows expressing the decoding fidelity as
\begin{gather}\label{eqn:rbim_replicated_subopt_fidelity}
{\calF}_\subopt^{(n)} = \frac{\sum_{k=0}^{n-1}\binom{n-1}{k}\Psi_{2k}}{\sum_{k=0}^{n}\binom{n}{k}\Psi_{2k}}\,, 
\quad \Psi_{2k} := \frac{\bfY_{2k}}{\bfY_0} \,.
\end{gather}

\subsection{Fidelity in the dual picture}\label{sec:replica_decoding_dual}
Alternatively, the replicated fidelities $\calF^{(n)}_\opt$ and $\calF^{(n)}_\subopt$ can be expressed in terms of the partition functions of the stat-mech model dual to the replicated RBIM.

Here, we identify the error configuration expansion of $\calQ_{\alpha,s}$ in Eq.~\eqref{eqn:syndrome_probability} with the high temperature expansion of a statistical mechanical model.
Specifically, we consider the amplitude
\begin{align}
\calZ_{\pm,\, s} := \calZ_{0,\,s} \pm \calZ_{1,\,s} \,,
\end{align}
and introduce Ising variables $\tau_i$ on the vertices of the original square lattice such that
\begin{align}\label{eqn:explicit_dual_partition_function}
\mathcal{Z}_{+,\,s} &= \sum_{\tau} e^{\sum_{\langle \bfr,\bfr'\rangle}\ri\theta_{\bfr\bfr'} \,\tau_\bfr\tau_{\bfr'} + \sum_{\bfr \in \partial} \ri\theta_\bfr \tau_\bfr} \prod_\bfr \tau_\bfr^{s_\bfr}\,,
\end{align}
where $\partial$ denotes the boundaries of the cylinder.
We note that since the error strings can terminate on the top and the bottom boundaries, the partition sum involves an additional boundary term, acting effectively as a $\bbZ_2$ symmetry-breaking field on the boundary spins.
The syndromes $s$ ``source'' the error chains at specific locations, which correspond to operator insertions in the partition sum. 

The partition sum $\mathcal{Z}_{-,\,s}$ takes a similar form.
The only difference is that homologically inequivalent error strings acquire a relative minus sign in the partition sum~\cite{aasen_mong_fendley_ising}.
One can incorporate this effect by flipping the sign of the coupling, i.e., $\theta \mapsto -\theta$, along a path that wraps around the periodic direction of the cylinder.

Next, we express the fidelity in terms of the replicated partition sums $\calZ_{\pm,\,s}$.
We identify the averaged partition sum with the partition function of a stat-mech model
\begin{align}
    \tilde \bfZ_0 := \sum_s \abs{\calZ_{+,\,s}}^{2n+2}\,. \label{eq:replicated_partition_dual}
\end{align}
We introduce the partition functions related to $\tilde\bfZ^{(2n)}_0$ by the insertions of symmetry defects
\begin{align}
    \tilde \bfZ_{2k} := \sum_s \abs{\calZ_{-,\,s}}^{2k}\abs{\calZ_{+,\,s}}^{2n+2-2k}\label{eq:dual_partition_func_with_defect}
    \,.
\end{align}

The replicated fidelity can be expressed in terms of the expectation values of defect insertions, i.e. the ratio between partition functions with and without symmetry defects
\begin{align}\label{eqn:dual_replicated_opt_fidelity}
{\calF}_\opt^{(n)} =\frac{1}{2} + 2\frac{\sum_{k=0}^{n-1} \binom{n-1}{k} \tilde\Phi_{2k+2}}{\sum_{k=0}^{n+1}\binom{n+1}{k}\tilde\Phi_{2k}}
\,, 
\quad \tilde\Phi_{2k} := \frac{\tilde\bfZ_{2k}}{\tilde\bfZ_0} \,.
\end{align}

For the suboptimal decoder, the averaged partition function in the presence of $2l$ defects in the first $2$ copies and $2k$ defects in the next $2n$ copies takes the form
\begin{equation}
    \tilde{\bfY}_{2l,\, 2k} :=  \sum_s \abs{\calZ_{+,\,s}}^{2-2l}\abs{\calZ_{-,\,s}}^{2l} \abs{\calY_{+,\,s}}^{2n-2k}\abs{\calY_{-,\,s}}^{2k} \,.
\end{equation}
Additionally, we define the averaged partition function with an odd number of defects in both the first two and the next $2n$ copies
\begin{equation}
\begin{aligned}
\tilde{\bfY}^{(R,\,R)}_{1,\,2k+1} &:=  \sum_s \calZ_{+,\,s}\calZ_{-,\,s}^* \calY_{+,\,s}\calY^*_{-,\,s} \abs{\calY_{+,\,s}}^{2n-2k-2}\abs{\calY_{-,\,s}}^{2k}
\,, \\
\tilde{\bfY}^{(R,\,L)}_{1,\,2k+1} &:=  \sum_s \calZ_{+,\,s}\calZ_{-,\,s}^* \calY_{-,\,s}\calY^*_{+,\,s} \abs{\calY_{+,\,s}}^{2n-2k-2}\abs{\calY_{-,\,s}}^{2k}
\,. 
\end{aligned}
\end{equation}
The two expressions for $\tilde{\bfY}^{(R,\, R)}$ and $\tilde{\bfY}^{(R,\,L)}$ differ by whether the defect has been inserted in the copy with ``even'' (corresponding to the ``bra'') or ``odd'' (corresponding to the ``ket'') replica index.

The replicated fidelity can be expressed as
\begin{align}\label{eqn:dual_replicated_subopt_fidelity}
    {\calF}_\subopt^{(n)}  =\frac{1}{2} + \frac{\sum_{k=0}^{n-1} \binom{n-1}{k}\left( \tilde\Psi^{(R,\, R)}_{1,\,2k+1} + \tilde\Psi^{(R,\,L)}_{1,\,2k+1} + \text{c.c.}\right)}{2\sum_{k=0}^{n}\binom{n}{k}\left(\tilde\Psi_{0,\,2k} + \tilde\Psi_{2,\,2k}\right)} \,,
\end{align}
where the twist expectation values are defined as
\begin{align}\label{eqn:dual_suboptimal_twist_expectation}
\tilde\Psi_{l,\,k}&  := \tilde{\bfY}_{l,\,k} / {\tilde{\bfY}_{0,\,0} }
\,.
\end{align}
The superscripts $L$ and $R$ denote whether an additional copy of $\calZ$ or $\calZ^*$ (also $\calY$ and $\calY^*$) is twisted, respectively, when $k$ ($l$) is odd.

Before proceeding, we remark on a crucial distinction between the $n$-th replicated partition functions $\bfZ_0$ and $\bfY_0$ for the optimal and suboptimal decoding.
Both of these quantities involve $n+1$ copies of the probability distribution $\calQ_{\alpha,\,s}$ and $\calP_{\alpha,\,s}$.
In optimal decoding, since the estimated distribution is the same as the true distribution $\calQ_{\alpha,\,s}$ of the syndromes, the replicated partition function $\bfZ_0$ exhibits an enlarged permutation symmetry $S_{n+1}$ over $n+1$ copies of the distribution, instead of the $S_n$ symmetry for $\bfY_0$.
The enlarged symmetry is a feature of the Bayesian inference problem and is linked to the Nishimori condition in the statistical mechanics problem that governs optimal decoding in the surface code subject to incoherent errors~\cite{le1988location,dennis2002topological,fan2023diagnostics}.
In the literature, to highlight this distinction, the optimal and suboptimal decoders are often associated with distinct replica limits $n \to 1$ and $n \to 0$, respectively.
In this work, we formulate the fidelities of both decoders in the $n \to 0$ limit and develop distinct effective theories for $\bfZ_0$ and $\bfY_0$.
Note that, due to the free fermion nature of our problem, the permutation symmetries of $\bfZ_0$ and $\bfY_0$ are enhanced to continuous symmetries.
However, they remain crucially distinct as shown in Sec.~\ref{sec:nlsm}.

\section{Effective non-linear sigma model}\label{sec:nlsm}
In this section, we derive a non-linear sigma model (NLsM) as an effective description for the replicated partition sums $\bfZ_0$ and $\bfY_0$ for the optimal and suboptimal decoding at a sufficiently large $\theta$ near $\pi/4$.  
Specifically, the replicated partition sum $\bfZ_0$, which governs optimal decoding, takes the form
\begin{align}
    \mathbf{Z}_{0} &= \int \calD Q \, 
    \exp \left( -\mathcal{S}_{\mathrm{eff}}[Q] \right)\,,\\
    \mathcal{S}_{\mathrm{eff}}[Q] &= -\frac{1}{2g_0}\int\rd^2 x \, \tr (\nabla Q)^2\,.
\end{align}
Here, the action exhibits an $\SO(2n+2)$ rotational symmetry.
The field $Q$ is a real orthogonal, anti-symmetric $(2n+2) \times (2n+2)$ matrix, which lives in the target space $\Gamma_{n+1} := \SO(2n+2)/\U(n+1)$.

In contrast, the replicated partition sum describing suboptimal decoding $\bfY_0$ takes the form
\begin{align}
\mathbf{Y}_{0} &= \int \calD Q \, 
    \exp \left( -\mathcal{S}_{\epsilon,\eff}[Q]\right)\,,\\
    \mathcal{S}_{\epsilon,\eff}[Q] &= -\frac{1}{2g_0}\int\rd^2 x \, \tr (\nabla Q)^2 + \epsilon^2 \int \rd^{2}x \,V[Q]\,,
\end{align}
with $Q\in \Gamma_{n+1}$, and a potential $V[Q]$ which explicitly breaks the $\SO(2n+2)$ symmetry of the action down to the subgroup $\SO(2)\times \SO(2n)$. 
The potential has a bare strength $\epsilon^2$ set by the proximity of the estimated rotation angle to its true value.
This potential is relevant under coarse-graining, and the effective theory at large scales is the NLsM with target space $\SO(2n)/\U(n)$.

Consequently, the replica limit $n \to 0$ for both decoders manifests as distinct limits $n \to 1$ and $n \to 0$ for the sigma model with target space $\Gamma_n := \SO(2n)/\U(n)$, leading to the distinct phase diagrams discussed in Sec.~\ref{sec:pdiagram_of_nlsm}.

We note that the same non-linear sigma model has been proposed as an effective description for the Chalker-Coddington network model in class D~\cite{zirnbauer1996riemannian,ludwig2016topological,ryu2010topological,gruzberg2001random,jian2022criticality}.
We here provide a microscopic derivation of the NLsM, which allows us to concretely analyze the physical quantities of interest in the effective theory in later sections (Sec.~\ref{sec:fidelity_predictions} and~\ref{sec:other_nlsm_predictions}).

Before proceeding, we provide an overview of our derivation and clarify the regime in which the effective NLsM is valid.
We start with the description of $\calZ_{\alpha,\,s}$ as the transition amplitude in the Chalker-Coddington network model (Sec.~\ref{sec:stat_mech_of_decoding}).
The network model has a special point at $\theta = \pi/4$, where the transfer matrix at each node is unitary in both spatial directions and acts as a SWAP gate on Majorana modes, $\hat{\sfT}\gamma_{2j-1}\hat{\sfT}^{-1}=\gamma_{2j}$ and $\hat{\sfT}\gamma_{2j}\hat{\sfT}^{-1}=-\gamma_{2j-1}$. 
The evolution, therefore, consists of two counter-propagating chiral Majorana modes and describes a ballistic metal due to the lack of backward scattering.

Moving away from $\theta=\pi/4$ introduces back-scattering and gives rise to a finite mean-free path $\lambda$ set by the density of back-scatterers in \eqref{eqn:network_model_node_transfer_matrices_form}. 
Parametrically, $\lambda \sim |\cos(2\theta)|^{-1}$ which diverges as $\theta\to \pi/4$. 
Our central claim is that, on scales large compared to this length ($L,\, T\gg \lambda$), the replicated partition function has an effective NLsM description. 
We emphasize that the NLsM description is not valid for the ballistic metal at $\theta=\pi/4$, where $\lambda \to \infty$.

The microscopic derivation of the sigma model starts from the special point at $\theta = \pi/4$, where the replicated partition sum is described by the field theory of $1+1$D non-interacting massless Majorana fermions.
Away from $\theta = \pi/4$, the back-scattering after averaging over disorder generates inter-replica interactions between chiral Majorana fermions.
To derive the effective field theory in a controlled way, we consider the coherent errors with rotation angle $\theta_{\ell}$ at each site drawn from a Gaussian distribution $p(\theta_\ell) = e^{-(\theta_\ell - \pi / 4)^2/(2g)}/\sqrt{2 \pi g}$ centered at $\pi/4$.
The inter-replica interaction after averaging over rotation angles can be decoupled using real anti-symmetric Hubbard–Stratonovich fields $Q$.
After integrating over the fermionic fields, we identify the saddle point of the matrix field $Q$ and establish the NLsM as the effective theory that characterizes the fluctuations around the saddle point.
The bare coupling $g_0 = 8g$ in the resulting sigma model is set by the variance $g$ of the rotation angle $\theta_\ell$, or equivalently, the back-scattering rate in the network model.

We note that a microscopic derivation of the NLsM has been carried out in various contexts. 
It is derived as an effective description of the 1+1D monitored free fermion dynamics~\cite{fava2023nonlinear}.
However, the derivation in Ref.~\cite{fava2023nonlinear} controlled in the large $N$ limit is technically different from our derivation controlled by the small parameter $g$.
Our derivation is, in spirit, more similar to the derivation of the sigma model for 2D disordered fermion systems, which is controlled in the weak disorder limit with a small scattering rate~\cite{altland2010condensed,lerner2003nonlinear}.

We expect the universal predictions of the sigma model at large scales to be insensitive to the coherent error model considered, in particular, to extend to the error model with uniform $\theta$.

The rest of this section is organized as follows.
Section~\ref{sec:nlsm_opt} and~\ref{sec:nlsm_subopt} derive the effective field theories for the optimal and suboptimal decoders, respectively.
Section~\ref{sec:pdiagram_of_nlsm} analyzes the RG flow of the effective NLsM near the weak coupling fixed point in two replica limits, $n\to 1$ and $n\to 0$, related to the optimal and suboptimal decoders.
We discuss the possible phase diagrams implied by this analysis.
We relegate the most technical steps of the derivation to Appendix~\ref{app:nlsm_derivation}.

\subsection{Effective theory for the optimal decoder}\label{sec:nlsm_opt}
We now derive the effective NLsM for the optimal decoder.
We first work in the RBIM picture in Sec.~\ref{sec:nlsm_opt_rbim} and comment on the modifications in the dual picture in Sec.~\ref{sec:nlsm_opt_dual}.
A comprehensive derivation is provided in Appendix~\ref{app:nlsm_derivation_opt}.

\subsubsection{Derivation in the RBIM picture}\label{sec:nlsm_opt_rbim}
To begin, we rewrite the replicated partition function such that the symmetry of the problem becomes apparent.
The rewriting makes use of the properties of the partition functions $\calZ(\eta)$; on the 2D square lattice with periodic boundary conditions, we have 
\begin{align}
    \calZ^*(\eta) &= \sum_\sigma e^{\sum_{\langle\bfr,\,\bfr'\rangle}\left(J_{\bfr\bfr'} + \frac{\ri\pi}{4} \right) \eta_{\bfr\bfr'} \sigma_\bfr \sigma_{\bfr'}} \nn \\
    &= \sum_\sigma e^{\sum_{\langle\bfr,\,\bfr'\rangle}\left(J_{\bfr\bfr'} - \frac{\ri\pi}{4} \right) \eta_{\bfr\bfr'} \sigma_\bfr \sigma_{\bfr'}}\prod_{\langle \bfr,\,\bfr'\rangle} \ri \eta_{\bfr\bfr'}\sigma_\bfr\sigma_{\bfr'}\nn\\
    &= \calZ(\eta) \prod_{\langle \bfr,\,\bfr'\rangle} \ri \eta_{\bfr\bfr'} \,.\label{eq:bipartite_transformation_rbim}
\end{align}
On the cylinder, $\calZ^*(\eta)$ is equal to $\calZ(\eta)$ up to modifications of the boundary conditions.
Specifically, the boundary state in the transfer matrix formulation of $\calZ(\eta)$ in  Eq.~\eqref{eqn:single_copy_rbim_prob_amplitude} is modified from $|+)^{\otimes L} \mapsto |-)^{\otimes L}$.

This property gives rise to an enhanced replica permutation symmetry in the partition function $\bfZ_0$ (\ref{eq:replicated_partition_rbim}).  On the cylinder, we may write

\begin{align}
\mathbf{Z}_0 = \sum_\eta \prod_{\langle \bfr,\,\bfr'\rangle}(\ri \eta_{\bfr\bfr'})^{n+1} \prod_{\sfa = 1}^{2n+2} \rbra{\psi_\sfa} \hat{\mathsf H}_{T}\hat{\mathsf T}_{T} \hat{\mathsf T}_{T-1} \cdots \hat{\mathsf T}_1 \rket{\psi_\sfa} 
    \,,\label{eqn:many_copy_rbim_prob_amplitude_before_averaging}
\end{align}
where $|\psi_\sfa) = |+)^{\otimes L}$, $|-)^{\otimes L}$ for odd and even replica index $\sfa$, respectively.  In this form, the partition function consists of $2n+2$ copies of a complex RBIM for each disorder realization $\eta$, which are identical in the bulk, and thus exhibit a $S_{2n+2}$ bulk permutation symmetry over all $2n+2$ copies. 

The partition function has a hidden continuous symmetry, which becomes clear after the Jordan-Wigner transformation.
In the fermion representation, the transfer matrix associated with the $\sfa$-th replica at each edge of the RBIM takes the form
\begin{align}
    \hat{\sfh}_{\bfr,\,\bfr+\hat{e}_x} &= e^{(J_{\bfr,\, \bfr+\hat{e}_x}-\frac{\ri\pi}{4})\eta_{\bfr,\,\bfr+\hat{e}_x} \ri\gamma_{2i}^{\sfa}\gamma_{2i+1}^{\sfa}} 
    \,, \\
    \hat{\sfv}_{\bfr,\,\bfr+\hat{e}_t} &=  (\ri\gamma_{2i-1}^\sfa\gamma_{2i}^\sfa)^{\frac{1-\eta_{\bfr,\,\bfr+\hat{e}_t}}{2}}e^{\ri\theta_{\bfr,\,\bfr+\hat{e}_t} \ri\gamma_{2i-1}^\sfa\gamma_{2i}^\sfa}\, .
\end{align}
Crucially, the transfer matrix on each edge is identical across all replicas.
This gives rise to a continuous $\O(2n+2)$ rotation symmetry of $\bfZ_0$ among fermionic modes, $\gamma^\sfa_i \mapsto \sum_\sfb O_{\sfa\sfb} \gamma_i^\sfb$, $O \in \O(2n+2)$, which includes the permutation symmetry identified above as a subgroup.
We remark that the $\O(2n+2)$ symmetry is non-local in the spin representation, but plays a crucial role in determining the decoding fidelity, which is associated with nonlocal observables in the RBIM.
The continuous symmetry hinges on the Gaussianity of the transfer matrix dynamics; a similar continuous symmetry for the entanglement dynamics in monitored free fermion systems has been identified in Ref.~\cite{bao2021symmetry,jian2022criticality}.

The fermion representation provides a starting point to derive an effective theory of $\bfZ_0$.
To enable a controlled microscopic derivation, we work with the error model with random rotation angles $\theta_\ell$ drawn independently from the Gaussian distribution $p(\theta_\ell)$ centered at $\pi/4$ and with variance $g$~\eqref{eq:angle_distribution}.
For simplicity of notation, we use $\bfZ_0$ to denote the partition function after averaging over random rotation angles $\theta_\ell$ from now on.

After averaging over $\theta_\ell$ and summing over $\eta$, the replica partition sum can be expressed in terms of a translationally invariant transfer matrix for all $2n+2$ replicas 
\begin{align}\label{eqn:many_copy_rbim_prob_amplitude}
    \bfZ_0 = (\Psi|\hat\bfH \hat{\bfT}^{T}|\Psi)\,,
\end{align}
where $|\Psi) = \bigotimes_{a}|\psi_{a}) $ is the boundary state, $\hat\bfT = \hat{\bfV}\hat{\bfH}$, $\hat{\bfH} = \prod_i \hat{\bfh}_i$, and $\hat{\bfV} = \prod_i \hat{\bfv}_i$.
The boundary state $|\Psi)$ for the partition function explicitly breaks the continuous symmetry.
In this section, we focus on deriving the effective theory for the bulk of the partition function. We will comment on the role of boundary states when analyzing the physical observables of the effective theory.

The transfer matrix at each site is given by 
\begin{align}
    \hat{\mathbf{h}}_i &= e^{-\frac{\ri\pi}{4}\sum_\sfa \ri\gamma_{2i}^{\sfa} \gamma_{2i+1}^{\sfa}+\frac{g}{2}\left(\sum_\sfa \ri\gamma_{2i}^{\sfa} \gamma_{2i+1}^{\sfa}\right)^2} K_{2i,\, 2i+1}
    \,,    \label{eqn:averaged_transfer_matrix_horizontal_rbim} \\
    \hat{\mathbf{v}}_i &= e^{\frac{\ri\pi}{4}\sum_{\sfa} \ri\gamma_{2i-1}^{\sfa}\gamma_{2i}^{\sfa}-\frac{g}{2}\left(\sum_{\sfa} \ri\gamma_{2i-1}^{\sfa}\gamma_{2i}^{\sfa}\right)^2} \tilde{K}_{2i-1,\, 2i}
    \,, \label{eqn:averaged_transfer_matrix_vertical_rbim}
\end{align}
where $K$ and $\tilde{K}$ are constraints
\begin{align}\label{eqn:constraints_in_rbim}
    K_{i,\, j} &= \frac{1 + \prod_{\sfa}\ri\gamma_{i}^{\sfa}\gamma_{j}^{\sfa}}{2} \,, \\ 
    \tilde{K}_{i,\, j} &= \frac{1 + (-1)^n\prod_{\sfa}\ri\gamma_{i}^{\sfa}\gamma_{j}^{\sfa}}{2} \,,
\end{align}
which arise from the ``disorder average'' over $\theta$ and $\eta$. In Appendix~\ref{app:local_constraint}, we show that the constraints define a set of stabilizers which generate a group invariant under the bulk dynamics.
Thus, the constraints can be imposed on the boundary state and ignored when deriving the effective field theory of the bulk.

In what follows, we derive an effective theory by first formulating the partition function $\bfZ_0$ using the fermion path integral (detailed in Appendix~\ref{app:fermion_path_integral}). 
At $g = 0$, the partition function is given by the path integral of $1+1$D massless Majorana fermions, which is a conformal field theory. 
For small $g$, we show that the partition function is governed by an effective NLsM.

When $g = 0$, i.e., the rotation angle $\theta_\ell = \pi/4$, the transfer matrix simply consists of SWAP gates; in the network model description, this limit describes a ballistic metal without back-scattering. 
For each replica, we thus have two chiral Majorana fermions propagating to the left and right, giving rise to a path integral 
\begin{align}
    \bfZ_0 &= \int \calD \chi_L\calD \chi_R \, e^{-\calS_0[\chi_L,\chi_R]}
\end{align}
with
\begin{align}
    \calS_0 &= \int \rd x \, \rd t \, \left(\chi_R \partial_+\chi_R + \chi_L \partial_-\chi_L\right) \,,
\end{align}
where $\chi_{L,R}$ is a $2n+2$ component real Grassmann field, and $\partial_\pm = \partial_t \pm \partial_x$.

When $g\ne 0$, the rotation angle $\theta_\ell$ deviates from $\pi/4$ and the Majorana fermions exhibit inter-replica interactions. To investigate whether these interactions can spontaneously break the $\SO(2n+2)$ replica symmetry, 
we introduce real anti-symmetric Hubbard-Stratonovich matrix fields to decouple the interaction and express the partition function as
\begin{align}
    \bfZ_0 &= \int \calD \chi \calD \sfQ_h \calD \sfQ_v \; e^{-\calS_0 - \calS_I - \frac{1}{g} \int\rd x\rd t \tr \sfQ_v^2 + \tr \sfQ_h^2}
\end{align}
where
\begin{align}
    \calS_I = \ri \int \rd x \, \rd t \,\left[ \chi_R (\sfQ_v+\sfQ_h) \chi_R + \chi_L (\sfQ_v-\sfQ_h)\chi_L\right] \,.
\end{align}
Here, $\sfQ_h$ and $\sfQ_v$ are the two matrix fields introduced to decouple the interaction associated with the horizontal and the vertical transfer matrix, respectively.

Integrating out the Majorana fermions (i.e. real Grassmann fields) yields an effective theory of the matrix fields.
For weak inter-replica interactions $g \ll 1$, the effective action has two sets of translationally invariant saddle points, each of which spontaneously breaks the replica symmetry:
\begin{equation}
\begin{aligned}
    &(1) \;\sfQ_h = (\ri g\pi/\sqrt{2})\sigma^y \otimes \mathds{1}_{n+1}\,,\quad \sfQ_v = 0\,; \\
    &(2) \;\sfQ_h = 0\,,\quad \sfQ_v = (\ri g\pi/\sqrt{2}) \sigma^y \otimes \mathds{1}_{n+1}\,.
\end{aligned}\label{eq:saddle_points}
\end{equation}
Any field configuration related to these two representative saddle points by an $\SO(2n+2)$ rotation, e.g. $\sfQ_h = O(\ri g\pi \sigma^y/\sqrt{2} \otimes \mathds{1}_{n+1}) O^T$ and $\sfQ_v = 0$ with $O \in \SO(2n+2)$, is also a saddle point.  Up to an overall re-scaling, these saddle points describe orthogonal, anti-symmetric matrices.  The two families of saddle points are related by a space-time rotation $\tau \mapsto -x$, $x \mapsto \tau$, which originates from the bulk rotational symmetry of the surface code.  
This transformation acts as $\partial_\pm \mapsto \pm \partial_\mp$ and $\chi_R \mapsto \chi_L$, $\chi_L \mapsto \ri \chi_R$, thus swapping the roles of the matrix fields $\sfQ_h \mapsto \sfQ_v$ and $\sfQ_v \mapsto - \sfQ_h$. 

We investigate Gaussian fluctuations around a given saddle point and show (see Appendix \ref{app:nlsm_derivation_effective_action}) that at a scale much greater than the mean-free path $\lambda$, the fluctuations are characterized by the NLsM with action
\begin{align}
    \calS_{\eff} = - \frac{1}{2 g_0} \int \rd x \, \rd t \tr \left(\nabla Q\right)^2 \,,\label{eq:nlsm_opt}
\end{align}
with bare coupling $g_0 = 8g$.
Here, $Q$ is an anti-symmetric matrix field, which is now normalized so that it is orthogonal ($Q^{T}Q = 1$).
This field can be explicitly parameterized as $Q = O (\ri\sigma^y \otimes \mathds{1}_{n+1}) O^T$ with $O \in \SO(2n+2)$.
Note that $Q$ itself is invariant under $O \mapsto O \cdot O_u$ with
\begin{align}\label{eqn:u_n_subgroup_definition}
    O_u = \frac{1}{2}\begin{pmatrix}
u+u^* & \ri u - \ri u^* \\
-\ri u + \ri u^* & u + u^*
    \end{pmatrix} \in \SO(2n+2) \,,
\end{align}
where $u \in \U(n+1,\mathbb{C})$.\footnote{Here, $O_u$ is a real orthogonal matrix that preserves the symplectic form $\ri\sigma^y \otimes \mathds{1}_{n+1}$. Such matrices form a non-normal subgroup of $\SO(2n+2)$ isomorphic to $\U(n+1)$, i.e. $\SO(2n+2,\, \mathbb{R}) \cap \Sp(2n+2,\, \mathbb{R}) \cong \U(n+1,\, \mathbb{C})$.}
The field $Q$, therefore, takes values in the coset space $\Gamma_{n+1} := \SO(2n+2)/\U(n+1)$.
We note that the action obtained has a global $\O(2n+2)$ symmetry $ Q\rightarrow O_1 QO_1^{T}$, $O_1\in \O(2n+2)$, which is consistent with the symmetry of the partition function identified in the fermion representation.

In principle, our derivation of the NLsM omits topological terms in the action. Since the target space satisfies $\pi_{2}(\O(2n)/\U(n))\cong\mathbb{Z}$ ($n\ge 2$), a $\Theta$-term is in general allowed~\cite{senthil2000quasiparticle}. 
The term is important at criticality and for distinguishing the two insulating phases, one of which carries a quantized thermal Hall conductivity. In addition, the fact that $\pi_{0}(\O(2n)/\U(n))\cong\mathbb{Z}_{2}$ permits ``domain walls'' across which the sigma-model field can alternate between the disconnected components of the target space which are distinguished by $\sgn[\Pf(Q)]$. In our analysis of the weak coupling fixed point $g\to 0$, in either replica limit, we neglect both of these contributions. We do not believe these contributions would affect the weak coupling analysis, though they will alter the global phase diagram or the physics in the vicinity of the decoding transition~\cite{putz_flow_to_nishimori}.

Before proceeding, we note that the property of $\calZ(\eta)$ in Eq.~\eqref{eq:bipartite_transformation_rbim} stems from the fact that RBIM lives on a lattice with an even coordination number.
This is the reason that the partition function exhibits an $\O(2n+2)$ symmetry.
Without this property, the partition function would exhibit an $\O(n+1) \times \O(n+1)$ symmetry and be governed by a sigma model with target space $\SO(n+1)$, which has implications for decoding of the surface code on non-bipartite lattices~\cite{forthcoming} (see discussion in Sec.~\ref{sec:conclusion_lattice}).
We note that, in a different context, effective sigma models also describe monitored free fermion dynamics~\cite{fava2023nonlinear,jian2022criticality,jian2023measurement,poboiko2023theory,fava2024monitored}.
There, the dependence of the sigma model target space on the bipartiteness of the lattice has been pointed out in Ref.~\cite{fava2023nonlinear}.

\subsubsection{Derivation in the dual picture}\label{sec:nlsm_opt_dual}
Following a similar procedure, we here show that the replicated partition function $\tilde\bfZ_0$ in the dual picture in Eq.~\eqref{eq:replicated_partition_dual} is also described by the NLsM with target space $\Gamma_{n+1}$.

Again, we start by rewriting the replicated partition function $\tilde\bfZ_0$ such that the symmetry becomes apparent.
This relies on the property of the partition function $\calZ_{+,\,s}$ in Eq.~\eqref{eqn:explicit_dual_partition_function},
\begin{align}
    \calZ^*_{+,\,s} &= \sum_{\tau} e^{-\sum_{\langle \bfr,\,\bfr'\rangle}\ri\theta_{\bfr\bfr'} \,\tau_\bfr\tau_{\bfr'} - \sum_{\bfr \in \partial} \ri\theta_\bfr \tau_\bfr} \prod_\bfr \tau_\bfr^{s_\bfr} \nn \\
    &= \sum_{\tau} e^{\sum_{\langle \bfr,\,\bfr'\rangle}\ri\theta_{\bfr\bfr'} \,\tau_\bfr\tau_{\bfr'} - \sum_{\bfr \in \partial} (-1)^{\bfr}\,\ri\theta_\bfr \tau_\bfr} \prod_\bfr (-1)^{\bfr}\tau_\bfr^{s_\bfr}\,.
\end{align}
We make use of the bipartiteness of the square lattice and redefine the spin variable on one sublattice, i.e., $\tau_\bfr \mapsto (-1)^{\bfr} \tau_\bfr$, where $(-1)^{\bfr} = \pm 1$ on two sublattices, respectively.
In a 2D system without boundaries, $\calZ_{+,\,s}^*$ equals $\calZ_{+,\,s}$ up to a sign depending on the syndrome configuration.

We can thus express the replicated partition function $\tilde\bfZ_0$ in a form that is symmetric among $2n+2$ replicas,
\begin{align}
\tilde{\mathbf{Z}}_0 &= \sum_{\tau,\,s} 
\prod_\bfr\left(\prod_{\sfa} \tau_{\bfr}^{\sfa} \right)^{s_\bfr}
e^{\ri\sum_{\sfa,\, \langle \bfr,\, \bfr'\rangle} (-1)^\sfa\theta_{\bfr\bfr'} \tau_\bfr^\sfa \tau_{\bfr'}^\sfa}
\\
&= \sum_{\tau,\,s} 
\prod_\bfr  \left( (-1)^{n\bfr} \prod_{\sfa} \tau_{\bfr}^{\sfa} \right)^{s_\bfr}
e^{\ri\sum_{\sfa, \, \langle \bfr, \, \bfr'\rangle} \theta_{\bfr\bfr'}\tau_\bfr^\sfa \tau_{\bfr'}^\sfa}
\,,
\end{align}
where we have omitted the boundary term.

The summation over syndrome $s_\bfr = 0,\, 1$ at site $\bfr$ imposes local constraints $\prod_\sfa \tau_\bfr^\sfa = (-1)^{n\bfr}$.
Since the boundary spin in the dual picture is pinned to be $ \pm 1$, these local constraints are equivalent to 
\begin{equation}\label{eqn:local_constraint_dual_picture}
\prod_{\langle \bfr, \bfr' \rangle}
    \frac{1 + (-1)^{n} \prod_{\sfa} \tau_\bfr^\sfa\tau_{\bfr'}^\sfa}{2}
    \,.
\end{equation}

Next, we express the partition function $\tilde\bfZ_0$ using the transfer matrix in the spatial direction (i.e. the compact direction of the cylinder), $\tilde \bfZ_0 = \tr \tilde{\bfT}^L$, where $\tilde{\bfT} = \tilde{\bfV} \tilde{\bfH}$, with $\tilde{\bfV} = \prod_{j=1}^{T} \tilde{\bfv}_j$ and $\tilde{\bfH} = \prod_{j=1}^{T}\tilde{\bfh}_j$.
In terms of Majorana fermions, the transfer matrices take the form
\begin{align}
    \tilde{\bfh}_{j} &=     \frac{1 + (-1)^n \prod_\sfa \ri \gamma_{2j}^\sfa\gamma_{2j+1}^\sfa}{2}
    e^{\ri\theta_{\bfr,\,\bfr+\hat{e}_x} \sum_\sfa  \ri \gamma_{2j}^\sfa\gamma^\sfa_{2j+1} }
    \,, \\
    \tilde{\bfv}_{j} &=\sum_{\eta = \pm 1} \eta^n e^{\eta(J_{\bfr,\,\bfr+\hat{e}_t}-\frac{\ri\pi}{4})\sum_\sfa \ri \gamma_{2j-1}^\sfa\gamma^\sfa_{2j}} \,,
\end{align}
where $\eta$ is introduced to simplify the expression. Summing over $\eta$ generates the constraint \eqref{eqn:local_constraint_dual_picture} associated with vertical bonds.
The transfer matrix again exhibits an $\O(2n+2)$ symmetry, corresponding to the rotation among $2n+2$ Majorana modes.

For the random rotation angles $\theta_\ell$ drawn from Gaussian distributions, the averaged transfer matrices for $g \ll 1$ are given by
\begin{align}
    \tilde{\bfh}_{j} &= 
    e^{-\frac{g}{2}\left(\sum_\sfa  \ri \gamma_{2j}^\sfa\gamma^\sfa_{2j+1}\right)^2} e^{\frac{\ri\pi}{4} \sum_\sfa  \ri \gamma_{2j}^\sfa\gamma^\sfa_{2j+1}}
    \tilde{K}_{2j,\, 2j+1}
    \,,
    \label{eqn:averaged_transfer_matrix_horizontal_dual}
    \\
    \tilde{\bfv}_{j} &= e^{\frac{g}{2}\left(\sum_\sfa \ri \gamma_{2j-1}^\sfa\gamma^\sfa_{2j}\right)^2}e^{-\frac{\ri\pi}{4} \sum_\sfa \ri \gamma_{2j-1}^\sfa\gamma^\sfa_{2j}}
    K_{2j-1,\, 2j}
    \,,
       \label{eqn:averaged_transfer_matrix_vertical_dual}
\end{align}
where $K$ and $\tilde{K}$ are given by~\eqref{eqn:constraints_in_rbim}.
The constraints $K$ and $\tilde{K}$ again commute with the bulk dynamics; one can combine the constraints for all steps and impose them at one specific time step of the transfer matrix dynamics.
We ignore the constraint when deriving the effective theory for the bulk of 2D partition function.
We note that with a slight abuse of notation, we use $\tilde{\bfh}_j$, $\tilde{\bfv}_j$ and $\tilde\bfZ_0$ to denote the transfer matrices and partition functions after averaging over random rotations.

From this point onward, the derivation of the effective theory for $\tilde{\bfZ}_0$ becomes essentially the same as that for $\bfZ_0$ in the RBIM picture.
For $g \ll 1$, we again obtain the non-linear sigma model with target space $\Gamma_{n+1}$ as an effective description of the fluctuation around the saddle point at the scale greater than the mean-free path $\lambda$.

\subsection{Effective theory for the suboptimal decoder}\label{sec:nlsm_subopt}
We now derive the effective theory for the partition function $\bfY_0$ associated with the suboptimal decoder.
To summarize, the suboptimal decoder's inaccurate estimate of the coherent rotation angles explicitly breaks the replica symmetry of the effective theory for the optimal decoder from $\O(2n+2)$ to $\O(2)\times \O(2n)$.

Consequently, at large scales, the partition function $\bfY_0$ for the suboptimal decoder is governed by the NLsM with the target space $\Gamma_n = \SO(2n)/\U(n)$.
A detailed microscopic derivation of the effective theory is provided in Appendix~\ref{app:nlsm_derivation_subopt}, and we present the steps in this derivation below.
The effective theory for $\tilde \bfY_0$ in the dual picture proceeds analogously and will not be presented here.

We outline the derivation of the effective theory for the suboptimal decoder.  First, the partition function $\bfY_0$ is expressed in the RBIM picture using the fermion transfer matrix,
\begin{align}
\mathbf{Y}_0 = \sum_\eta \prod_{\langle \bfr,\,\bfr'\rangle}(\ri \eta_{\bfr\bfr'})^{n+1} \prod_{\sfa = 1}^{2n+2} \rbra{\psi_\sfa} \hat{\mathsf H}^\sfa_{T}\hat{\mathsf T}^\sfa_{T} \hat{\mathsf T}^\sfa_{T-1} \cdots \hat{\mathsf T}^\sfa_1 \rket{\psi_\sfa} 
    \,,\label{eqn:many_copy_rbim_prob_amplitude_subopt_before_averaging}
\end{align}
The transfer matrix for $\bfY_0$ consists of transfer matrices $\hat\sfT^\sfa$ for $2n+2$ copies of Majorana fermions [similar to Eq.~\eqref{eqn:many_copy_rbim_prob_amplitude_before_averaging}].
Here we make the replica index explicit because the single-copy transfer matrices $\hat{\sfT}^\sfa$ are not identical for all replicas, since the estimated rotation angle $\theta'_\ell \neq \theta_\ell$. As a result, the fermionic representation only exhibits an explicit $\O(2) \times \O(2n)$ symmetry.
The $\O(2)$ and $\O(2n)$ symmetries describe rotations within the first two and the next $2n$ replicas, respectively, corresponding to the true syndrome distribution $\calQ_{\alpha,\,s}$ and the decoder's estimate $\calP_{\alpha,\,s}$ in Eq.~\eqref{eqn:rbim_replicated_subopt_fidelity}, respectively.

To obtain a concrete microscopic derivation, we consider the suboptimal decoder defined by the relation between $\theta'_\ell$ and $\theta_\ell$ in Eq.~\eqref{eq:model_suboptimal_decoder}.
Nevertheless, we expect the universal predictions of the resulting effective theory to apply broadly to any suboptimal decoding scheme with small angle miscalibration. The derivation follows a similar procedure as that for $\bfZ_0$ in Sec.~\ref{sec:nlsm_opt_rbim} (as detailed in Appendix~\ref{app:nlsm_derivation_subopt}).

In the path integral formulation of the partition function $\bfY_0$, we decouple the inter-replica interaction by introducing anti-symmetric Hubbard–Stratonovich matrix fields $Q$. 
Integrating over the fermion fields yields symmetry-breaking saddle points.  
When the mismatch $\theta'_\ell \ne \theta_{\ell}$ is small ($\epsilon \ll 1$), fluctuations about a given saddle point are described by a non-linear sigma model supplemented by a symmetry-breaking potential, whose bare strength is parametrically weak.
In the case that the fluctuation out of the reduced subspace is small,  the action takes the form
\begin{align}
\calS_{\epsilon,\eff} = -\frac{1}{2g_0}\int \rd x \rd t \, \left[\tr(\nabla  Q)^2 +\frac{\epsilon^2 \pi^2 g_0^2}{16} \tr Q^2_{\text{off-diag}}\right]\,,
\end{align}
where $Q \in \Gamma_{n+1}$, and $Q_{\text{off-diag}} \notin \Gamma_1 \times \Gamma_n$ is the matrix field in the off-diagonal blocks between the first two and the next $2n$ replicas.

This symmetry-breaking potential has scaling dimension two and is therefore relevant at the metallic fixed point $g = 0$.
At length scale $L$, its renormalized strength grows as $\calO(\epsilon^2 L^2)$, capturing how the distinction between optimal and suboptimal decoding becomes increasingly important under coarse-graining. At a sufficiently large scale, the flow confines the theory to the NLsM with a reduced target space $\Gamma_1\times \Gamma_n = \Gamma_n$, giving the long-distance description quoted above.

\subsection{Phase diagram of the sigma model}\label{sec:pdiagram_of_nlsm}
The optimal and suboptimal decoders are governed by the non-linear sigma model with target space $\SO(2n)/\U(n)$ in the limits $n \to 1$ and $n \to 0$, respectively.
In this section, we discuss the renormalization group flows of the NLsM near its weak coupling fixed point, and the implications for the phase diagram for optimal and suboptimal decoding.

A key point is that the stability of the weak coupling $g=0$ fixed point depends on the replica limit.  
The perturbative beta function for $g$ has been computed for the NLsM with target space $\SO(2n)/\U(n)$ within the $d=(2+\varepsilon)$ expansion, and for an arbitrary number of replicas; here, we quote the known beta functions when  $\varepsilon\rightarrow 0$~\cite{hikami_three_loop_beta,wegner_four_loop_beta,evers2008anderson} and in the appropriate replica limits.  
In the limit $n\to 1$ associated with the optimal decoder, the weak coupling fixed point is unstable with perturbative beta-function 
\begin{align}
\frac{\rd g_{R}}{\rd \ln L} = 4g_{R}^3 + \calO(g_R^4) \,, \label{eq:n=1_RG_flow}
\end{align}
so that the renormalized stiffness at scale $L$ is $g^{-1}_{R}(L) = \sqrt{1/g^2_0 - 8\ln L}$, where $1/g_0$ is the non-universal bare stiffness. 
This renormalization group (RG) flow is valid until a scale at which the renormalized stiffness is $O(1)$.
By contrast, in the limit $n\to 0$ relevant for suboptimal decoding, this coupling constant is \emph{marginally irrelevant}
\begin{align}
\frac{\rd g_{R}}{\rd\ln L} = -2g_{R}^2 + \calO(g_R^3) \,.\label{eq:n=0_RG_flow}
\end{align}
Integrating the beta function yields a renormalized stiffness that grows with scale $g_{R}^{-1} = g_0^{-1} + 2\ln L$. 
The stable weak coupling fixed point at $g_R = 0$ is known as the {thermal metal} fixed point; this terminology originates from studies of disordered fermion systems in two dimensions, where the NLsM arises as an effective theory with coupling inversely proportional to the thermal conductance in a system of size $L$, i.e., $G(L) \sim 1/g_{R}(L)$ grows logarithmically with $L$~\cite{senthil2000quasiparticle}. 

Beyond the metallic phase, the sigma model also exhibits gapped phases, which are reached if the renormalized coupling $g_{R}$ grows under coarse-graining. In the fermion language, these are the two localized phases of symmetry class D—trivial and topological superconducting phases—which are distinguished by the topological $\Theta$-term in the infrared. In the surface code decoding problem, these phases map onto the decodable and non-decodable phases which are familiar from maximum-likelihood decoding with incoherent Pauli noise.

These renormalization group considerations suggest distinct phase diagrams as the bare coupling is increased:
\begin{itemize}
\item For the optimal decoder, associated with the limit $n \to 1$, although the metallic fixed point is unstable, increasing the variance $g$ of the rotation angle can either lead to the insulating phase describing a quantum memory or induce a phase transition between two insulating phases.
In later sections, we numerically simulate the fidelity (Sec.~\ref{sec:fidelity_predictions_opt_numerics}) and various other quantities associated with the optimal decoder and observe no signature of a transition when tuning $\theta$.
We therefore believe that the surface code is always in the decodable insulating phase when $\theta < \pi/4$.

\item For the suboptimal decoder, associated with the limit $n \to 0$, the stable metallic fixed point indicates the existence of a metal phase and a metal-to-insulator transition when increasing $g$~\cite{chalker2001thermal}.
When the rotation angle $\theta$ is uniform in the system, we expect a phase transition at an intermediate angle $\theta'_c$.
In Sec.~\ref{sec:conductance}, we numerically simulate the conductance in the associated network model to estimate $\theta'_c$.
\end{itemize}

In the following sections, we predict distinct scalings of various physical quantities based on the qualitatively different RG flows in the vicinity of the metallic fixed points associated with the optimal and suboptimal decoders.
We verify our predictions using large-scale numerical simulations based on the free fermion representation of our problem.

Verifying the predictions of the RG flows near the metallic fixed point requires careful analysis.
The system is at the metallic fixed point when $g_R \to 0$ or $\theta \to \pi/4$ in the error model with uniform rotation.
However, there is also a diverging length scale, the \emph{mean-free path} $\lambda$, associated with this limit.
Right at $\theta = \pi/4$, the network model has no backward scattering ($\lambda \to \infty$), describing a ballistic metal.
The sigma model description is only valid when $g_0$ is small but non-vanishing (or equiv. $\theta$ is close to but not $\pi/4$) and at a scale $L \gg \lambda$.
The presence of this length scale $\lambda$ requires a large-scale simulation to verify the predictions of the NLsM.
Besides, the optimal and the suboptimal decoders are distinguished by the marginal RG flows, which only lead to notable distinctions at large scales.

We note that a NLsM in the appropriate replica limit appears as the effective description of various other physics problems.
For two-dimensional systems of disordered fermions in class D, the effective theory is always formulated as the $n \to 0$ limit of the NLsM with target space $\SO(2n)/\U(n)$~\cite{evers2008anderson}.
The RBIM with real couplings on the Nishimori line also maps to the Chalker-Coddington network model in class D, however, it is formulated as the $n \to 0$ limit of the NLsM with target space $\SO(2n+1)/\U(n)$~\cite{gruzberg2001random}.
Moreover, the NLsM is developed as an effective description for the monitored free fermion dynamics in 1+1D~\cite{fava2023nonlinear,jian2023measurement,poboiko2023theory,fava2024monitored}.
In that case, the NLsM with target space $\SO(2n)/\U(n)$ in the limit $n \to 1$ is identified as the effective theory in various setups~\cite{fava2023nonlinear,wang_self_dual_cho_fisher,putz_flow_to_nishimori}.

\section{Predictions of decoding fidelity}\label{sec:fidelity_predictions}
In this section, we analyze the decoding fidelity based on the effective non-linear sigma model description.
The effective theory predicts an unstable ``metal'' fixed point ($g_R = 0$) associated with an optimal decoder and a stable metal phase for the suboptimal decoders, leading to distinct scalings of the decoding fidelity in the vicinity of the metallic fixed point, i.e., $g_R \ll 1$.
We verify these predictions with large-scale numerical simulation of the optimal and suboptimal decoding fidelity using the numerical algorithm detailed in Appendix~\ref{app:syndrome_sampling}.

We note that, for simplicity, we analyze the decoding fidelity for the surface code on the torus, while our numerical simulations are carried out on the cylinder.
However, we believe that the scaling of the fidelity in both cases is qualitatively the same.
In the RBIM picture, the fidelity is related to the twist inserted in the longitudinal direction; the boundary condition does not affect the scaling of the twist expectation value as long as the system size is large. 
In the dual picture, the fidelity is related to the twist inserted in the periodic direction of the cylinder.
Here, the boundary condition breaks the $\O(2n)$ symmetry, and the twist acts non-trivially on the boundary state.
The symmetry-breaking boundary condition stems from the boundary condition in each individual replica before the Jordan-Wigner transformation, which breaks $\bbZ_2$ symmetry.
Such a boundary state changes under the twist (i.e., the symmetry defect) which maps $\sigma^{\sfa}$ to $-\sigma^{\sfa}$.
Thus, we believe that the expectation value of the twist defect inserted along the periodic direction will exhibit the same scaling as that on the torus.

\subsection{Fidelity of the optimal decoder}
We start with the fidelity of the optimal decoder. 
We first analyze the replicated fidelity $\calF_\opt^{(n)}$ based on the effective NLsM with target space $\Gamma_{n+1} = \SO(2n+2)/\U(n+1)$ and then take the replica limit $n \to 0$.
The coupling in the sigma model is marginally relevant in this limit, and our prediction is valid up to a scale where $g_R = \calO(1)$.

The replicated fidelity can be expressed in terms of the partition functions with symmetry defect insertions as in Eq.~\eqref{eqn:rbim_replicated_opt_fidelity} [Eq.~\eqref{eqn:dual_replicated_opt_fidelity}].
The expectation value $\Phi_{2k}$ ($\tilde \Phi_{2k}$) of defect insertion maps to the twist expectation value in the non-linear sigma model as shown in Appendix~\ref{app:nlsm_derivation_symmetry_defects}.
In particular, in the sigma model for $\bfZ_0$ in the RBIM picture, $ \Phi_{2k} $ maps to the expectation value of inserting a twist in the vertical direction (illustrated in Fig.~\ref{fig:twist_intro}); the twist acts on the matrix field as 
\begin{equation}
\begin{aligned}
    Q \mapsto \Lambda_{2k} Q \Lambda_{2k}, \quad \Lambda_{2k} = \begin{pmatrix}
    -\mathds{1}_{2k} & \\
    & \mathds{1}_{2n+2-2k}
\end{pmatrix} \,.
\end{aligned}
\end{equation}
Similarly, the twist $\tilde \Phi_{2k}$ in the sigma model derived from the dual picture is inserted in the horizontal direction.
We predict the scaling of the replicated decoding fidelity $\calF_\opt^{(n)}$  in Eq.~\eqref{eqn:n_rep_fidelity} based on the scaling of the twist expectation value in the NLsM.

The twist expectation value in the NLsM exhibits distinct scalings in the regime $\kappa = T/L \gg 1$ and $\kappa \ll 1$.
In what follows, we discuss these two regimes separately.

\subsubsection{\texorpdfstring{$\kappa \gg 1$}{κ >> 1}}
When $\kappa \gg 1$, we coarse-grain the sigma model up to scale $L$ and obtain an effective one-dimensional sigma model with action
\begin{align}
    \calS_{\eff} = -\int_0^\kappa \rd t \frac{1}{2g_R}\tr(\partial_t Q)^2 \,,
\end{align}
where $g_R = g_R(L)$ is the renormalized coupling at scale $L$.

In the RBIM picture, the twist insertion leads to a modified action with a local potential
\begin{align}
\calS^\Lambda_{\eff} = -\int_0^{\kappa} \rd t \, \left[ \frac{1}{2g_R} \tr (\partial_t Q)^2 + \frac{L}{2g_0} \tr (Q - \Lambda Q \Lambda)^2 \right] \,,
\end{align}
where $\Lambda = \Lambda_{2k}$, and we suppress the subscript for simplicity of presentation.
The local potential is relevant and becomes a local constraint $Q = \Lambda Q \Lambda$ at large scales.

The one-dimensional sigma model has a finite correlation length $\xi \sim 1/g_R$ and becomes disordered in the limit $\kappa \gg \xi$.
This leads to distinct predictions of the twist expectation value in two limits.
In the regime where $\kappa \gg 1/g_R$, the effective 1D model consists of decoupled spins at the scale of $1/g_R$, leading to
\begin{align}
    \Phi_{2k} = \left(\frac{2\text{Vol}\left(\Gamma_k \times \Gamma_{n+1-k}\right)}{\text{Vol}\left(\Gamma_{n+1}\right)} \right)^{\calO(\kappa g_R)}
    \,& &(\kappa \gg 1/g_R) \,.
\end{align}
We compute the volume of the target space $\Gamma_n$ in Appendix~\ref{app:target_space_volume}.
In the opposite regime, $\kappa \ll 1/g_R$, the twist expectation value is governed by the quadratic part of the NLsM action and exhibits a scaling (as shown in Appendix~\ref{app:twist_field})
\begin{align}
    \Phi_{2k} \sim \left(\frac{1}{g_R\kappa}\right)^{k(n+1-k)}\, & &(\kappa \ll 1/g_R) \,.
\end{align}
The results of the twist expectation value lead to the following qualitative predictions for $\calF_\opt^{(n)}$:
\begin{itemize}
    \item The fidelity $\calF_\opt^{(n)}$ increases with the renormalized coupling $g_R$. 
    One can increase $g_R$ either by increasing the overall scale for fixed aspect ratio $\kappa$ or by increasing the bare coupling $g$.
    
    \item The fidelity $\calF_\opt^{(n)}$ increases with $T$ when $L$ is fixed.
\end{itemize}

These qualitative predictions for $\calF_\opt^{(n)}$ can also be obtained in the dual picture, in which the twist is inserted in the horizontal direction.
The twisted partition function is described by a modified 1D action with a boundary term
\begin{align}
    &\tilde\calS^\Lambda_{\eff} = \nn \\
    &\quad -\int_0^{\kappa} \rd t \frac{1}{2g_R}\tr(\partial_t Q)^2 + \frac{L}{2g}\tr (Q(0) - \Lambda Q(1/L) \Lambda)^2 \,. 
\end{align}
This boundary term is relevant and imposes the boundary condition $Q(0^-) = \Lambda Q(0^+) \Lambda$ in the thermodynamic limit.
In the limit $\kappa \gg 1/g_R$, the twist does not modify the partition function up to an exponentially decaying correction, i.e.,
\begin{align}
    \tilde\Phi_{2k} = 1 - e^{-\calO(\kappa g_R)} & & (\kappa \gg 1/g_R) \,.
\end{align}
In the opposite limit, $\kappa \ll 1/g_R$, we numerically simulate the twist expectation value and obtain an empirical scaling (see Appendix~\ref{app:twist_field})
\begin{align}
    \tilde\Phi_{2k} = \tilde \Phi_{2n+2-2k} = e^{-\frac{1}{\kappa g_R} (\alpha_{n+1} + \beta_{n+1} k)} & & (\kappa \ll 1/g_R) \,.
\end{align}

Next, we attempt to take the replica limit $n \to 0$ to obtain the scaling of the decoding fidelity $\calF_\opt$ for the optimal decoder:
\begin{itemize}
\item In the regime $\kappa \gg 1/g_R(L)$, the replicated fidelity takes the form $\calF_\opt^{(n)} = 1 - e^{-\calO(\kappa g_R(L))}$.
We expect that the fidelity in the replica limit $n \to 0$ scales as
\begin{align}
    \calF_\opt = 1 - e^{-\calO(\kappa g_R(L))} \,,\label{eq:fid_kappa_greater_than_1/gR}
\end{align}
where $g_R(L)$ is governed by the RG flow in Eq.~\eqref{eq:n=1_RG_flow}.
This predicts that the decoding infidelity $1-\calF_\opt$ decays exponentially in the aspect ratio, and the decay coefficient increases with the scale for fixed $\kappa$, governed by the marginal RG flow.

\item In the regime $\kappa \ll 1/g_R(L)$, we obtain the decoding fidelity from the dual picture,\footnote{We note that in principle, one can also obtain the same fidelity in the replica limit from the RBIM picture. However, we do not know how to perform the analytic continuation.}
\begin{align}\label{eq:opt_decoding_kappa_large_ thermal_metal}
    \calF_\opt = \frac{1}{2} + A e^{-\frac{\beta_1}{\kappa g_R(L)}} \,,
\end{align}
where $A$ and $\beta_1 = \lim_{n \to 1} \beta_n$ are $\calO(1)$ numbers and we keep the leading order in $e^{-\beta_1/(\kappa g_R(L))}$.
This result relies on the empirical scaling of the twist expectation value; the detailed derivation is provided in Appendix~\ref{app:fidelity_replica_limit}.
\end{itemize}

\subsubsection{\texorpdfstring{$\kappa \ll 1$}{κ << 1}}
When $\kappa \ll 1$, we coarse-grain the sigma model up to scale $T$ and obtain the effective 1D sigma model with action
\begin{align}
    \calS_{\eff} = - \int_{0}^{1/\kappa} \rd x \frac{1}{2g_R} \tr (\partial_x Q)^2 \,,
\end{align}
where $g_R = g_R(T)$ is the renormalized coupling at scale $T$.

In contrast to when $\kappa \gg 1$, the twist defect in the RBIM picture modifies the boundary coupling in the effective 1D model, while the twist in the dual picture manifests as a local potential.
Specifically, in the RBIM picture, we obtain the twist expectation value
\begin{align}
    \Phi_{2k} &= 1 - e^{-\calO(\frac{g_R}{\kappa})} & & (1/\kappa \gg 1/g_R) \,, \nn\\
    \Phi_{2k} &= \Phi_{2n+2-2k} = e^{-\frac{\kappa}{g_R} (\alpha_{n+1} + \beta_{n+1} k)} & & (1/\kappa \ll 1/g_R) \,.
\end{align}
Whereas in the dual picture, we have
\begin{align}
    \tilde{\Phi}_{2k} &= \left(\frac{2\text{Vol}\left(\Gamma_k \times \Gamma_{n+1-k}\right)}{\text{Vol}\left(\Gamma_{n+1}\right)} \right)^{\calO(g_R/\kappa)} & & (1/\kappa \gg 1/g_R) \,, \nn\\
    \tilde{\Phi}_{2k} &\sim \left(\frac{\kappa}{g_R}\right)^{k(n+1-k)} & & (1/\kappa \ll 1/g_R) \,.
\end{align}

The twist expectation values lead to the following predictions for the replicated fidelity $\calF_\opt^{(n)}$:
\begin{itemize}
    \item The fidelity $\calF_\opt^{(n)}$ decreases with increasing $g_R$. 
    The renormalized coupling $g_R$ increases with bare coupling or the overall scale when the aspect ratio $\kappa$ is fixed.
    
    \item The fidelity $\calF_\opt^{(n)}$ increases with $T$ when $L$ is fixed.
\end{itemize}

\begin{figure*}[t]
\centering
\includegraphics[width=.99 \textwidth  ]{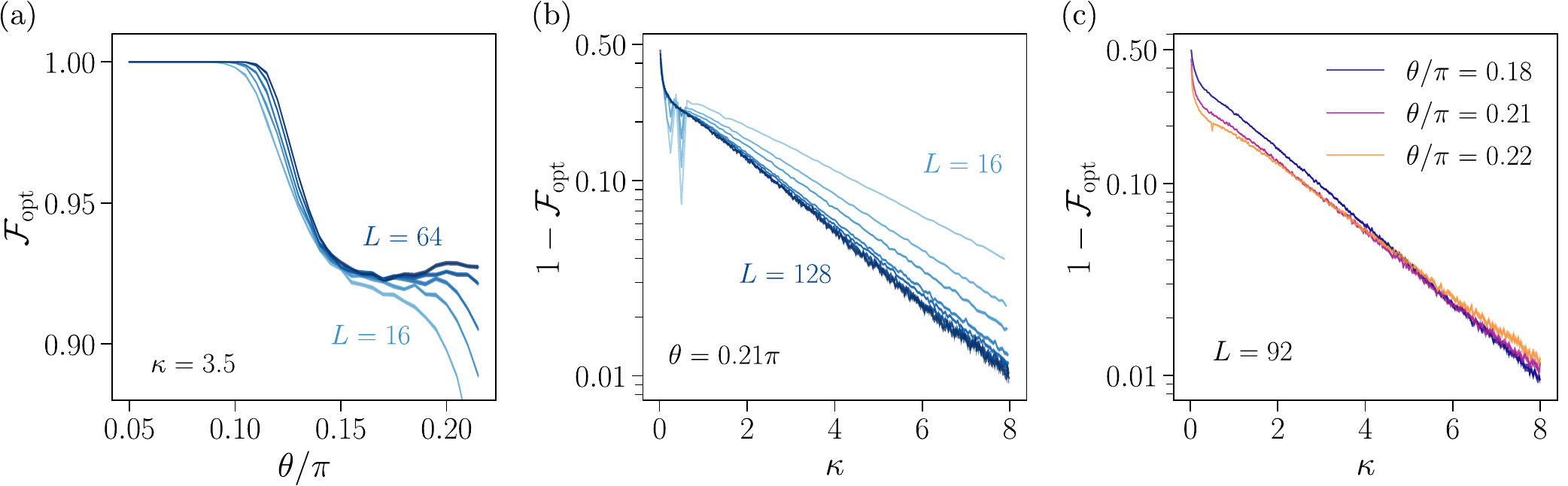}
    \caption{
    Fidelity of the optimal decoder.
    (a) $\calF_\mathrm{opt}$ as a function of $\theta$ at a fixed $\kappa = 3.5$ for various $L$.
    For small $\theta$, the fidelity is very close to $1$ while for larger $\theta$, it increases with $L$, consistent with a decodable phase.
    (b, c) Decoding infidelity $1 - \calF_\mathrm{opt}$ as a function of $\kappa$.
    The results suggest that $\calF_\mathrm{opt}$ increases when increasing $\kappa$ for fixed $L$, or increasing $L$ for fixed $\kappa > 1$.
    Furthermore, increasing $g_R^{-1}(L)$ through $\theta$ leads to increased fidelity for $\kappa < 1$ but decreased fidelity for $\kappa > 1$.
    These behaviors are consistent with our predictions.
    Numerical results are averaged over $900$ to $34000$ syndrome realizations.  
    The error bars are indicated by the linewidth.
    }
    \label{fig:fid_opt_combined}
\end{figure*}

In the replica limit $n \to 0$, we obtain the scaling of the fidelity in two regimes:
\begin{itemize}
\item When $1/\kappa \gg 1/g_R(T)$, the replicated fidelity is close to unity up to an exponentially small correction, i.e., $\calF_\opt^{(n)} = 1/2 + e^{-\calO(g_R(T)/\kappa)}$.
In the replca limit, we expect the scaling
\begin{align}\label{eq:opt_fidelity_small_kappa_1}
    \calF_\opt = \frac{1}{2} + e^{-\calO(g_R(T)/\kappa)} \,.
\end{align}

\item When $1/\kappa \ll 1/g_R(T)$, we obtain the decoding fidelity in the replica limit from the RBIM picture
\begin{align}
    \calF_\opt = 1 - A e^{-\frac{\beta_1\kappa}{g_R(T)}} \,, \label{eq:opt_fidelity_kappa_smaller_1}
\end{align}
where $A$ is a constant. 
Again, this result relies on the empirical scaling of the twist expectation value (see Appendix~\ref{app:fidelity_replica_limit} for detailed derivation).
\end{itemize}

A notable prediction of our theory is that the fidelity of the optimal decoder is qualitatively distinct in the regimes $\kappa \gg 1$ and $\kappa \ll 1$.
In particular, increasing the renormalized coupling $g_R$ leads to an increased fidelity when $\kappa \gg 1$ but a decreased fidelity when $\kappa \ll 1$.

We remark that close to the metallic fixed point, i.e., $\kappa,\, 1/\kappa \ll 1/g_R$, our results in the replica limit rely on the empirical scaling of the twist expectation values.
The scaling of the fidelity extracting in this way raises a puzzle regarding its dependence on $\kappa$.
First of all, the effective theory for the optimal decoder is a conformal field theory at the metallic fixed point with a marginally relevant perturbation.
This fixed point can be accessed by sending the bare coupling $g_0 \to 0$ while simultaneously considering larger system sizes $L \gg 1/g_0$ such that the sigma model remains valid a valid description.
At the metallic fixed point, the fidelity, which is related to the twist expectation value, naturally depends on the aspect ratio.
In fact, at the weak coupling fixed points of the sigma models with $n \geq 2$, the twist expectation value should decay exponentially in the aspect ratio~\cite{cardy1984conformal}, giving rise to a fidelity which increases with the aspect ratio.
However, we obtain from Eq.~\eqref{eq:opt_decoding_kappa_large_ thermal_metal} and~\eqref{eq:opt_fidelity_kappa_smaller_1}, that as we approach the metallic fixed point $g_R \to 0$, the fidelity for $\kappa \gg 1$ is $1/2$ while it is $1$ for $\kappa \ll 1$.
Together, these results suggest that the fidelity decays with the aspect ratio.

We emphasize that the decreasing fidelity cannot arise when the decoding problem admits a description in terms of a classical statistical-mechanics model, as in the surface code with incoherent Pauli errors, where the decoding fidelity is always a monotonically increasing function of $\kappa$. 
By contrast, non-monotonic dependence on aspect ratio is possible in principle when the path-integral description involves complex amplitudes. Indeed, as we show in Sec.~\ref{sec:ballistic_metal}, the decoding fidelity is an oscillatory function of the aspect ratio exactly at $\theta=\pi/4$ (the ``ballistic metal''). 
Thus, the decaying fidelity with the aspect ratio may be a special feature of the theory in the replica limit.
Alternatively, it may be attributed to the assumption of the empirical scaling as one takes the replica limit.
We leave a detailed analysis of the universal function describing the aspect-ratio-dependence of the optimal decoding fidelity at the metallic fixed point for future work.

\subsubsection{Numerical results}\label{sec:fidelity_predictions_opt_numerics}

We perform large-scale numerical simulations to determine the phase diagram and verify our analytical predictions.
We simulate the decoding fidelity $\calF_\opt$ in the surface code under coherent errors with a uniform rotation angle $\theta$ throughout the system (presented in Fig.~\ref{fig:fid_opt_combined}).
The fidelity is almost perfect for small $\theta$ and starts to deviate from unity in a finite-size system when $\theta$ becomes large [Fig.~\ref{fig:fid_opt_combined}(a)].
Importantly, we do not observe signatures of a potential phase transition when tuning $\theta$, indicating that a decodable phase may persist up to $\theta = \pi/4$.
According to our analysis in Sec.~\ref{sec:pdiagram_of_nlsm}, a transition between two insulating phases when tuning $\theta$ remains possible in theory.
In the case that such a transition exists, the decoding fidelity in the surface code with a fixed aspect ratio would be scale-free at the transition point; however, this is not observed numerically.


We further investigate the fidelity in the regime of large $\theta$ and compare it with the predictions of the sigma model.
Our results are consistent with the marginally relevant RG flow near the metallic fixed point, verifying the prediction that a metallic phase does not exist.
The infidelity $1-\calF_\opt$ decays exponentially in $\kappa$ for a fixed system size $L$, as $\kappa$ is proportional to the code distance $T = \kappa L$ [Fig.~\ref{fig:fid_opt_combined}(b, c)].
Close to the metallic fixed point at $\theta = \pi/4$, for a large $\kappa \gg 1,1/g_R$, the decay coefficient is controlled by the renormalized coupling $g_R(L)$ as in Eq.~\eqref{eq:fid_kappa_greater_than_1/gR}.
Our numerical results are consistent with this prediction; we observe that the decay coefficient increases as the system size increases [Fig.~\ref{fig:fid_opt_combined}(b)] or the rotation angle $\theta$ reduces [bare coupling increases, Fig.~\ref{fig:fid_opt_combined}(c)].

We note that the NLsM has a marginal RG flow near the metallic fixed point, which predicts that the infidelity should decay slowly with the system size when $\kappa \gg 1$, consistent with the results in Fig.~\ref{fig:fid_opt_combined}(b).
The decay of infidelity within the accessible system sizes is slower than an exponential decay in $L$, which would be observed in the small $\theta$ regime when we are deep in the decodable phase (or would also be observed in the decodable phase of the surface code with incoherent errors).

Moreover, the sigma model predicts ``a trend reversal'' in the fidelity as a function of $g_R$.
The infidelity decreases with $g_R$ for $\kappa \gg 1$, while it increases with $g_R$ for $\kappa \ll 1$.
In Fig.~\ref{fig:fid_opt_combined}(c), our numerical study verifies this prediction, as $\theta$ directly tunes the bare coupling, thus controlling $g_R$ at a fixed scale.
We note that in Fig.~\ref{fig:fid_opt_combined}(b), we observe the decrease of infidelity with $L$ at large $\kappa$, while the signal is not clear for small $\kappa$.
This is because, for $\kappa \ll 1$,  $T = \kappa L$ becomes comparable to the mean-free path $\lambda$ when the system size $L$ is small.
The sigma model is not a valid description in this regime, and we observe oscillations in the infidelity [Fig.~\ref{fig:fid_opt_combined}(b)] as expected on scales below the mean-free path (see Sec. \ref{sec:ballistic_metal}).

We further note that the fidelity in Fig.~\ref{fig:fid_opt_combined}(a) appears to be independent of $\theta$ when $\theta$ is large and system size $L = 64$.
We attribute this to a finite-size effect, which should be absent at larger scales that are inaccessible in our numerics.
The trend reversal demonstrated in Fig.~\ref{fig:fid_opt_combined}(c) suggests that for a fixed system size, there is an intermediate regime of $\kappa$ for which the fidelity depends weakly on $\theta$.
For $L = 64$, $\kappa = 3.5$ happens to be in this regime, leading to the behavior in Fig.~\ref{fig:fid_opt_combined}(a).
As the renormalized coupling $g_R$ grows with scale, flowing away from the metallic fixed point, the NLsM will cease to be a valid long-wavelength description, and we believe that the system will eventually settle into a decodable phase, in which the fidelity approaches unity, though we are unable to simulate large enough system sizes to directly see this behavior.

\begin{figure*}[t!]
\centering
\includegraphics[width=.99 \textwidth  ]{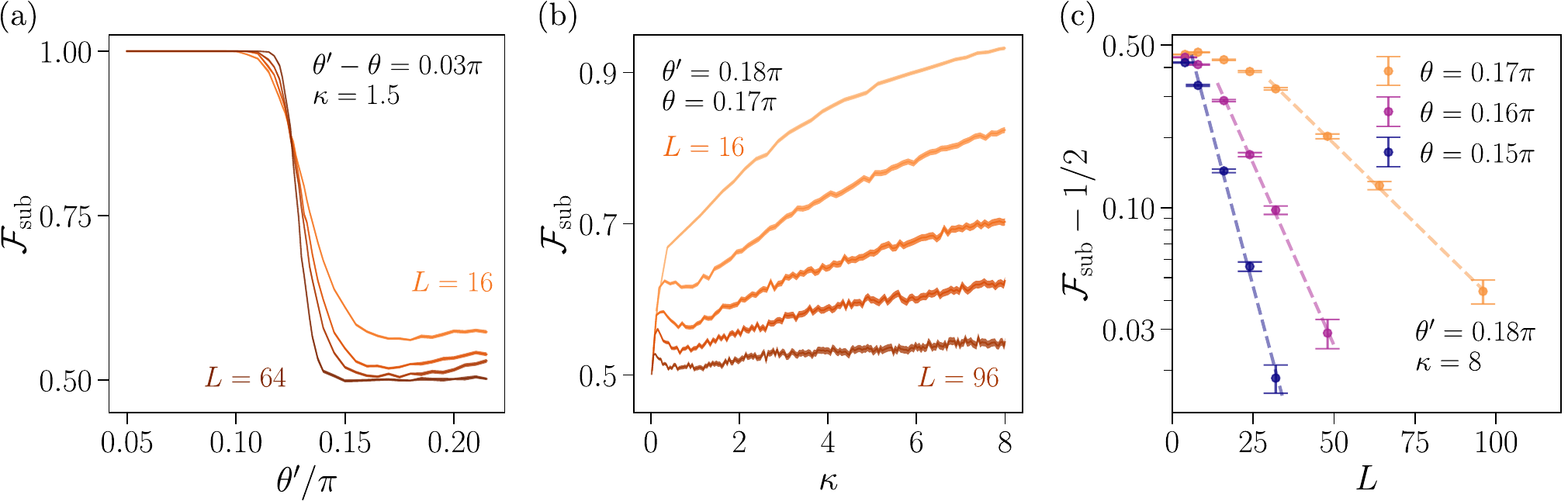}
    \caption{
    Fidelity of the suboptimal decoder $\calF_\mathrm{sub}$.
    (a) $\calF_\subopt$ as a function of rotation angle $\theta'$ at a fixed aspect ratio $\kappa = 1.5$ for various $L$. The curves for various system sizes cross at $\theta'_c \sim 0.13\pi$, indicating a decoding transition.
    (b) Fidelity as a function of $\kappa$ with $\theta' = 0.18\pi$, $\theta= 0.17\pi$.
    (c) Fidelity as a function of $L$ for a fixed $\kappa = 8$, $\theta' = 0.18\pi$, and various $\theta$. $\calF_\mathrm{sub}$ decays exponentially to $1/2$ in $L$, at a rate which increases with $\theta' - \theta$.
    The results are averaged over $2000$ to $27000$ samples.  The error bars in panels (a) and (b) are indicated by the linewidth.
    }
    \label{fig:fid_sub_combined}
\end{figure*}

\subsection{Fidelity of the suboptimal decoder}\label{sec:fidelity_subopt}
The fidelity of the suboptimal decoder takes a simpler form in the thermal metal phase, which is non-decodable and has fidelity $1/2$ up to a correction exponentially decaying in the system size. 
In what follows, we provide analytical reasoning and numerical simulations to demonstrate this result.

Recall that the replica NLsM for the suboptimal decoder describes an order parameter field  $Q\in \Gamma_{n+1}$, with massive fluctuations out of the reduced target space $\Gamma_{n}$.  
In the thermodynamic limit, this mass constrains $Q$ to the subspace $\Gamma_n$.
In this case, the twist expectation value in the RBIM picture [as in Eq.~\eqref{eqn:rbim_replicated_subopt_fidelity}] has the property that
\begin{align}
    \Psi_{2k} = \Psi_{2n-2k} \,,
\end{align}
which immediately implies that the decoding fidelity $\calF_\subopt^{(n)} = 1/2$ for any $n$ as well as in the replica limit $n \to 0$.

The finite-size decoding fidelity is governed by the renormalization group (RG) eigenvalue of the perturbation that suppresses fluctuations out of the reduced target space $\Gamma_{n}$ at the thermal-metal fixed point.
The associated potential has bare strength $m^2 = \calO(\epsilon^2)$ and is relevant under RG.
After coarse-graining to scale $L$, it renormalizes to $\epsilon^2 L^2$.  
This leads to a decoding fidelity which approaches $1/2$ up to an exponentially small finite-size correction (as shown in Appendix~\ref{app:nlsm_derivation_subopt}) 
\begin{align}\label{eq:suboptimal_fidelity}
    \calF^{(n)}_\subopt = \frac{1}{2} + e^{-\calO\left(L\abs{\epsilon}\sqrt{ g_0/g_R(L)}\right)} \,,
\end{align}
when $\epsilon^2 L^2 \gg 1$.
We expect this scaling in the replica limit $n \to 0$ with $g_R(L)$ governed by Eq.~\eqref{eq:n=0_RG_flow}.

To justify this scaling, it is convenient to analyze $\calF_\subopt^{(n)}$ in the dual picture [as in Eq.~\eqref{eqn:dual_replicated_subopt_fidelity}].
Here, the defect twists an odd number of replicas out of both the first $2$ and next $2n$ replicas.
Twisting an odd number of replicas in the reduced subspace $\Gamma_n$ changes the Pfaffian of $Q \in \Gamma_n$, i.e., $\Pf(Q) = -\Pf(\Lambda Q \Lambda)$ and forces the field configuration to vary in the massive direction.
This leads to an excess free energy $\calO(m(L)/g_R(L))$ controlled by the potential term $m(L) = \calO(L\abs{\epsilon}\sqrt{g_0 g_R(L)})$ at scale $L$.
This results in twist expectation values that decay exponentially in the system size, and hence the claimed scaling of the suboptimal decoding fidelity.

The fidelity being $1/2$ when the potential is large has a physical interpretation.
The replicated versions of the fidelity and infidelity ($1 - \calF_\subopt$) differ by inserting a twist defect in the first two replicas.
In the limit that the potential is infinite, the first two replicas and the next $2n$ replicas are decoupled in the field theory, and the partition function is invariant under such a twist.
Hence, both the fidelity and the infidelity are $1/2$ in this limit.

We perform large-scale numerical simulations to verify these predictions (Fig.~\ref{fig:fid_sub_combined}).
We simulate the fidelity of the suboptimal decoder in the surface code with uniform coherent rotations $\theta$ throughout the system.
In Fig.~\ref{fig:fid_sub_combined}(a), we consider a particular suboptimal decoder in which the estimated rotation angle $\theta'$ has a finite offset from the true rotation angle $\theta$, i.e., $\theta' = \theta + 0.03\pi$.
We observe that the fidelity $\calF_\subopt$ undergoes a phase transition at $\theta'_c \sim 0.13\pi$. 
For $\theta' < \theta'_c$, $\calF_\subopt$ increases with system size to unity, while $\calF_\subopt$ decays with $L$ to $1/2$ when $\theta' > \theta'_c$.

In the non-decodable phase, we show that the fidelity decays with the system size $L$ and increases with $\kappa$.\footnote{The non-monotonic behavior at small $\kappa$ occurs when $T = \kappa L$ is comparable to the crossover scale $1/\epsilon$, below which the suboptimal and the optimal decoder are indistinguishable.}
In particular, $\calF_\subopt - 1/2$ exhibits an exponential decay with a decay coefficient that increases with $\epsilon$ [Fig.~\ref{fig:fid_sub_combined}(c)].

\begin{figure*}[t]
\centering
\includegraphics[width=.95 \textwidth  ]{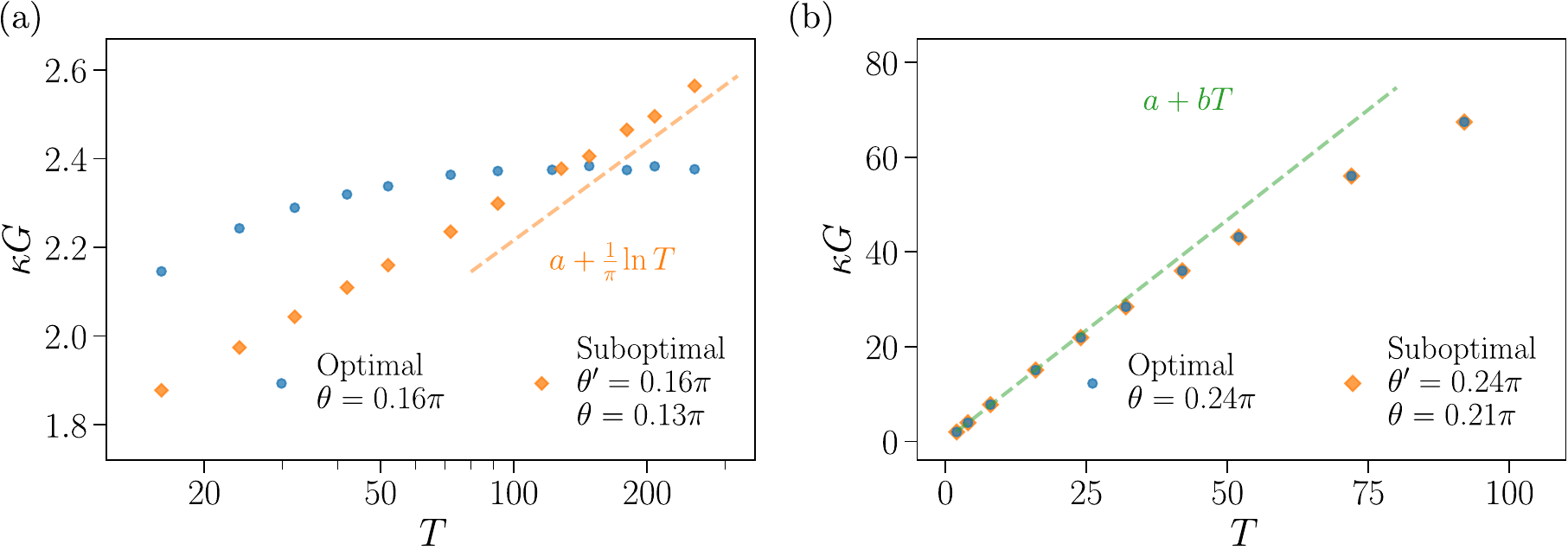}
    \caption{The conductivity $\kappa G$ of the network model 
    corresponding to the optimal (blue) and suboptimal (orange) decoder.
    The decoder's estimate of the coherent rotation angle ($\theta$ for the optimal and $\theta'$ for the suboptimal) is $0.16\pi$ (a) and $0.24\pi$ (b).
    For the suboptimal decoder, $\theta' -\theta = 0.03 \pi$ is fixed.
    In (a), the conductivity of the suboptimal decoder approaches the $\pi^{-1}\ln T $ scaling predicted by the NLsM in the thermal metal phase while the optimal decoder appears to plateau.
    Numerically distinguishing the two decoders is challenging due to proximity to the ballistic metal at $\theta = \pi /4$ giving rise to a large crossover scale below which $\kappa G$ increases rapidly as illustrated in (b).
    Here, the blue markers are on top of the orange markers.
    Data is presented with $500$ to $15000$ samples and with geometry $T = L/4$.
    All errorbars are within the marker size.}
    \label{fig:compare_decoder_conductance}
\end{figure*}

\section{Predictions of other physical quantities}\label{sec:other_nlsm_predictions}
In addition to the decoding fidelity, we utilize the effective non-linear sigma model to predict the scaling of various other physical observables in the statistical mechanical model describing the decoding problem.
These quantities serve as probes which can more sensitively discriminate the behavior of the sigma model describing the optimal and suboptimal decoders.

Specifically, we consider the thermal conductance in the network model formulation, the free energy cost of inserting a symmetry defect in the partition function $\calZ$ ($\calY$ for the suboptimal decoder), and the purification entropy in the dynamics defined by the transfer matrix.
We verify the predictions with numerical simulations in the case that the rotation angle $\theta$ is spatially uniform.
For the suboptimal decoder, we set the decoder's estimate $\theta'$ to differ from the true rotation angle $\theta$ by a constant offset, which is also spatially uniform.
We provide additional numerics in Appendix~\ref{app:additional_numerics_non-uniform} to verify that our predictions hold when the coherent rotation angles are drawn from a Gaussian distribution.

\subsection{Conductance}\label{sec:conductance}
The first quantity we consider is the thermal conductance $G$ (or equivalently, conductivity $\kappa G$) of the network model, the scaling of which is governed by the beta function of the effective NLsM~\cite{evers2008anderson}.
We demonstrate numerical evidence of distinct scalings for the network models associated with the optimal and suboptimal decoders.
In particular, for the suboptimal decoder, we observe quantitative agreement with the NLsM prediction within the thermal metal phase, while it is absent for the optimal decoder.
Furthermore, the conductance is expected to be a scale-free quantity at the metal-to-insulator transition, allowing us to estimate the critical rotation angle $\theta'_c$ of the suboptimal decoder.

The network model in class D describes a 2+1D dirty superconductor. 
Its thermal conductance characterizes the energy transport and is given by the Landauer formula~\cite{landauer_formula,d_fisher_landauer_formula,pichard_thesis}
\begin{equation}\label{eqn:conductance_definition}
    G = \tr
    \frac{2}{\mathsf{T}^\dagger \mathsf{T} + (\mathsf{T}^\dagger \mathsf{T})^{-1} + 2}
    = 
    \tr \mathsf{t}^\dagger \mathsf{t}
    \,,
\end{equation}
where $\mathsf{t}$ is the transmission block of the single particle transfer matrix $\mathsf{T}$ defined in Appendix~\ref{app:network_model_conductance}.\footnote{The transfer matrix of the $\U(1)$ network model is slightly different from Eq.~\eqref{eqn:network_model_node_transfer_matrices_form}.
The complex fermion for two copies of the Majorana network is defined as $c_{2i-1} = (\gamma_{2i-1}^{(1)} + \ri \gamma_{2i-1}^{(2)})/\sqrt{2}$ and $c_{2i} = (\gamma^{(2)}_{2i} - \ri \gamma_{2i}^{(1)})/\sqrt{2}$.
In terms of the complex fermions, the transfer matrices associated with horizontal and vertical bonds are given by
\begin{equation}\label{eqn:u1_network_node_transfer_matrices}
\begin{aligned}
    \sfh_{\bfr,\,\bfr+\hat{e}_x} &= \frac{\ri}{\sin2\theta} \begin{pmatrix}
        -\cos2\theta & \eta \\
        \eta & -\cos2\theta
    \end{pmatrix},\\
    \sfv_{\bfr,\,\bfr+\hat{e}_t} &= \eta\begin{pmatrix}
        -\cos2\theta & \ri\sin2\theta \\
        \ri\sin2\theta & -\cos2\theta
    \end{pmatrix}
    \,.
\end{aligned}
\end{equation}
These transfer matrices are equivalent to those for Majorana modes in Eq.~\eqref{eqn:network_model_node_transfer_matrices_form} up to local transformations.
}
We note that, technically, the Landauer formula computes the electrical conductance in the system with $\U(1)$ symmetry.
In practice, we consider two identical copies of class D networks, which is equivalently a network model of complex fermions with a $\U(1)$ symmetry.
The thermal conductance of the two copies is related to the electrical conductance of the complex fermion by the Wiedemann-Franz law~\cite{cho_fisher,merz_chalker,senthil2000quasiparticle}.

The conductance in a network with a fixed aspect ratio $\kappa = T / L$ exhibits distinct scalings with the system size in the three regimes considered here.
First, at the ballistic metal $\theta = \pi /4$ ($\theta' = \pi /4$ for the suboptimal decoder), the transfer matrix~\eqref{eqn:network_model_node_transfer_matrices_form} is unitary.
Thus, the conductance is linear in the system size at a fixed aspect ratio $\kappa$, i.e. $G \sim L$.

Next, in the thermal metal phase with $\theta < \pi /4$, the system is characterized by the effective NLsM at scales much greater than the mean-free path.
When $\theta$ is close to $\pi / 4$, the conductivity is proportional to the NLsM coupling $\kappa G = 1 / (2 \pi  g_R)$ and the two decoders are distinguished through the perturbative flow of the sigma model coupling constant.
Specifically, for the suboptimal decoder, we expect a stable metal phase in this regime with conductivity
\begin{align}
    \kappa G = \frac{1}{2\pi g_0} + \frac{1}{\pi}\ln L \,,
\end{align}
to leading order, with $1/(2 \pi  g_0)$ being the bare conductivity.
The prefactor $1/\pi$ is a universal number which follows directly from the perturbative RG flow equations.
For the optimal decoder, the metal phase is unstable, and the marginally relevant flow of $g_R$ predicts a decreasing conductivity with system size.

Finally, for small $\theta$, both decoders are in an insulating phase, with conductance decaying as $G  \sim e^{- L / \xi}$ with localization length $\xi$.

These predictions are consistent with our numerical simulation of the conductivity $\kappa G$ shown in Fig.~\ref{fig:compare_decoder_conductance}.
In Fig.~\ref{fig:compare_decoder_conductance}a, we observe that the conductivity of the suboptimal network model 
approaches the $\pi^{-1} \ln L$ scaling predicted by the perturbative RG.
In contrast, such scaling is absent for the optimal network, which exhibits a conductance that appears to plateau in the numerically accessible system sizes.
We remark that we do not observe the slow decrease of conductance with scale as predicted by the marginally relevant flow of the NLsM coupling.
This may be attributed to a very large crossover scale originating from proximity to the ballistic metal fixed point, beyond which the conductance begins to decrease.
The ballistic scaling of the conductance is presented in Fig.~\ref{fig:compare_decoder_conductance}b, demonstrating the mean-free path below which the conductance of both decoders grows rapidly with system size.

\begin{figure}[t]
    \centering
    \includegraphics[width=.95\linewidth]{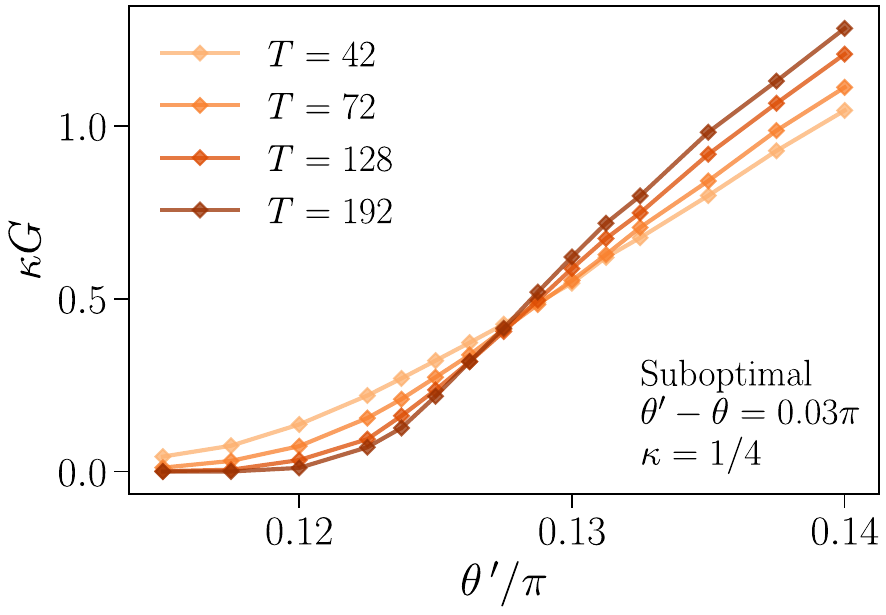}
    \caption{Conductivity as a function of $\theta'$ in the network model associated with the suboptimal decoder. 
    The system size $T$ is increased from lighter to darker color with fixed aspect ratio $\kappa = 1/4$ and $\theta = \theta' - 0.03\pi$.
    We estimate a critical rotation angle $\theta'_c = 0.127(3)\pi$, although we do not observe a scale-free critical conductivity due to finite-size effects.
    The results are averaged over $600$ to $3000$ samples, with all error bars within the size of the markers.}
    \label{fig:phase_transition_conductance}
\end{figure}

The conductivity at an Anderson metal-to-insulator transition is expected to be scale-invariant and take a universal value depending on only the symmetry class and geometry~\cite{slevin_ohtsuki_universality,slevin_ohtsuki_kawarabayashi_topology_dependence,braun_boundary_conditions}.
In Fig.~\ref{fig:phase_transition_conductance} we demonstrate the conductance of the suboptimal network when plotted against $\theta'$ for fixed $\theta = \theta' - 0.03\pi$.
However, we do not observe a perfect scale-invariant critical conductance, which we attribute to strong finite-size effects in the available system sizes.
Nevertheless, Fig.~\ref{fig:phase_transition_conductance} yields an estimate of the critical rotation angle $\theta'_c \sim 0.127(3)\pi$, which is consistent with the previous estimate using the decoding fidelity in Sec.~\ref{sec:fidelity_subopt}.
Furthermore, the value of the average conductivity in the neighborhood of $\theta'_c$ is consistent with previous numerical studies of Anderson localization in class D, which found $\kappa G_c \sim 0.41$~\cite{medvedyeva_effective_mass}.

We note that $\pi^{-1}\ln L$ conductivity has also been numerically observed in the network model associated with a particular decoding problem~\cite{behrends2024surface}.
Specifically, one obtains the syndromes from the surface code corrupted by incoherent errors and performs maximum-likelihood decoding by assuming that the syndromes are caused by coherent errors.
The network model description in this case involves disorder (i.e., bond variables $\eta_{\bfr, \, \bfr'}$) that are independently random.
This is different from the disorder in our network model associated with the suboptimal decoder, which is correlated.
Our analytic and numerical results suggest that the $\pi^{-1}\ln L$ scaling of conductance is a general feature of network models associated with the suboptimal decoding problem.

\begin{figure*}[t]
\centering
\includegraphics[width=.95 \textwidth  ]{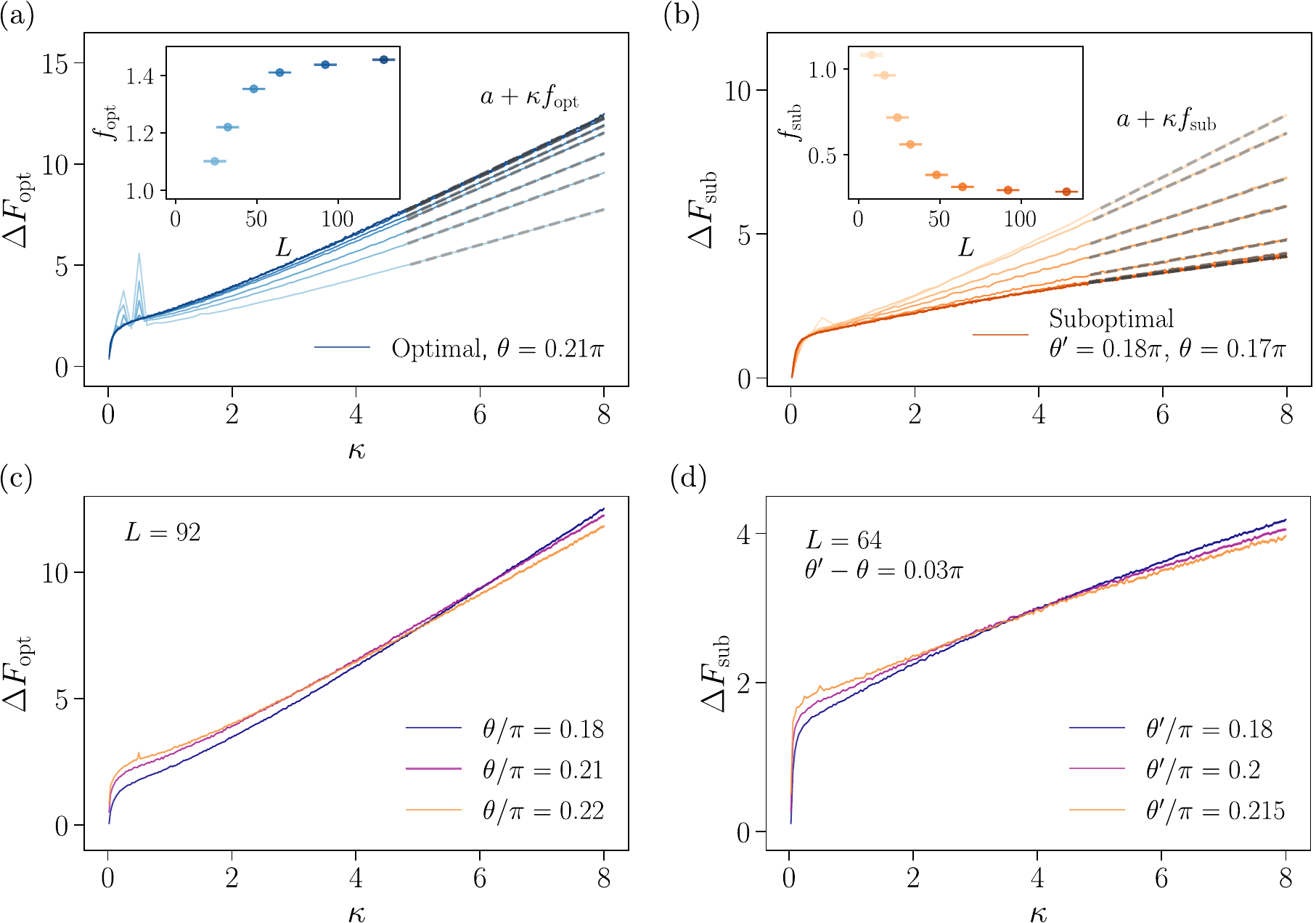}
    \caption{
    The defect free energy $\Delta F$ against aspect ratio $\kappa = T / L$ for the optimal (a) and suboptimal (b) decoder.
    Data is presented for $L = 8, 16, 24, 32, 48, 64, 92, 128$ from lighter to darker color.
    We find that $\Delta F$ is an increasing function of $\kappa$, which agrees with our analytic prediction.
    In the insets, we extract the slope $f$ through a linear fit for $\kappa > 1$.
    The NLsM predicts the slope to be proportional to $g_R(L)$, which is consistent with the numerical result that it increases with $L$ for the optimal and decreases for the suboptimal.
   In (c) and (d), we plot $\Delta F$ against $\kappa$ for fixed $L$ and for different values of $\theta$ (optimal) or $\theta'$ (suboptimal).
    Increasing $\theta$ ($\theta'$) decreases $g_R(L)$ by tuning $g$, leading to increased $\Delta F$ for $\kappa < 1$ and decreased $\Delta F$ for $\kappa > 1$ as predicted by the NLsM.
    Data generated with $6000$ to $27000$ samples.
    The error bars are indicated by the linewidth.
    }
    \label{fig:combined_dfe}
\end{figure*}

\subsection{Defect free energy}\label{sec:predictions_dfe}
Another physical quantity to study is the excess free energy for inserting a symmetry defect in the RBIM.
For the optimal decoder, this quantity is closely related to the decoding fidelity.
However, it exhibits different finite-size scalings close to the metallic fixed point, which distinguish the marginal RG flows for the optimal and suboptimal decoders.
We verify our predictions with extensive numerical simulations.

We introduce the excess free energy associated with the optimal and the suboptimal decoder,
\begin{align}
    \Delta F_\opt :=& \sum_s \calQ_s \abs{\ln \frac{\calQ_{0,s}}{\calQ_{1,s}}} \,, \\
    \Delta F_\subopt :=& \sum_s \calQ_s \abs{\ln \frac{\calP_{0,s}}{P_{1,s}}} \,.
\end{align}
The defect free energy exhibits distinct scalings in the usual decodable and non-decodable phase of the surface code with incoherent errors.
Specifically, $\Delta F = \calO(L)$ is linear in the system size in the decodable phase, as the probability of one homology class is exponentially larger than the other, while it is $\calO(1)$ in the non-decodable phase as the two homology classes are equally probable.

The defect free energy has an upper and a lower bound, which are more amenable for analytical studies,
\begin{align}
    \Delta F_{\opt/\subopt}^- \leq \Delta F_{\opt/\subopt} \leq \Delta F_{\opt/\subopt}^- + \ln 2 \,,
\end{align}
where 
\begin{align}
\Delta F_\opt^- &:= \sum_s \calQ_s \ln \frac{\calQ_{0,s}^2 + \calQ_{1,s}^2}{2\calQ_{0,s}\calQ_{1,s}} \,, \\
\Delta F^-_{\subopt} &:= \sum_s \calQ_s \ln \frac{\calP_{0,s}^2 + \calP_{1,s}^2}{2\calP_{0,s}\calP_{1,s}} \,.
\end{align}
These quantities $\Delta F^-_{\opt/\subopt}$ can be formulated as the replica limit $n \to 0$ of the following sequences,
\begin{align}
    \Delta F_{\opt}^{(n)} &= \frac{1}{n}\ln \frac{\sum_{s} \calQ_{s} (\calQ_{0,s}^2 + \calQ_{1,s}^2)^n}{\sum_s \calQ_s (2\calQ_{0,s}\calQ_{1,s})^n} \,, \\
    \Delta F_{\subopt}^{(n)} &= \frac{1}{n}\ln \frac{\sum_{s} \calQ_{s} (\calP_{0,s}^2 + \calP_{1,s}^2)^n}{\sum_s \calQ_s (2\calP_{0,s}\calP_{1,s})^n} \,.
\end{align}
In this way, we can relate the defect free energies for the optimal and suboptimal decoder to the twist expectation values in the sigma model
\begin{align}
    \Delta F_\opt^{(n)} &= \frac{1}{n} \ln \frac{\sum_{k = 0}^n \binom{n}{k} \Phi_{4k} }{2^n \Phi_{2n}} \,, \\
    \Delta F_{\subopt}^{(n)} &= \frac{1}{n} \ln \frac{\sum_{k = 0}^n \binom{n}{k} \Psi_{4k} }{2^n \Psi_{2n}} \,.
\end{align}
Here, $\Phi_{4k}$ is associated with twisting $4k$ replicas in the sigma model with target space $\SO(4n+2)/\U(2n+1)$.
The field theory for the suboptimal decoder is the sigma model with a potential term which breaks the symmetry down to $\SO(2) \times \SO(4n)$.
The twist expectation value $\Psi_{2n}$ is associated with twisting $4k$ out of the last $4n$ replicas.
In what follows, we use the twist expectation values analyzed in Appendix~\ref{app:twist_field} to make qualitative predictions of the defect free energies near the metallic fixed point.

We start with the defect free energy associated with the optimal decoder.
For fixed $L$, the twist expectation value $\Phi_{4k}$ always increases with $T$.
In the regime $\kappa \gg 1/g_R \gg 1$, $\Phi_{4k} = e^{-\calO(\kappa g_R)}$ decays exponentially except when $k = 0$.
This gives rise to $\Delta F_\opt = \calO(\kappa g_R)$ that is linear in $\kappa$ (when $L$ is fixed), as shown in Fig.~\ref{fig:combined_dfe}(a).
The coefficient of linear scaling is set by the renormalized coupling $g_R$ in the NLsM governed by the marginally relevant RG flow in Eq.~\eqref{eq:n=1_RG_flow}.
This predicts an increasing linear coefficient with the overall scale $L$ as shown in the inset of Fig.~\ref{fig:combined_dfe}(a).

The twist expectation value exhibits distinct scaling in the regime $\kappa \ll 1$, where it increases with the overall scale $L$.
Thus, we expect a ``reversed trend'' for the defect free energy; it decreases with $L$ when $\kappa \ll 1$ is fixed. 
However, for small aspect ratio, $T = L \kappa$ becomes comparable to the mean-free path for the system sizes simulated in Fig.~\ref{fig:combined_dfe}(a); the data presented for the small system sizes are not described by the sigma model.
To observe the reversed trend, we choose a large system size and vary the rotation angle $\theta$ to tune the bare coupling of the NLsM.
In Fig.~\ref{fig:combined_dfe}(c), we indeed observe that $\Delta F_\opt$ decreases with $g_R$ for small $\kappa$ and increases with $g_R$ for large $\kappa$.

For the defect free energy associated with the suboptimal decoder, we consider a scale $L \gg 1/\epsilon$ at which the suboptimal and optimal decoder become distinguishable, i.e., the potential term constrains the matrix field to the reduced target space $\Gamma_{2n}$.
Thus, $\Psi_{4k}^{(4n+2)} = \Phi_{4k}^{(4n)}$, where we use the superscript to label the total number of replicas.
In this case, $\Delta F_{\subopt}$ depends on $\kappa$ and $g_R$ in a similar way as in $\Delta F_{\opt}$.
First, for fixed $L$, $\Delta F_\subopt$ increases with $\kappa$.
At large $\kappa$, the twist expectation value is exponentially small and $\Delta F_\subopt = \calO(\kappa g_R)$ is linear in $\kappa$ with the linear coefficient set by the renormalized coupling.\footnote{Note that this scaling is valid when $\kappa \gg 1/g_R$ and does not hold in the thermodynamics limit when $g_R \to 0$.}
Crucially, the renormalized coupling $g_R$ is governed by the marginally irrelevant RG flow in Eq.~\eqref{eq:n=0_RG_flow}.
This indicates that the linear coefficient decreases with scale as shown in Fig.~\ref{fig:combined_dfe}(b).
We do not observe a clear signature for $\Delta F_{\subopt}$ increasing with scale $L$ at small $\kappa \ll 1$ since $T$ is again comparable to the mean-free path.
In Fig.~\ref{fig:combined_dfe}(d), we fix a large system size and observe that  $\Delta F_{\subopt}$ is controlled by the bare coupling set by $\theta$, the dependence on which exhibits the ``reversed trend.''

We note that for the suboptimal decoder, the slope of the defect free energy $\Delta F_\subopt$ as a function of $\kappa$ saturates to a constant as the system size increases, as shown in the inset of Fig.~\ref{fig:combined_dfe}(b).
This is because the effective theory flows to a conformal field theory (CFT) in the thermodynamic limit (approaching the thermal metal fixed point), at which the excess free energy of a symmetry defect in the vertical direction of a cylinder approaches a scale-invariant value given by $\Delta F_\subopt = 2\pi \Delta_\mu \kappa$ with $\Delta_\mu$ the scaling dimension of the defect operator~\cite{cardy1984conformal}.
The dependence of $\Delta F_\subopt$ on the aspect ratio at the thermal metal fixed point is distinct from that of the decoding fidelity, which is $1/2$ regardless of $\kappa$.

For the optimal decoder, the slope of $\Delta F_\opt$ as a function of $\kappa$ appears to increase slowly with $L$ for accessible system sizes [inset of Fig.~\ref{fig:combined_dfe}(a)].
We believe that this is the behavior at an intermediate scale, before the crossover to the proximity of the insulating fixed point. 

\begin{figure}[t]
\centering
\includegraphics[width=.95 \linewidth  ]{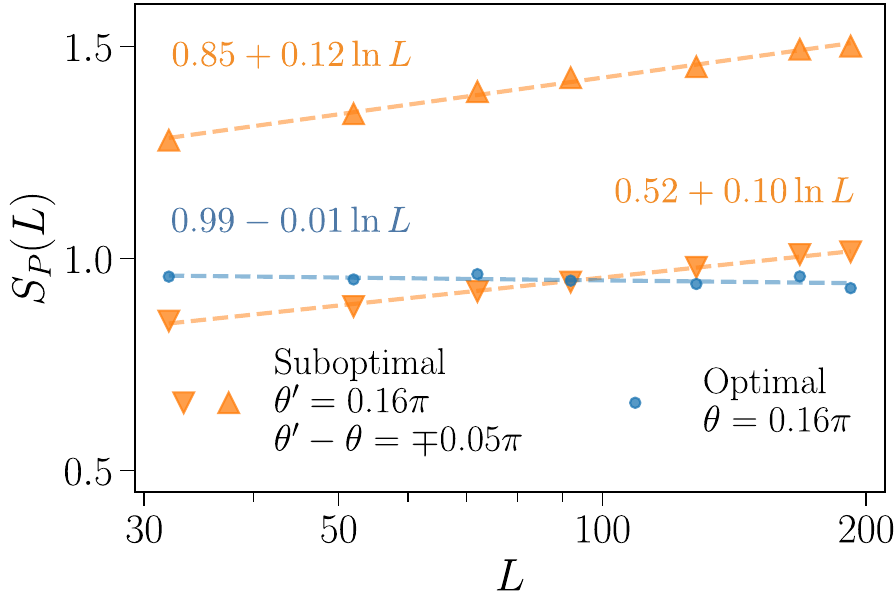}
    \caption{
    Scaling of purification entropy $S_P$.
    The entropy $S_P$ is computed in a 1D fermion system initialized in a maximally-mixed state and evolved under the 1+1D dynamics associated with the decoding algorithm after time $T=2L$.
    The blue dots and the orange upwards (downwards) facing triangles represent the results for the optimal decoder at $\theta = 0.16\pi$ and the suboptimal at $\theta' = 0.16\pi$ with $\theta' - \theta = -0.05\pi$ ($+0.05\pi$), respectively.
    At the numerically accessible systems sizes,
    $S_P \sim a + b \ln L$ with $b >0$ for the suboptimal decoders and $b < 0$ for the optimal decoder, consistent with the predictions from the sigma model.
    The data are generated with $600$ to $1000$ samples, with all error bars within the marker size.
    }
    \label{fig:combined_nlsm_predictions}
\end{figure}
\subsection{Purification dynamics}\label{sec:purification_dynamics}
Universal features of the optimal and suboptimal decoders are also revealed by the 1+1D purification dynamics associated with the decoding algorithm used to compute the coset probability $\calQ_{\alpha,s}$ and $\calP_{\alpha,s}$.
Specifically, in the decoding problem, the classical computer runs an algorithm that computes the coset probability based on the observed syndromes $s$ in the quantum device by simulating a 1+1D dynamics (described in Ref.~\cite{bravyi_coherent} and reviewed in Appendix~\ref{app:syndrome_sampling}).
We show that the purification entropy in this dynamics discriminates between the two decoders.
Namely, the dynamics purifies in the shortest time when the decoding algorithm knows the coherent rotation angle.
This suggests a practical way to characterize the error model of the underlying quantum device.

Specifically, the decoding algorithm simulates a particular $1+1$D free fermion dynamics involving unitaries that scramble, and measurements that generically purify a mixed initial state.\footnote{Technically, the $1+1$D dynamics defined by the decoding algorithm involves unitaries, measurements and ancillas.
At the end of each time step, the measured degrees of freedom are discarded and replaced by the ancillas.
We consider the purification dynamics in the state with a fixed number of fermions after the ancillas have replaced the measured fermions.
}
The unitaries applied to the state are determined by the estimated rotation angle $\theta'_\ell$ whereas the measurement outcomes are ``post-selected'' to be consistent with the syndrome measurements on the quantum device which experiences the true rotation angle $\theta_\ell$.
We show in Appendix~\ref{app:syndrome_sampling_dynamics} that these dynamics are equivalent to contracting the transfer matrix of the class D network model.
The subsequent mapping to the non-linear sigma model predicts distinct scalings of the purification entropy for the optimal and suboptimal decoders, which we verify by numerical simulation.

We begin with the purification dynamics of a maximally-mixed initial state evolved under the Gaussian dynamics.
The central quantity of interest is the purification entropy, i.e., the von Neumann entropy of the final state, averaged over trajectories
\begin{align}
    S_P = -\sum_s \calQ_s \tr \rho_s \ln\rho_s
    \,,
\end{align}
with $\rho_s = \varrho_s / \tr \varrho_s$ where $\varrho_s$ is the unnormalized free fermion state after $T$ time steps under dynamics with rotation angle $\theta_\ell$ ($\theta'_\ell$) for the optimal (suboptimal) decoder starting from the maximally-mixed state.
Following~\cite{jian2020measurement}, this quantity can be formulated as the $k\to 0$, $n\to1$ limit of the following replica sequence
\begin{align}
S_P^{(n,k)} = \frac{1}{(1-n)k}\ln\frac{\sum_s \calQ_s (\tr\varrho^n_s)^k}{\sum_s \calQ_s (\tr\varrho_s)^{nk}}
\,.
\end{align}
This can be expressed in terms of transition amplitudes of the averaged transfer matrix $\hat{\bfT}$~\eqref{eqn:many_copy_rbim_prob_amplitude} acting on $2nk + 2$ replicas
\begin{align}\label{eqn:purification_entropy_transition_amplitude}
    S_P^{(n,k)} = \frac{1}{(1-n)k}
    \ln\frac{(\bbC_{n,k}|\hat{\bfT}_{T} \cdots \hat{\bfT}_1|\mathds{1}_{n,k})}
    {(\mathds{1}_{n,k}|\hat{\bfT}_{T} \cdots \hat{\bfT}_1|\mathds{1}_{n,k})}
    \,.
\end{align}
The transition amplitude is between the initial and final states $|\mathds{1}_{n,k})$ and $|\bbC_{n,k})$, which are chosen to yield $\calQ_s$ in the first $2$ replicas and $(\tr \varrho_s^n)^k$ in the final $2nk$ replicas.
For the optimal decoder, the coherent rotation angle is $\theta_\ell$ in all $2nk + 2$ replicas, leading to a $\O(2nk+2)$ symmetry.
On the other hand, the rotation angle $\theta'_\ell$ differs for the last $2nk$ replicas of the suboptimal decoder, leading to a reduced $\O(2) \times \O(2nk)$ symmetry.
Again, this leads to distinct limits $nk \to 1$ ($nk \to 0$) in the effective NLsM with target space $\Gamma_{nk} = \SO(2nk) / \U(nk)$ describing the optimal (suboptimal) decoder.

The purification entropy in the 1+1D monitored dynamics of free fermions with $\bbZ_2$ parity symmetry has been analyzed in Ref.~\cite{fava2023nonlinear} using an effective NLsM derived for a different microscopic model.
On general grounds, we expect that the initial and final boundary states~\eqref{eqn:purification_entropy_transition_amplitude} break the $\SO(2nk)$ symmetry and favor a particular matrix field $Q$ in the coset space $\SO(2nk) / \U(nk)$.
Accordingly, the purification entropy maps to the free energy cost of inserting a domain wall around the spatial direction.
For a system of size $L$ undergoing dynamics for time $T > L$, the free energy cost implies the scaling
\begin{align}\label{eqn:purification_entropy}
    S_P \propto \frac{1}{g_R(L)}\frac{L}{T}
\end{align}
for the purification entropy, up to some non-universal constant.

The purification entropy should distinguish the suboptimal decoder from the optimal decoder in the regime of large coherent rotation $\theta$.
For the optimal decoder, the marginally relevant flow of the coupling $1/g_R(L) = (1/g_0^2 - 8\ln L)^{1/2}$ gives rise to the purification entropy  $S_P \propto L(1-4g_0^2\ln L)/T$ close to the thermal metal fixed point $g_0^2 \ln L \ll 1$.
On the other hand, the coupling is marginally irrelevant for the suboptimal decoder, leading to $S_P \sim L(\ln L)/T$.
Thus, in both cases, we expect $S_P \propto \kappa (a + b \ln L)$ with $b > 0$ for the suboptimal and $b < 0$ for the optimal decoder.
This functional form of the scaling is consistent with numerical calculations of $S_P$ for a maximally-mixed initial state under the free fermion dynamics defined by the decoding algorithm.
In Fig.~\ref{fig:combined_nlsm_predictions}, we find that $b > 0$ for two suboptimal decoders with different values of $\theta' - \theta$ while $b < 0$ for the optimal decoder.
The small value of $b$ for the optimal decoder is consistent with the fact that $g_0^2 \ln L$ is small. 

Finally, we comment that measurements are typically more effective at purifying a mixed state when the outcomes are selected according to the Born probabilities of the underlying state compared to when they are randomly selected independent of the state (commonly referred to as ``forced'' or ``post-selected'').
This effect has previously been studied for various $0+1$D models in Refs.~\cite{fidkowski_how_quantum_memories_forget,deluca_universality_classes_purification,giachetti_deluca_elusive_phase_transition,bulchandani_sondhi_chalker_purification}.
In our case, $\calQ_s$ is a probability amplitude between two particular pure states~\eqref{eqn:single_copy_rbim_prob_amplitude} and is not directly related to the Born probability $\tr \varrho_s$ of the state undergoing the purification dynamics.
Therefore, the different scalings of the purification entropy with system size in~\eqref{eqn:purification_entropy} for the optimal and suboptimal decoders is a direct consequence of the distinct symmetries of the bulk dynamics and subsequent RG flows in the two replica limits of the effective NLsM.

\section{Ballistic metal at \texorpdfstring{$\theta = \pi /4$}{θ=π/4}}\label{sec:ballistic_metal}
So far, we have studied the NLsM that emerges as an effective description of the decoding problem at scales large compared to the mean-free path $\lambda$.
However, exactly at $\theta = \pi /4$, the mean-free path diverges and the decoder is no longer described by a NLsM at any scale.
Instead, this point corresponds to a ``ballistic metal'' distinct from the ``thermal metal'' fixed point of the NLsM at weak coupling.
In this section, we provide a microscopic derivation of the fidelity of the optimal decoder for completeness.

When $\theta = \pi / 4$, the error channel takes the initial logical state $\ket{+}_L$ to
\begin{equation}
    \ket{+}_L \mapsto \prod_\ell e^{\ri \frac{\pi}{4} Z_\ell} \ket{+}_L \,,
\end{equation}
followed by Pauli-stabilizer measurements and Pauli recovery, all of which are Clifford gates~\cite{gottesman_stabilizer_codes}.
This implies that, for fixed syndrome measurement outcomes $s$, the effective gate $V_s$ applied on the logical subspace is also Clifford and takes the form
\begin{equation}
    V_s =  e^{\ri m_s \frac{\pi}{4} X_L} e^{\ri n_s\frac{\pi}{4} Z_L}\,,
\end{equation}
with $n_s, \, m_s \in \mathbb{Z}$.
The phase gate $e^{\ri m_s \frac{\pi}{4} X_L}$ enforces certain reality conditions depending on the parity of the code distance~\cite{bravyi_coherent} but does not affect the overall fidelity.
On the other hand, the value of the integer $n_s$ determines whether one can recover the encoded state by applying Pauli operators.
Specifically, when $n_s$ is odd, the code remains in an equal superposition of both logical states post-recovery and fidelity is $1/2$, while $n_s$ even results in fidelity $1$.
In fact, it is a special feature of the $\theta = \pi / 4$ point that for fixed code geometry, the fidelity is either $1/2$ or $1$ independently of the syndromes that are actually observed.
As a result, $\calF_\mathrm{ML}$ and $\calF_\mathrm{opt}$ are equivalent and we may refer to either as $\calF$.

\begin{figure}[t]
\centering
\includegraphics[width=.95 \linewidth  ]{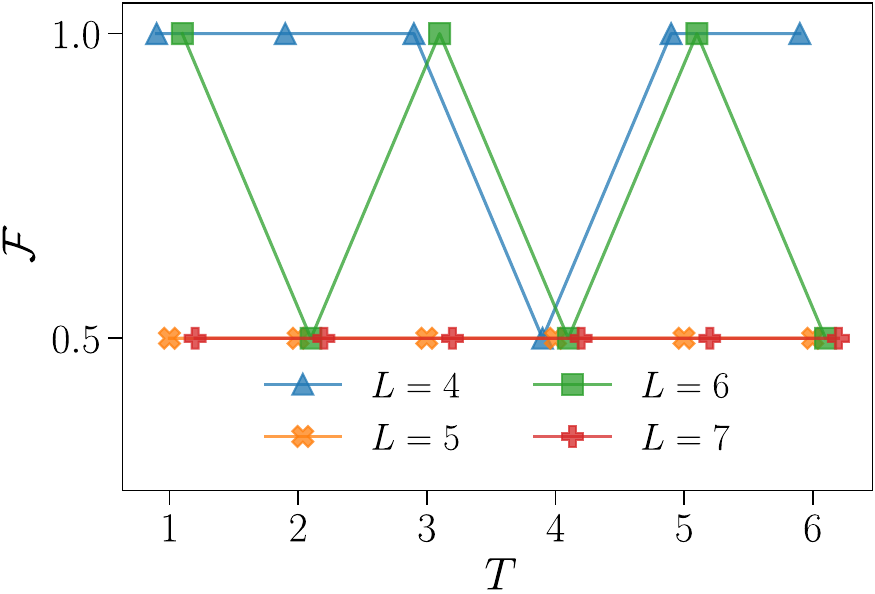}
    \caption{
    Fidelity $\calF$ at $\theta = \pi/4$ for various system sizes $L$ as a function of the cylinder length $T$.
    The results are averaged over $20$ syndrome realizations; however, we have observed that $\calF$ is independent of syndrome configuration and is a function of only $L$ and $T$.
    }
    \label{fig:ballistic_metal}
\end{figure}

The dependence of $\calF$ on both dimensions of the code $L$, $T$ is more complicated and generally oscillates between $1$ and $1/2$.
Our main result is that, for the cylindrical geometry with circumference $L$ and height $T$, the fidelity is
\begin{equation}\label{eqn:pi_four_oscillating_fid}
    \calF = \begin{cases}
        1  , &  L / \gcd(L, T) \equiv 0 \mod 2 \\
        1/2 ,  & L / \gcd(L, T) \equiv 1 \mod 2
    \end{cases}
    \,.
\end{equation}
That is, $\calF = 1/2$ whenever $T$ is a multiple of the largest power of $2$ that divides $L$ and $\calF = 1$ otherwise.
The fidelity $\calF$ for various $L$, $T$ is illustrated in Fig.~\ref{fig:ballistic_metal}.

\begin{figure}[t]
    \centering
    \includegraphics[width=.45\linewidth]{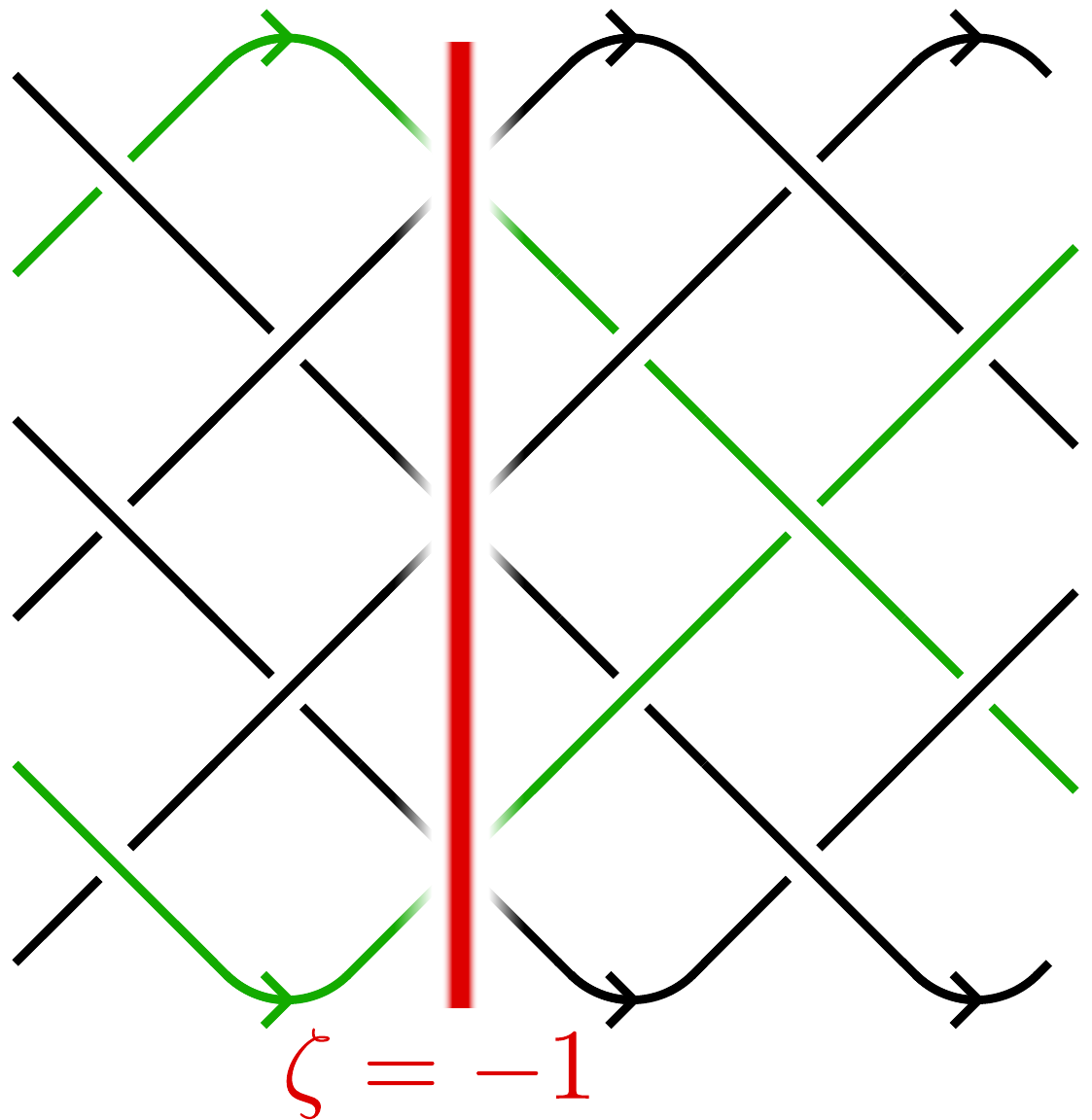}
    \caption{SWAP-gate circuit on the cylinder for ($L$, $T$) $=$ ($3$, $3$).
    Periodic boundary conditions are imposed in the horizontal direction, with the left boundary identified with the right.
    A $\pi$-phase is acquired by the Majorana worldline (green) across a symmetry defect (red).
    In this case, the winding number of the worldline is even, leading to $\calF = 1/2$.}
    \label{fig:majorana_circuit}
\end{figure}

This result has an intuitive explanation (see Appendix~\ref{app:pi_four_point} for a detailed derivation).
The decoding fidelity is determined by the relative values of $\calQ_{\alpha, \, s}$, which in turn are related to the transition amplitude of the RBIM transfer matrix~\eqref{eqn:single_copy_rbim_prob_amplitude}
\begin{equation}
    \calQ_{\alpha | s} \propto |\calZ_{\alpha, \,s}|^2  = |(\psi_0| V_{\alpha,\, s} |\psi_0)|^2 \,.
\end{equation}
In the Majorana formulation, the circuit dynamics, described by $V_{\alpha,\, s}$, consists entirely of SWAP-gates.
The boundary conditions, given by $|\psi_0)$, are paired states and can be viewed as imposing ``reflecting'' boundary conditions for the propagation of the left- and right-moving chiral Majoranas.
The worldline of one such Majorana is pictured in Fig.~\ref{fig:majorana_circuit}; upon returning to the initial location, the final amplitude is either zero or non-zero depending on the total phase accrued as it winds the cylinder.
Importantly, $\calZ_{0, \, s}$ and $\calZ_{1, \, s}$ are distinguished by the insertion of a symmetry defect across which the Majorana acquires a $\pi$-phase.
As a result, if $L$ and $T$ are such that the winding number of any given worldline is even, the symmetry defect has no effect and $\calQ_{1,\,s} = \calQ_{0,\,s}$ such that $\calF = 1/2$.
On the other hand, when the winding number is odd, $\calQ_{1,\,s}$ is non-zero whenever $\calQ_{0,\,s}$ is zero, and vice versa, leading to $\calF = 1$.
The parity of this winding number is related to the order of $T$ as an element of $\mathbb{Z}_L$, leading to our result~\eqref{eqn:pi_four_oscillating_fid}.

We note that a similar calculation can be carried out for the surface code on a torus or on a rectangle with open boundaries. 

\section{Discussion}\label{sec:conclusion}
In this work, we study decoding in the surface code subject to single-qubit coherent errors creating electric anyon excitations.
We focus on decoders that first estimate, for each set of measured syndromes, the total weight of error processes in each homology class and then apply a recovery operator that is a product of single-qubit Pauli operators.
In this setting, we derive an effective non-linear sigma model with target space $\SO(2n)/\U(n)$ that governs the performance of decoding.

Our central result is that optimal and suboptimal decoding are governed by different replica limits, $n \to 1$ and $n \to 0$, respectively. 
This leads to qualitatively distinct decoding phase diagrams—for the optimal decoder, the metallic fixed point is unstable, whereas for suboptimal decoders there is a stable thermal metal phase. 
The resulting field theoretic description allows us to predict the scaling of the decoding fidelity, as well as other observables in the associated statistical-mechanics formulation of the decoding problem. 
Our derivation of the NLsM is based on a perturbative expansion about the metallic fixed point at the maximally coherent rotation angle, and the stability analysis of this fixed point provides a useful organizing principle for the global decoding phase diagram.
Similar analysis of fixed points at the maximum error rate has been carried out for the surface code under Pauli-Y decoherence~\cite{hauser2026information} and fractional quantum Hall states under density dephasing~\cite{wang2025fractional}. 

\subsection{Distinct sigma models and phases of the triangular-lattice surface code}\label{sec:conclusion_lattice}
The emergence of the $\O(2n)$ symmetry in the replicated partition function and the resulting $\SO(2n)/\U(n)$ target space for the NLsM relies crucially on the bipartiteness of the lattice on which the syndromes reside.
Specifically, when deriving the NLsM in the dual picture [see Sec.~\ref{sec:nlsm_opt_dual}], this bipartiteness gives rise to a sublattice transformation of the dual Ising spins, which directly relates the partition sum $\calZ_{+,\,s}$ to its complex conjugate $\calZ_{+,\,s}^{*}$ (\ref{eq:bipartite_transformation_rbim}).
This leads to an enlarged $\O(2n)$ replica symmetry.
The enlarged symmetry also appears in the derivation of the NLsM in the RBIM picture [see Sec.~\ref{sec:nlsm_opt_rbim}].
In that case, the Ising spins live on sites of the lattice dual to the one on which the syndromes reside; the enlarged replica symmetry is present if this lattice has an even coordination number, which is equivalent to the bipartiteness of the syndrome lattice.

From another perspective, the error strings compatible with a fixed syndrome configuration $s$ on a bipartite lattice necessarily have a total length that only differs by an even number.
Thus, the amplitudes $A(\calC) = (\ri \tan\theta)^{|\calC|}$ and $A^{*}(\calC)$ of a fixed error string $\calC$ in the partition sums $\calZ_{\alpha,\,s}$ and $\calZ_{\alpha,\,s}^*$ are related through $A(\calC) = f_s A^*(\calC)$ by an overall factor $f_s$ which only depends on the syndrome configuration but not the choice of error string $\calC$.
This leads to the enlarged replica symmetry.

By contrast, if the syndromes reside on a non-bipartite lattice, the partition sums $\calZ_{\alpha,\,s}$ and $\calZ^*_{\alpha,\,s}$ are not related by a local transformation.
As an example, one can consider the triangular-lattice surface code with coherent error $e^{\ri\theta_\ell Z_\ell}$ creating syndromes that reside on vertices of the triangular-lattice.
In this case, the replicated partition function would only exhibit an $(\O(n) \times \O(n)) \rtimes \bbZ_2$ symmetry.
This replica symmetry is spontaneously broken at a saddle point with a non-vanishing Hubbard-Stratonovich matrix field $Q = (0, -o^T; o, 0)$, where $o \in \O(n)$.
This order parameter field $Q$ is invariant under an orthogonal transformation $Q = O Q O^T$ of the form $O = \mathrm{diag}(o_1, o \, o_1\, o^T)$ for any $o_1 \in \O(n)$.
As a result, the fluctuations around the saddle point are characterized by the $\SO(n)$ NLsM (neglecting disconnected components of the target space).
This sigma model has been well-studied in the context of Anderson localization in symmetry class DIII~\cite{AZ_10fold}.
It is well-known that the $\SO(n)$ NLsM admits a stable thermal metal phase, present for \emph{both} $n \to 0$ and $n \to 1$~\cite{evers2008anderson} replica limits, in contrast to that for class D.
This indicates that the optimal decoder can also witness a thermal metal phase.
The consequences of this observation for the universal properties of decoding will be discussed elsewhere~\cite{forthcoming}.

\subsection{Outlook}
Our results also open several directions for future work.
First, the decoding phase diagram of the two-dimensional surface code bears a strong resemblance to that of the monitored dynamics of the one-dimensional repetition code~\cite{hauser2025information}.
This raises the possibility that the repetition code, when subjected to weak stabilizer measurements and coherent single-qubit rotations preserving the underlying $\mathbb{Z}_{2}$ symmetry, may admit an analogous effective field theory description.

More broadly, our theory reveals universal differences between optimal and suboptimal decoders.
These differences may provide experimentally useful signatures of the underlying error model in a quantum device and may also guide the design of decoders that infer or learn the error model directly from observed syndrome data.

\begin{acknowledgments}
We thank Jan Behrends, Benjamin B\'eri, Matthew P. A. Fisher, Jacob Hauser, Andreas W. W. Ludwig, Adam Nahum, Rushikesh A. Patil, Simon Trebst, and Guo-Yi Zhu for helpful discussions.
We thank Xudong Dai for helpful comments on an earlier version of the manuscript.
S.W.Y. and S.V. acknowledge the support of the National Science Foundation under Grant No. DMR–2441671. 
Y.B. is supported in part by grant NSF PHY-2309135 and the Gordon and Betty Moore Foundation Grant No. GBMF7392 to the Kavli Institute for Theoretical Physics (KITP).
This research was done using services provided by the OSG Consortium~\cite{osg_1,osg_2,osg_3,osg_4}, which is supported by the National Science Foundation awards No. 2030508 and No. 2323298.

\emph{Note added.}--- Upon completion of the current manuscript, we became aware of a related work, which also discusses how the lattice structure affects optimal decoding in the surface code with coherent errors~\cite{chaoming_forthcoming}.

\end{acknowledgments}

\newpage
\appendix
\section{Performance of the probabilistic decoder}\label{app:performance_prob_decoder}
In this appendix, we show that the probabilistic decoder and the maximum-likelihood decoder exhibit the same decoding threshold provided that the decoding infidelity below the threshold is asymptotically zero.

On one hand, the decoding infidelity of the ML decoder has an upper bound
\begin{align}
    \Delta_{\mathrm{ML}} = 1 - \calF_{\mathrm{ML}} &= \sum_s \calQ_s \left(1 - e^{-\mathrm{H}^{(\infty)}[\calQ_{\alpha|s}]}\right) \nonumber \\
    &\le \sum_s \calQ_s \left(1 - e^{-\mathrm{H}^{(2)}[\calQ_{\alpha|s}]}\right) = \Delta_2\,,
\end{align}
where $\mathrm{H}^{(2)}[\calQ_{\alpha|s}] = -\ln \sum_\alpha {\calQ}_{\alpha|s}^2$ and $\mathrm{H}^{(\infty)}[{\calQ}_{\alpha|s}] = -\ln \max_\alpha{\calQ}_{\alpha|s}$
are the R\'enyi entropy of order-$2$ and order-$\infty$ for the conditional distribution
$\{\calQ_{\alpha|s} = \calQ_{\alpha, s}/\calQ_s\}$, respectively.
On the second line, we use the monotonicity of R\'enyi entropy as a function of index $\alpha$, $\mathrm{H}^{(2)} \ge \mathrm{H}^{(\infty)}$.

On the other hand, one can lower bound the infidelity $\Delta_{\mathrm{ML}}$ of the ML decoder using that of the probabilistic decoder $\Delta_2$:
\begin{align}
    \Delta_{\mathrm{ML}} &= \sum_s \calQ_s \left(1 - e^{-\mathrm{H}^{(\infty)}[\calQ_{\alpha|s}]}\right) \nonumber \\
    &\ge \sum_s \calQ_s \frac{1 - 2^{-K}}{K\ln 2} \mathrm{H}^{(\infty)}[\calQ_{\alpha|s}] \nonumber \\
    &\ge \frac{1 - 2^{-K}}{2K\ln 2} \sum_s \calQ_s  \mathrm{H}^{(2)}[\calQ_{\alpha|s}]\nonumber \\
    &\ge \frac{1 - 2^{-K}}{2K\ln 2} \sum_s \calQ_s \left(1 - e^{-\mathrm{H}^{(2)}[\calQ_{\alpha|s}]}\right) \nonumber \\
    &= \frac{1 - 2^{-K}}{2K\ln 2} \Delta_2\,.
\end{align}
For the first inequality, we use the fact that the maximum value of $\mathrm{H}^{(\infty)} = K\ln 2$ with $K$ being the number of logical qubits.
For the second inequality, we use $2\mathrm{H}^{(\infty)} \ge \mathrm{H}^{(2)}$.

We thus obtain both an upper and a lower on the infidelity of the ML decoder
\begin{align}
    \frac{1 - 2^{-K}}{2K\ln 2} \Delta_2 \le \Delta_{\ML} \le \Delta_2\,.
\end{align}
Assuming that, in the thermodynamic limit, the infidelity $\Delta_\ML$ of the ML decoder vanishes below the threshold, then the ML decoder and the probabilistic decoder exhibit the same threshold.

\section{Statistical Kramers-Wannier Duality at \texorpdfstring{$\theta = \pi / 4$}{θ=π/4}}\label{app:kw_duality}
In this appendix, we demonstrate that $\mathcal{Q}_{\alpha, \, s} = |\mathcal{Z}_{\alpha, \, s}|^2$ 
exhibits a statistical Kramers-Wannier symmetry at $\theta = \pi /4$.
The coset probability may be written in terms of a bra and ket copy of
the transfer matrices in
Eq.~\eqref{eqn:complex_coupling_RBIM_transfer_matrix_spin_representation}.
At the $\theta = \pi / 4$ ($J=0$) point, this takes the following form in the fermion representation
\begin{align}
    \hat{\mathsf{v}}^{(2)}_{\bfr, \, \bfr + \hat{e}_t} &=
    \exp \ri \frac{\pi}{4}(2 - \eta_{\bfr, \, \bfr +
    \hat{e}_t})
    \left(
    \ri\gamma_{2i-1} \gamma_{2i}
    -
    \ri\tilde{\gamma}_{2i-1} \tilde{\gamma}_{2i}
    \right)
\end{align}
and
\begin{align}
    \hat{\mathsf{h}}^{(2)}_{\bfr, \, \bfr + \hat{e}_x} &=
    \exp
    - 
    \frac{\ri \pi}{4}
\eta_{\bfr,\, \bfr + \hat{e}_x}
\left(
    \ri\gamma_{2i} \gamma_{2i+1}
    -
    \ri\tilde{\gamma}_{2i} \tilde{\gamma}_{2i+1}
    \right)
    \,.
\end{align}
In this appendix, we adopt the convention that $\gamma$ ($\tilde\gamma$) is the
Majorana mode on the ket (bra) copy.
In the first line, an $\eta$-dependent phase cancels between the ket and bra
copy of the transfer matrix.

The row transfer matrix then takes the form
\begin{widetext}
\begin{equation}
    \hat{\mathsf{T}}^{(2)}_j = \prod_{i = 1}^L \left(-\gamma_{2i-1}
    \tilde\gamma_{2i-1}\gamma_{2i} \tilde\gamma_{2i}\right)
    \prod_{i=1}^L
    e^{\eta_{\bfr, \, \bfr + \hat{e}_t}\frac{\pi}{4}(\gamma_{2i-1}\gamma_{2i} - \tilde{\gamma}_{2i-1}
    \tilde{\gamma}_{2i})}
    \prod_{i=1}^L
    e^{\eta_{\bfr, \, \bfr + \hat{e}_x}\frac{\pi}{4}(\gamma_{2i}\gamma_{2i+1} - \tilde{\gamma}_{2i}
    \tilde{\gamma}_{2i+1})}
    \,.
\end{equation}
    
\end{widetext}
We observe that this is invariant under the Kramers-Wannier symmetry which
translates Majoranas by one site (for example $\gamma_{2i} \mapsto
\gamma_{2i+1}$) provided that the syndrome configuration $s \mapsto s'$ transforms according to $\eta_{\bfr,\, \bfr + \hat{e}_t} \mapsto
\eta_{\bfr, \, \bfr + \hat{e}_x }$ and $\eta_{\bfr, \, \bfr + \hat{e}_x} \mapsto
\eta_{\bfr + \hat{e}_x, \, \bfr + \hat{e}_x + \hat{e}_t}$.
Thus, the Kramers-Wannier transformation applied to both the bra and ket $\calQ_{\alpha, \, s} \mapsto \calQ_{\alpha, \, s'}$ is a ``weak'' symmetry of the disorder-averaged ensemble, for instance, leaving the replicated partition function $\mathbf{Z}_0$ in Eq.~\ref{eq:replicated_partition_rbim} invariant.
This demonstrates the statistical Kramers-Wannier symmetry of $\calQ_{\alpha,\, s}$.

\begin{widetext}
    
\section{Derivation of the non-linear sigma model}\label{app:nlsm_derivation}
In this appendix, we provide an explicit derivation of the non-linear sigma model for the optimal decoder, mapping $\mathbf{Z}_0$ to the partition function of the NLsM with target space $\SO(2n)/\U(n)$ (Appendix~\ref{app:nlsm_derivation_opt}).
The symmetry defect in the replicated RBIM is identified with the twist defect that acts on the coarse-grained matrix fields in the NLsM.
With small modifications, we also derive an effective field theory for the suboptimal decoder (Appendix~\ref{app:nlsm_derivation_subopt}).

\subsection{Non-linear sigma model for the optimal decoder}\label{app:nlsm_derivation_opt}

\subsubsection{The local constraint}\label{app:local_constraint}
We begin by addressing the role of the local constraints $K$ and $\tilde{K}$  in~\eqref{eqn:averaged_transfer_matrix_horizontal_rbim},~\eqref{eqn:averaged_transfer_matrix_vertical_rbim}.
We focus on the RBIM since the discussion in the dual picture is analogous.
When the partition function is written in terms of a transfer matrix as in~\eqref{eqn:many_copy_rbim_prob_amplitude}, the constraints correspond to the insertion of a series of projection operators at each time step $j$~\eqref{eqn:constraints_in_rbim}.
These take the form
\begin{align}
\prod_i
 \frac{1 + (-1)^n\calO_{2i-1, 2i}}{2}
    \frac{1 + \calO_{2i, 2i+1}}{2}
    \,,
\end{align}
where we defined $\calO_{i, i'} = \prod_\sfa \ri \gamma_{i}^\sfa \gamma_{i'}^\sfa$ in the Majorana representation.

At each time step, the total set of projectors fix a stabilizer group.
We now show that the size of this group does not grow under the dynamics defined by the transfer matrix such that the projections can be deferred to the final time step.
Specifically, note that $\calO_{i, i'}$ commutes with all the terms in the transfer matrix, except possibly the SWAPs.
Thus, it suffice to show that the stabilizer group does not grow under the SWAP dynamics.

First,
the total stabilizer group at time $j=1$ is generated by the set
\begin{equation}\label{eqn:stab_group_rbim}
        \bigcup_i \left\{ (-1)^n \calO_{2i-1, 2i}
    \,,
    \,\,
    \calO_{2i, 2i+1}
    \right\}
    \,.
\end{equation}
We now argue that this group does not grow under the dynamics.
When the constraints are pushed to the next time step $j=2$, the generator $(-1)^n \calO_{2i-1, 2i}$ is invariant under half the SWAPs $\prod_i e^{\frac{\ri\pi}{4}\sum_\sfa \ri \gamma_{2i-1}^\sfa\gamma_{2i}^\sfa}$ but is transformed under the action of the other half $\prod_i e^{-\frac{\ri\pi}{4}\sum_\sfa \ri \gamma_{2i}^\sfa\gamma_{2i+1}^\sfa}$ into 
$(-1)^n \calO_{2i-2, 2i+1}$.
However, these generators are redundant since
\begin{equation}
    \calO_{2i-2, 2i-1}
    \times (-1)^n \calO_{2i-1, 2i} \times
    \calO_{2i, 2i+1}
    =
    (-1)^n
    \calO_{2i-2, 2i+1}
    \,.
\end{equation}
Similarly, the generator $\calO_{2i, 2i+1}$ becomes $\calO_{2i-1, 2i+2}$ under the action of $\prod_j e^{\frac{\ri\pi}{4}\sum_\sfa \ri \gamma_{2i-1}^\sfa\gamma_{2i}^\sfa}$ which is redundant since
\begin{equation}
(-1)^n
    \calO_{2i-1, 2i}
    \times
    \calO_{2i, 2i+1}
    \times
(-1)^n
    \calO_{2i+1, 2j+2}
=
    \calO_{2i-1, 2i+2}
    \,,
\end{equation}
and thus the stabilizer group is invariant.

\subsubsection{Fermion path integral}\label{app:fermion_path_integral}
We formulate the replicated partition function $\mathbf{Z}_0$ in terms of the fermion path integral.
We start with $\bfZ_0$ given by the fermion transfer matrix as in~\eqref{eqn:many_copy_rbim_prob_amplitude}.
After averaging over $\eta$ and $\theta$, each transfer matrix $\hat{\mathbf{T}}$ consists of transfer matrices associated with horizontal and vertical bonds~\eqref{eqn:averaged_transfer_matrix_horizontal_rbim}~\eqref{eqn:averaged_transfer_matrix_vertical_rbim}, $\hat{\bfT} = \prod_i \hat{\bfv}_i \prod_i \hat{\bfh}_i$, 
\begin{align}
    \hat{\bfh}_i &= e^{-\frac{\ri \pi}{4}\sum_\sfa \ri \gamma_{2i}^\sfa\gamma_{2i+1}^\sfa + \frac{g}{2}\left(\sum_\sfa \ri \gamma_{2i}^\sfa\gamma_{2i+1}^\sfa\right)^2}, \\
    \hat{\bfv}_i &= e^{\frac{\ri \pi}{4}\sum_\sfa \ri \gamma_{2i-1}^\sfa\gamma_{2i}^\sfa - \frac{g}{2}\left(\sum_\sfa \ri \gamma_{2i-1}^\sfa\gamma_{2i}^\sfa\right)^2}.
\end{align}
We have omitted the constraints as they can be imposed on the boundaries as discussed in the previous section.

We first introduce $2n$ fictitious Majorana modes $\eta_{i}^\sfa$ at each location $i$ and rewrite the partition function as the transfer matrix involving the physical and the fictitious Majoranas,
\begin{align}
    \bfZ_0 = (\Psi | \hat{\mathbf{H}} \hat{\mathbf{T}}^T| \Psi) 
\end{align}
where $\hat{\mathbf{T}} = \prod_i \hat{\mathbf{v}}_i \prod_i \hat{\mathbf{h}}_i $,
\begin{align}
    \hat{\mathbf{h}}_i 
    &= e^{-\frac{\ri\pi}{4} \left(\sum_\sfa \ri \gamma_{2i}^\sfa\gamma_{2i+1}^\sfa 
    + \ri \eta_{2i}^\sfa\eta_{2i+1}^\sfa\right) + \frac{g}{2}\left(\sum_\sfa \ri \gamma_{2i}^\sfa\gamma_{2i+1}^\sfa\right)^2} 
    = e^{-\frac{\ri\pi}{2}\sum_{\sfa} \ri c^{\sfa,\dagger}_{2i} c^\sfa_{2i+1} - \ri c^{\sfa,\dagger}_{2i+1} c_{2i} 
    + \frac{g}{2}\left(\sum_\sfa \ri \gamma_{2i}^\sfa\gamma_{2i+1}^\sfa\right)^2} \,, \\
    \hat{\mathbf{v}}_i 
    &= e^{\frac{\ri\pi}{4} \left(\sum_\sfa \ri \gamma_{2i-1}^\sfa\gamma_{2i}^\sfa + \ri \eta_{2i-1}^\sfa\eta_{2i}^\sfa\right) 
    - \frac{g}{2}\left(\sum_\sfa \ri \gamma_{2i-1}^\sfa\gamma_{2i}^\sfa\right)^2} 
    = e^{\frac{\ri\pi}{2}\sum_{\sfa} \ri c^{\sfa, \dagger}_{2i-1} c^\sfa_{2i} - \ri c^{\sfa,\dagger}_{2i} c^\sfa_{2i-1} 
    - \frac{g}{2}\left(\sum_\sfa \ri \gamma_{2i-1}^\sfa\gamma_{2i}^\sfa\right)^2} \,.
\end{align}
Here, $c_i^\dagger := (\gamma_{i}+\ri\eta_{i})/2$, and $c_i := (\gamma_{i}-\ri\eta_{i})/2$. 
We note that the fictitious Majoranas undergo a swap dynamics and only contribute to the partition function by a constant factor.
The transfer matrix for each time step then takes the form
\begin{align}
    \hat{\mathbf{T}} = e^{\sum_i
    \frac{g}{2}\left(\sum_\sfa \ri \gamma_{2i-1}^\sfa\gamma_{2i+2}^\sfa\right)^2 
    -\frac{g}{2}\left(\sum_\sfa \ri \gamma_{2i-1}^\sfa\gamma_{2i}^\sfa\right)^2 }
    e^{\frac{\ri\pi}{2}\sum_{i,\sfa}  \ri c^{\sfa,\dagger}_{2i-1} c^\sfa_{2i} - \ri c^{\sfa,\dagger}_{2i} c^\sfa_{2i-1} }
    e^{-\frac{\ri\pi}{2}\sum_{i,\sfa} \ri c^{\sfa,\dagger}_{2i} c^\sfa_{2i+1} - \ri c^{\sfa,\dagger}_{2i+1} c_{2i}^\sfa} 
    \,.
\end{align}

To express the replicated partition function in terms of the path integral, we insert the resolution of identity in terms of the fermion coherent state between the transfer matrix $\hat{\mathbf{T}}$ for two consecutive time steps for each replica $\sfa$,
\begin{align}
    \mathds{1} \propto \prod_{i} \int \rd \psi^\sfa_{i,t}\rd\bar{\psi}^\sfa_{i,t} e^{-\bar{\psi}^\sfa_{i,t}\psi^\sfa_{i,t}} \ketbra{\psi^\sfa_{i,t}}{\bar{\psi}^\sfa_{i,t}}
    \,.
\end{align}
where $\ket{\psi^\sfa_i} := e^{-\psi^\sfa_i c_i^{\sfa,\dagger}}\ket{0}$.
This allows us to express the partition function as $\bfZ_0 := \int \calD\psi \, \calD{\bar \psi}  \, e^{-\calS_0[\psi,\bar \psi] - \calS_I[\psi,\bar{\psi}]}$.

The action involves two parts.
The non-interacting part $\calS_0$ of the action (quadratic in Grassmann fields) is given by
\begin{align}
    \calS_0[\psi,\bar{\psi}] &= 
    \sum_{i,t,\sfa} 
    \bar{\psi}_{2i-1,t}^\sfa \psi_{2i-1,t}^\sfa 
    +\bar{\psi}_{2i,t}^\sfa \psi_{2i,t}^\sfa 
    + \bar{\psi}_{2i-1,t}^\sfa \psi_{2i+1,t-1}^\sfa 
    + \bar{\psi}_{2i+2,t}^\sfa\psi_{2i,t-1}^\sfa
    \,,
\end{align}
Here, we use the fact that the swap gate (for each replica $\sfa$) acts on the fermion coherent state as
\begin{align}
    e^{-\frac{\ri\pi}{2} (\ri c_{2i}^{\dagger} c_{2i+1} - \ri c_{2i+1}^{\dagger} c_{2i})} \ket{\psi_{2i},\psi_{2i+1}} 
    &= e^{\psi_{2i}c_{2i+1}^\dagger} e^{-\psi_{2i+1}c_{2i}^\dagger}\ket{0} 
    = \ket{-\psi_{2i+1},\psi_{2i}}, \\
    e^{\frac{\ri\pi}{2} (\ri c_{2i-1}^\dagger c_{2i} - \ri c_{2i}^\dagger c_{2i-1})} \ket{\psi_{2i-1},\psi_{2i}} 
    &= e^{\psi_{2i-1}c_{2i}^\dagger} e^{-\psi_{2i}c_{2i-1}^\dagger}\ket{0} 
    = \ket{\psi_{2i},-\psi_{2i-1}}
    \,.
\end{align}
We now introduce the real Grassmann fields
\begin{align}
    \chi_{i,t}^\sfa := \frac{\psi_{i,t}^\sfa + \bar{\psi}_{i,t}^\sfa}{\sqrt{2}} \,,\quad 
    \zeta_{i,t}^\sfa := \frac{\ri\psi_{i,t}^\sfa - \ri\bar{\psi}_{i,t}^\sfa}{\sqrt{2}} \,.
\end{align}
In this way, the non-interacting part of the action takes the form
\begin{align}
    \calS_0[\chi,\zeta] 
    = \sum_{i,t,\sfa} &\ri \zeta_{2i-1,t}^\sfa\chi_{2i-1,t}^\sfa + \ri \zeta_{2i,t}^\sfa\chi_{2i,t}^\sfa \\
    &+ \frac{(\chi_{2i-1,t}^\sfa + \ri \zeta_{2i-1,t}^\sfa)(\chi_{2i+1,t-1}^\sfa - \ri\zeta_{2i+1,t-1}^\sfa)}{2} 
    + \frac{(\chi_{2i+2,t}^\sfa + \ri \zeta_{2i+2,t}^\sfa)(\chi_{2i,t-1}^\sfa - \ri\zeta_{2i,t-1}^\sfa)}{2} \nonumber \,.
\end{align}
The interacting part $\calS_I$ of the action is given by
\begin{align}
    \calS_I := \calS_{I, \, v} + \calS_{I, \,h} \,, \quad 
    \calS_{I, \,v}[\chi] := \frac{g}{2}\sum_{i,t}\left(\sum_\sfa \ri \chi_{2i-1,t}^\sfa\chi_{2i,t}^\sfa\right)^2 \,, \quad 
    \calS_{I, \, h}[\chi] := -\frac{g}{2}\sum_{i,t}\left(\sum_\sfa \ri \chi_{2i-1,t}^\sfa\chi_{2i+2,t}^\sfa\right)^2
    \,.
\end{align}
Here, we have dropped the constant term and treated the interaction as a strictly local term in spacetime, which is valid for small $g$.
The interacting part of the action consists of $\calS_{I, \,v}$ and $\calS_{I, \,h}$ associated with the vertical and the horizontal transfer matrices, representing repulsive and attractive inter-replica interactions, respectively.

Next, we introduce the real Grassmann fields for chiral Majorana modes on the lattice
\begin{align}
    \chi_{R,i,t}^\sfa := (-1)^t \chi_{2i,t}^\sfa \,, \quad
    \chi_{L,i,t}^\sfa := (-1)^t \chi_{2i+1,t}^\sfa \,,\label{eq:chiral_majorana_lattice}
\end{align}
with a similar definition for $\zeta_{L/R}$, which allows rewriting the action as
\begin{gather}
    \begin{aligned}
    \calS_0[\chi_L,\chi_R,\zeta_L,\zeta_R] 
    &= \sum_{i,t,\sfa} \ri \zeta_{R,i,t}^\sfa\chi_{R,i,t}^\sfa + \ri \zeta_{L,i,t}^\sfa\chi_{L,i,t}^\sfa \nonumber\\
    &\quad - \frac{(\chi_{L,i-1,t}^\sfa + \ri \zeta_{L,i-1,t}^\sfa)(\chi_{L,i,t-1}^\sfa - \ri\zeta_{L,i,t-1}^\sfa)}{2} 
    - \frac{(\chi_{R,i+1,t}^\sfa + \ri \zeta_{R,i+1,t}^\sfa)(\chi_{R,i,t-1}^\sfa - \ri\zeta_{R,i,t-1}^\sfa)}{2} \nonumber \\
    &= -\frac{1}{2} \sum_{i,t,\sfa} \left[\chi_{R,i+1,t}^\sfa \chi_{R,i,t-1}^{\sfa} + \chi^\sfa_{L,i,t}\chi^\sfa_{L,i+1,t-1} + 
    (\chi \leftrightarrow \zeta) \right] \nonumber \\
    &\quad +\frac{1}{2}\sum_{i,t,\sfa} \Big[\ri (\zeta_{R,i+1,t}^\sfa-\zeta_{R,i,t-1}^\sfa) \chi_{R,i+1,t}^\sfa + \ri \zeta_{R,i+1,t}^\sfa(\chi_{R,i+1,t}^\sfa-\chi_{R,i,t-1}^\sfa) \nonumber \\
    &\quad + \ri (\zeta_{L,i-1,t}^\sfa - \zeta_{L,i,t-1}^\sfa)\chi_{L,i-1,t}^\sfa + \ri \zeta_{L,i-1,t}^\sfa (\chi_{L,i-1,t}^\sfa - \chi_{L,i,t-1}^\sfa) \Big] \,, \end{aligned}\\
    \calS_{I, \,v}[\chi_L,\chi_R] := \frac{g}{2}\sum_{i,t}\left(\sum_\sfa \ri \chi_{L,i-1,t}^\sfa\chi_{R,i,t}^\sfa\right)^2 \,, \quad
    \calS_{I, \,h}[\chi_L,\chi_R] := -\frac{g}{2}\sum_{i,t}\left(\sum_\sfa \ri \chi_{L,i-1,t}^\sfa \chi_{R,i+1,t}^\sfa\right)^2 \,.\label{eq:S_int_discrete} 
\end{gather}
\begin{equation*}
\begin{tikzpicture}
\foreach \i in {0,3,6}{
\draw[color=red, line width=2pt] (\i,0) -- (\i+1.8,0);
}
\foreach \i in {0,3}{
\draw[color=blue, line width=2pt] (4.8+\i,0) arc[start angle=60, end angle=120, radius=4.8];
}
\draw[color=blue, line width=2pt] (6,0) arc[start angle=120, end angle=90, radius=4.8];
\draw[color=blue, line width=2pt] (1.8,0) arc[start angle=60, end angle=90, radius=4.8];
\foreach \i in {0,3,6}{
\filldraw[fill=white, draw=black, line width=2pt] (\i, 0) circle (1.5mm);
\filldraw[fill=white, draw=black, line width=2pt] ({1.8+\i}, 0) circle (1.5mm);
}
\node at (0,-0.5) {$\chi_{L,i-1}$};
\node at (1.8,-0.5) {$\chi_{R,i}$};
\node at (3,-0.5) {$\chi_{L,i}$};
\node at (4.8,-0.5) {$\chi_{R,i+1}$};
\node at (6,-0.5) {$\chi_{L,i+1}$};
\node at (7.8,-0.5) {$\chi_{R,i+2}$};
\end{tikzpicture}
\end{equation*}
Here, we illustrate the repulsive $\calS_{I,v}$ and the attractive interaction $\calS_{I,h}$ with red and blue lines, respectively.
One can rearrange the chiral Majorana modes such that the interactions couples the nearest neighbors.
In this form, the action is equivalent to that of a one-dimensional chain of Majorana modes with alternating repulsive and attractive bond interactions.

We then go to the continuum using the following identification with $a$ being the lattice spacing,
\begin{align}
    \chi_{R/L}(x,t) = \frac{1}{\sqrt{a}}\chi_{R/L,i,t}, \quad (\text{same for $\zeta$}).
\end{align}
In what follows, we set $a = 1$.
In this way, the action takes the form
\begin{align}
    \calS_0[\chi_L,\chi_R] = \frac{1}{2} \int \rd x\rd t \, \sum_{\sfa}\chi_R^\sfa \partial_+ \chi_R^\sfa + \chi_L^\sfa \partial_- \chi_L^\sfa + (\chi\leftrightarrow \zeta) \,,
\end{align}
where $\partial_\pm = \partial_t \pm \partial_x$.
We have dropped a total derivative term in the continuum, which can be ignored in the action.
In the current form, the physical and the fictitious fields are decoupled.
In what follows, we drop the terms associated with fictitious Majorana, as we are only interested in physical correlations.

\subsubsection{Saddle point}\label{app:nlsm_derivation_saddle}
The NLsM can now be derived from the fermion path integral.
In particular, we introduce the Hubbard-Stratonovich matrix fields to decouple the interaction, solve for the saddle point of the matrix fields, and obtain the sigma model, which describes the low-energy fluctuation around the saddle point.

First, we introduce Hubbard-Stratonovich matrix fields $\sfQ_h$ and $\sfQ_v$ to decompose the fermion interaction associated with the horizontal and the vertical transfer matrices,
\begin{align}
e&^{-\frac{g}{2}\left(\sum_\sfa \ri \chi_{L,i}^\sfa\chi_{R,i+1}^\sfa\right)^2} = \int \rd \sfQ_{v;\, i} \, 
e^{\sum_{\sfa\sfb}-\frac{1}{g}(\sfQ^{\sfa\sfb}_{v; \,i})^2
+(\chi_{R,i+1}^\sfa\chi_{R,i+1}^\sfb+\chi_{L,i}^\sfa\chi_{L,i}^\sfb)\ri \sfQ_{v; \,i }^{\sfa\sfb}}
\,,\\
e&^{\frac{g}{2}\left(\sum_\sfa \ri \chi_{L,i-1}^\sfa\chi_{R,i+1}^\sfa\right)^2} = \int \rd \sfQ_{h; \,i} \, 
e^{\sum_{\sfa\sfb}-\frac{1}{g}(\sfQ_{h; \,i}^{\sfa\sfb})^2 +( \chi_{R,i+1}^\sfa\chi_{R,i+1}^\sfb 
-  \chi_{L,i-1}^\sfa\chi_{L,i-1}^\sfb) \ri \sfQ^{\sfa\sfb}_{h; \,i} }
\,.\label{eq:HS_decoupling}
\end{align}
For simplicity, we have suppressed the indices $t$ that label the time step.
These matrix fields are $2n\times 2n$ and are chosen to be real and anti-symmetric.
We note that the Hubbard-Stratonovich decoupling is not unique, and we choose to decouple in this channel as it produces a non-trivial saddle point.

We can now go to the continuum and express the partition function as $\bfZ_0 := \int \calD [\sfQ,\chi] \, e^{-\calS[\sfQ,\chi]}$ with action
\begin{align}
    \calS[\sfQ,\chi] 
    = & \int \rd t \rd x \,
    \frac{1}{2}\sum_{\sfa}
    \chi_R^\sfa\partial_+\chi_R^\sfa 
    + \chi_L^\sfa\partial_-\chi_L^\sfa 
    +\frac{1}{g} 
    \sum_{\sfa,\sfb} 
    (\sfQ^{\sfa\sfb}_{h})^2
    +
    (\sfQ^{\sfa\sfb}_{v})^2
    \nn\\
    &
    -\sum_{\sfa, \sfb} \chi_{R}^\sfa 
    \left(\ri \sfQ^{\sfa\sfb}_{v}  
    + \ri \sfQ^{\sfa\sfb}_{h} \right)
    \chi_{R}^\sfb
    -\sum_{\sfa, \sfb}
    \chi_{L}^\sfa
    \left(
    \ri \sfQ^{\sfa\sfb}_{v}-\ri \sfQ^{\sfa\sfb}_{h}  \right)
    \chi_{L}^\sfb \,, \\
    =& 
    \int\rd t \, \rd x \, 
    \frac{1}{2} 
    \sum_\sfa \chi^\sfa \slashed\partial\chi^\sfa 
    +\frac{1}{g} \sum_{\sfa\sfb} 
    (\sfQ^{\sfa\sfb}_{h})^2
    +(\sfQ^{\sfa\sfb}_{v})^2
    -
    \sum_{\sfa\sfb} \chi^\sfa
    \left(\ri \sfQ^{\sfa\sfb}_{v} \sigma_0 
    + \ri \sfQ^{\sfa\sfb}_{h} \sigma_z\right)
    \chi^\sfb 
    \,,
\end{align}
where we have suppressed space-time indices on the fields.
On the second line, we introduce the spinor field $\chi := (\chi_R; \chi_L)$.
The matrices $\sigma_0$ and $\sigma_z$ act in this spinor space with $\slashed{\partial} = \partial_t \sigma_0 + \partial_x \sigma_z$.

We obtain an effective action for the matrix fields by integrating out the Grassmann fields $\chi$,
\begin{align}
    \calS_{\text{eff}} = -\frac{1}{2}\Tr\ln\left( \frac{\slashed\partial}{2}  - \ri \sfQ_{v} \sigma_0 - \ri \sfQ_{h} \sigma_z \right)
    -  \frac{1}{g} 
    \int\rd t \, \rd x \,
    \tr \sfQ_{v}^2 + \tr \sfQ^2_{h} \,,
\end{align}
with $\Tr(\cdot)$ representing the trace over both spatial and internal indices.

The saddle point of the effective action satisfies a set of matrix-valued equations,
\begin{align}
    \frac{2}{g} \sfQ_v + \frac{1}{2}\int \rd k \rd \omega \, \frac{-\ri}{\ri\omega/2 + \ri k/2 - \ri \sfQ_{v} - \ri \sfQ_h} + \frac{-\ri}{\ri\omega/2 - \ri k/2 - \ri \sfQ_{v} + \ri \sfQ_h} = 0 
    \,,
    \nn\\
    \frac{2}{g} \sfQ_h + \frac{1}{2} \int \rd k \rd \omega \, \frac{-\ri}{\ri\omega/2 + \ri k/2 - \ri \sfQ_{v} - \ri \sfQ_h} + \frac{\ri}{\ri\omega/2 - \ri k/2 - \ri \sfQ_{v} + \ri \sfQ_h} = 0 
    \,.
    \label{eq:saddle_point_equations}
\end{align}
Here, we have set the cutoff $2\pi/a$ to unity such that $Q$ represents the matrix field in both real and momentum space.
We obtain translationally invariant saddle points of the action with 
\begin{itemize}
    \item $\sfQ_h = \ri \sfg\Sigma_y$, $\sfQ_v = 0$
    \item $\sfQ_h = 0$, $\sfQ_v = \ri \sfg \Sigma_y$
\end{itemize}
where $\sfg = g \pi/\sqrt{2}$, $\Sigma_y = (0, -\ri\mathds{1}_n; \ri\mathds{1}_n, 0)$ is the Pauli-Y matrix in the $2n$-dimensional replica space. 
The two sets of saddle points are related by a space-time rotation ($x \mapsto \tau, \tau \mapsto -x$) which exchanges the matrix fields ($\sfQ_h \mapsto \sfQ_v$, $\sfQ_v \mapsto - \sfQ_h$).
Moreover, these equations depend only on the spectrum of the matrix fields; for example, any $\sfQ_v = O \ri \sfg \Sigma_y O^T$, $\sfQ_h = 0$ related by $O \in O(2n)$ conjugation is also a saddle point.

\subsubsection{Fluctuations around the saddle point: NLsM}\label{app:nlsm_derivation_effective_action}
We now derive the NLsM which characterizes the fluctuations around the saddle point.
We will focus the analysis on the case where $\sfQ_v$ has a non-trivial expectation value.
However, most of the following analysis and the final result remains unchanged if we had picked the other saddle.
We start with a canonical choice of the saddle $\sfQ_v = \ri \sfg \Sigma_y$, where the fermion Green's function is
\begin{align}\label{eqn:free_fermion_greens_function_saddle}
    G_{R,L}(k,\omega) = \frac{1}{\ri\omega/2 \pm \ri k/2 + \sfg \Sigma_y} \,.
\end{align}
Since only $\sfQ_v$ is non-vanishing at this saddle, we drop the subscript $\sfQ := \sfQ_v$.

To derive the NLsM, we perform a gradient expansion within the saddle point manifold $Q^2 = - \mathds{1}$ where $Q(x) = \sfQ(x) / \sfg = O(x) \ri \Sigma_y O^T(x)$ is the normalized orthogonal anti-symmetric matrix field parameterized by $O \in \SO(2n)$.
The effective action up to the second order in $\sfg$ is given by
\begin{align}
    \calS_{\text{eff}}[Q] 
    &= -\frac{1}{2}\Tr\ln\left(G_R^{-1}+O^T\left[\frac{\partial_+}{2}, O\right]\right) -\frac{1}{2}\Tr\ln\left(G_L^{-1}+O^T\left[\frac{\partial_-}{2}, O\right]\right) \nn \\
    &= \frac{1}{4}\Tr\left( G_R O^T\left[\frac{\partial_+}{2}, O\right] G_R O^T\left[\frac{\partial_+}{2}, O\right] \right) + \frac{1}{4}\Tr\left( G_L O^T\left[\frac{\partial_-}{2}, O\right] G_L O^T\left[\frac{\partial_-}{2}, O\right] \right) \nn \\
    &= \frac{1}{4}\int \rd^2 p_{1,2} \, \tr \big( G_R(p_1) A_+(p_2)G_R(p_1-p_2)A_+(-p_2) + G_L(p_1) A_-(p_2)G_L(p_1-p_2)A_-(-p_2) \big) \nn \\
    &\approx \frac{1}{4}\int \rd^2 p_{1,2} \, \tr \big(G_R(p_1) A_+(p_2)G_R(p_1)A_+(-p_2) + G_L(p_1) A_-(p_2)G_L(p_1)A_-(-p_2)\big)
    \,,
\end{align}
where $A_{\pm}(p)$ is the Fourier transformation of $O^T[\partial_\pm/2, O]$.
On the second line, we have dropped the constants from the zeroth order terms in the expansion of the logarithm.
On the last line, we assume that $O(x)$ is slowly-varying in spacetime, i.e., we take the approximation of small momentum transfer, $p_2+p_3 \ll \lambda^{-1} \sim \sfg$. 
Integrating over $p_1$ yields the action in the real space
\begin{align}
    \calS_{\text{eff}}[Q]&= \frac{1}{8 g} \Tr\left[\Sigma_y, O^T\left[\frac{\partial_+}{2}, O\right]\right]^2 + \frac{1}{8 g} \Tr\left[\Sigma_y, O^T\left[\frac{\partial_-}{2}, O\right]\right]^2
    = -\frac{1}{16 g}  \int \rd x \rd t \, \tr(\partial_t Q)^2 + \tr(\partial_x Q)^2
    \,.\label{eq:sigma_model_action}
\end{align}
This effective action describes the NLsM with target space $\SO(2n)/\U(n)$, which is the manifold of the saddle point.
In the standard notation of the sigma model, the bare coupling $g_0 = 8 g$.

Finally, we remark that the fluctuation around the saddle point with non-vanishing $\sfQ_h$ is also described by the NLsM.
In this work, the symmetry defect we consider acts on the saddle points with non-vanishing $\sfQ_v$ and $\sfQ_h$ in the same way.
We therefore do not distinguish these two sets of saddle points in our discussion.

\subsubsection{Symmetry defects in the sigma model}\label{app:nlsm_derivation_symmetry_defects}
Having established the mapping between the partition function $\bfZ_0$ and the effective sigma model, we now show how a symmetry defect insertion in $\bfZ_{2k}$ in Eq.~\eqref{eq:dual_partition_func_with_defect} modifies the sigma model.

We first derive the fermion path integral with the symmetry defect.
To begin, the symmetry defect in $\bfZ_{2k}$ flips the sign of horizontal couplings in the first $2k$ replicas.
The modified horizontal transfer matrix takes the form
\begin{align}
    \hat{\bfh}_i^{\Lambda} = e^{-\frac{\ri \pi}{4}\sum_\sfa \Lambda_\sfa \ri \gamma_{2i}^\sfa\gamma_{2i+1}^\sfa + \frac{g}{2}\left(\sum_\sfa \Lambda_\sfa \ri \gamma_{2i}^\sfa\gamma_{2i+1}^\sfa\right)^2} \,,
\end{align}
where $\Lambda_\sfa = -1$ for the first $2k$ replicas and $\Lambda_\sfa = +1$ otherwise.
The defect $\Lambda_\sfa$ in the SWAP part of the transfer matrix modifies the boundary conditions of the Majorana fermions.
The defect inserted in the horizontal bond also affects the decoupling of the interacting part,
\begin{align}
e&^{\frac{g}{2}\left(\sum_\sfa \Lambda_\sfa \ri \chi_{L,i-1}^\sfa\chi_{R,i+1}^\sfa\right)^2} = \int \rd \sfQ_{h; \,i} \, 
e^{\sum_{\sfa\sfb}-\frac{1}{g}(\sfQ^{\sfa\sfb}_{h; \, i})^2 
+(\chi_{R,i+1}^\sfa\chi_{R,i+1}^\sfb -\Lambda_\sfa\Lambda_\sfb\chi_{L,i-1}^\sfa\chi_{L,i-1}^\sfb)\ri \sfQ_{h; \, i}^{\sfa\sfb}}
\,.
\end{align}
This leads to a modified Lagrangian at the spacetime location where the defect is inserted
\begin{align}\label{eqn:modified_lagrangian_twist_nlsm_derv}
    \calL =&
    \frac{1}{2}\sum_{\sfa}
    \chi_R^\sfa\partial^\Lambda_+\chi_R^\sfa 
    + \chi_L^\sfa\partial^\Lambda_-\chi_L^\sfa 
    +\frac{1}{g} 
    \sum_{\sfa,\sfb} 
    (\sfQ^{\sfa\sfb}_{h})^2 
    + 
    (\sfQ^{\sfa\sfb}_{v})^2
    -\sum_{\sfa, \sfb} \chi_{R}^\sfa 
    \left(\ri \sfQ^{\sfa\sfb}_{h} 
    + \ri \sfQ^{\sfa\sfb}_{v} \right)
    \chi_{R}^\sfb
    -\sum_{\sfa, \sfb}
    \chi_{L}^\sfa
    \left(\ri \sfQ^{\sfa\sfb}_{h}\Lambda_\sfa\Lambda_\sfb
    -\ri \sfQ^{\sfa\sfb}_{v}  \right)
    \chi_{L}^\sfb \nn \\
    =& 
    \frac{1}{2} 
    \sum_\sfa \chi^\sfa \slashed\partial^\Lambda\chi^\sfa 
    +\frac{1}{g} \sum_{\sfa\sfb} 
    (\sfQ^{\sfa\sfb}_{h})^2 
    +(\sfQ^{\sfa\sfb}_{v})^2 
    -
    \sum_{\sfa\sfb} \chi^\sfa
    \left(\ri \sfQ^{\sfa\sfb}_{h} (\sigma_z)^{\frac{1-\Lambda_\sfa\Lambda_\sfb}{2}} 
    + \ri \sfQ^{\sfa\sfb}_{v} \sigma_z \right)
    \chi^\sfb 
    \,.
\end{align}
The gradient operator $\partial_\pm^\Lambda$ is modified due to the flipped boundary conditions in the first $2k$ replicas.
This modified kinetic term assigns an additional minus sign to the fermion field when moving across the symmetry defect.

As before, we obtain the effective action for the matrix field by integrating out the fermions.
Around the saddle point with non-vanishing $\sfQ_h$, the fluctuations can then be captured by the effective action for $\sfQ := \sfQ_h$
\begin{align}
    \calS_{\eff}^\Lambda &= -\frac{1}{2}\Tr\ln\left( \frac{\partial^\Lambda_+}{2} - \ri \sfQ \right) -\frac{1}{2}\Tr\ln\left( \frac{\partial^\Lambda_-}{2} - \ri \sfQ + (\ri \sfQ - \ri \Lambda \sfQ \Lambda)\delta(x) \right)
    -  \frac{1}{g} 
    \int\rd t \, \rd x \,
    \tr \sfQ^2 \,.
\end{align}

The symmetry defect modifies the action in two ways: it twists the boundary condition and rotates the matrix field at the location $x = 0$ of defect insertion by $Q(0) \mapsto \Lambda Q(0) \Lambda$.
Note that this local rotation is only present in the action around the saddle with non-vanishing $Q_h$; it simply moves the twist by one lattice site and does not affect the free energy.
Thus, the only non-trivial effect of the symmetry defect is twisting the boundary condition in the sigma model, which affects both saddles (with non-vanishing $\sfQ_v$ and $\sfQ_h$) in the same way.
Across the twist defect, it is energetically favorable to align the matrix field $Q(0^-)$ with $\Lambda Q(0^+) \Lambda$. 

\subsection{Effective field theory for the suboptimal decoder}\label{app:nlsm_derivation_subopt}
We consider the suboptimal decoder, in which the estimated rotation angle $\theta'$ is related to the rotation angle $\theta$ in the surface code by $(\pi/4 - \theta')(1+\epsilon) = \pi/4 - \theta$.
In what follows, we show that the partition function $\bfY_0$ is described by the field theory of the matrix field $Q \in \SO(2n)/\U(n)$.
The fluctuation out of the saddle point manifold becomes massive with a mass $\calO(\epsilon^2)$.
The corresponding potential term is relevant.
In the thermodynamic limit, the partition sum is governed by the sigma model with target space $\SO(2n)/\U(n)$.

For the suboptimal decoder, the replicated transfer matrix for $\bfY_0$ is given by
\begin{align}
\hat{\bfh}_{\epsilon, i} &= e^{-\frac{\ri\pi}{4}\sum_\sfa\ri \gamma^\sfa_{2i} \gamma^\sfa_{2i+1} + \frac{g}{2}
\left(\sum_{\sfa = 1}^{2n+2}\ri \gamma_{2i}^\sfa \gamma_{2i+1}^\sfa + \epsilon \sum_{\sfb = 1}^{2} \ri \gamma_{2i}^\sfb \gamma_{2i+1}^\sfb\right)^2} \,, \\
\hat{\bfv}_{\epsilon,i} &= e^{\frac{\ri\pi}{4}\sum_{\sfa} \ri\gamma_{2i-1}^{\sfa}\gamma_{2i}^{\sfa}-\frac{g}{2}\left(\sum_{\sfa=1}^{2n+2} \ri\gamma_{2i-1}^{\sfa}\gamma_{2i}^{\sfa} + \epsilon \sum_{\sfb = 1}^2 \ri\gamma_{2i-1}^{\sfa}\gamma_{2i}^{\sfa}\right)^2} \,.
\end{align}

We can similarly construct the fermion path integral representation of $\bfY_0$.
We again introduce an anti-symmetric matrix field $\sfQ$ to decouple the interaction part of the Lagrangian using the Hubbard-Stratonovich transformation [similar to Eq.~\eqref{eq:HS_decoupling}],
\begin{align}
e&^{-\frac{g}{2}\left(\sum_{\sfa=1}^{2n+2} \ri \chi_{L,i}^\sfa\chi_{R,i+1}^\sfa 
+ \epsilon\sum_{\sfb = 1}^2 \ri \chi_{L,i}^\sfb\chi_{R,i+1}^\sfb\right)^2} = \int \rd \sfQ_{v;\, i} \, 
e^{\sum_{\sfa\sfb}-\frac{1}{gD_\sfa D_\sfb}(\sfQ^{\sfa\sfb}_{v; \,i})^2
+(\chi_{R,i+1}^\sfa\chi_{R,i+1}^\sfb+\chi_{L,i}^\sfa\chi_{L,i}^\sfb)\ri \sfQ_{v; \,i }^{\sfa\sfb}}
\,,\\
e&^{\frac{g}{2}\left(\sum_{\sfa = 1}^{2n+2} \ri \chi_{L,i-1}^\sfa\chi_{R,i+1}^\sfa 
+ \epsilon\sum_{\sfb = 1}^2 \ri \chi_{L,i-1}^\sfb\chi_{R,i+1}^\sfb\right)^2} = \int \rd \sfQ_{h; \,i} \, 
e^{\sum_{\sfa\sfb}-\frac{1}{gD_\sfa D_\sfb}(\sfQ_{h; \,i}^{\sfa\sfb})^2 +( \chi_{R,i+1}^\sfa\chi_{R,i+1}^\sfb 
-  \chi_{L,i-1}^\sfa\chi_{L,i-1}^\sfb) \ri \sfQ^{\sfa\sfb}_{h; \,i} }
\,.
\end{align}
where $D_\sfa = 1+\epsilon$ for $\sfa = 1,2$ and $D_\sfa = 1$ otherwise. 

Integrating over the fermionic degrees of freedom yields an effective action for the matrix field
\begin{align}
    \calS_{\epsilon,\text{eff}} &= -\frac{1}{2}\Tr\ln\left( \frac{\slashed\partial}{2}  - \ri \sfQ_{v} \sigma_0 - \ri \sfQ_{h} \sigma_z \right)
    +  \frac{1}{g} 
    \int\rd t \, \rd x \,
    \sum_{\sfa\sfb} \frac{1}{D_\sfa D_\sfb} \left[(\sfQ_{v}^{\sfa\sfb})^2 + (\sfQ_{h}^{\sfa\sfb})^2\right] \, \nn \\
    &= -\frac{1}{2}\Tr\ln\left( \frac{\slashed\partial}{2}  - \ri \sfQ_{v} \sigma_0 - \ri \sfQ_{h} \sigma_z \right)
    -  \frac{1}{g} 
    \int\rd t \, \rd x \,
    \tr(\sfQ_{v}D^{-1}\sfQ_v D^{-1}) + \tr(\sfQ_{h}D^{-1} \sfQ_h D^{-1}) \,.
\end{align}
The saddle point of the action is determined by a set of equations,
\begin{align}
\frac{2}{g} D^{-1} \sfQ_v D^{-1} + \frac{1}{2}\int \rd k \rd \omega \, \frac{-\ri}{\ri\omega/2 + \ri k/2 - \ri \sfQ_{v} - \ri \sfQ_h} + \frac{\ri}{\ri\omega/2 - \ri k/2 - \ri \sfQ_{v} + \ri \sfQ_h} = 0 
\,,
\nn\\
\frac{2}{g} D^{-1} \sfQ_h D^{-1} + \frac{1}{2} \int \rd k \rd \omega \, \frac{-\ri}{\ri\omega/2 + \ri k/2 - \ri \sfQ_{v} - \ri \sfQ_h} + \frac{-\ri}{\ri\omega/2 - \ri k/2 - \ri \sfQ_{v} + \ri \sfQ_h} = 0 
\,.
\label{eq:saddle_point_equations_suboptimal}
\end{align}
where $D = \mathrm{diag}(D_\sfa)$ is a diagonal matrix.
The saddle points, i.e. the solutions to these equations, are given by 
\begin{itemize}
    \item $\sfQ_h = \ri \sfg D \Sigma_y D$, and $\sfQ_v = 0$
    \item $\sfQ_h = 0$, and $\sfQ_v = \ri \sfg D \Sigma_y D$
\end{itemize}
along with all the other solutions related by $\O(2) \times \O(2n)$ rotations.
Here, $\sfg = \pi g/\sqrt{2}$.
We note that the diagonal matrix $D$ breaks the symmetry of the path integral from $\O(2n+2)$ to $\O(2) \times \O(2n)$.

Next, we derive an effective theory by expanding the action around the saddle point.
We note that, for $\epsilon > 0$, the saddle points are characterized by the anti-symmetric matrix field $\sfQ$, which belongs to the target space $\Gamma_1 \times \Gamma_n = \Gamma_n$.
In the case of the optimal decoder (i.e. $\epsilon = 0$), the saddle point belongs to a larger target space $\Gamma_{n+1} = \SO(2n+2)/\U(n+1)$.
A non-vanishing $\epsilon$ leads to massive fluctuations out of the reduced target space $\Gamma_n$.

We consider the saddle point with non-vanishing $\sfQ_v = \ri \sfg D \Sigma_y D$.
For simplicity, we suppress the subscript $v$ in the rest of the derivation.
A perturbed field configuration takes the form 
\begin{align}
    \sfQ(x) = \sfQ_0 + \delta \sfQ(x), \quad \sfQ_0 = \ri \sfg D\Sigma_y D, \quad \delta \sfQ = \begin{pmatrix}
        \delta \sfQ_{11} & \delta \sfQ_{12} \\
        \delta \sfQ_{21} & \delta \sfQ_{22}
    \end{pmatrix},
\end{align}
where the subscript $1,2$ label the two subspaces of the first two and the next $2n$ replicas.
Up to the second order in $\delta \sfQ$, the perturbed action takes the form
\begin{align}
    \calS_{\epsilon, \eff} &= -\frac{1}{4}\left[\Tr(G_L\delta\sfQ G_L\delta\sfQ) +\Tr(G_R\delta\sfQ G_R\delta\sfQ)\right] - \frac{1}{g} \int\rd t\rd x \tr (\sfQ(x)D^{-1}\sfQ(x)D^{-1}) - \tr (\sfQ_0 D^{-1} \sfQ_0 D^{-1}),
\end{align}
where $G_{L/R}$ are the Green's functions at the saddle point.

We analyze these terms by going to momentum space,
\begin{align}
    \Tr (G_L\delta\sfQ G_L\delta\sfQ) &= \int \rd^2 p \rd^2 p' \tr \left(G_{L,p} \delta\sfQ_{p'} G_{L,p-p'} \delta\sfQ_{-p'}\right)\\
    &= \int \rd^2 p \rd^2 p' \tr \left(\frac{1}{\ri\omega/2 + \ri k/2 + \sfg D \Sigma_y D} \delta\sfQ_{p'} \frac{1}{\ri (\omega-\omega')/2 + \ri (k-k')/2 + \sfg D \Sigma_y D} \delta\sfQ_{-p'} \right)
\end{align}
Here, $\tr$ represents the trace over the internal space.
We expand this term in $p' = (\omega', k')$ and evaluate the integral order-by-order. 
At the zeroth-order, we have
\begin{align}
    \text{(0-th)}=& \int \rd^2 x \frac{\pi}{ 2\sqrt{2}(1+\epsilon)^2\sfg} \tr [ \Sigma_{y,1}, \delta \sfQ_{11}(x)]^2 + \frac{\pi}{2\sqrt{2}\sfg} \tr[\Sigma_{y,2}, \delta \sfQ_{22}(x)]^2 \nn \\
    &+ \frac{2\sqrt{2}\pi}{(1 + (1+\epsilon)^2)\sfg}\left(\tr (\Sigma_{y,1} \delta \sfQ_{12}(x) \Sigma_{y,2} \delta \sfQ_{21}(x)) - \tr(\delta \sfQ_{12}(x) \delta \sfQ_{21}(x))\right). %
\end{align}
At the second-order, we obtain
\begin{align}
    \text{(2-nd)} =& \int \rd^2 p \rd^2 p' \, \tr \left(\frac{1}{\ri\omega/2 + \ri k/2 + \sfg D\Sigma_yD} \delta\sfQ_{p'} \frac{-(\omega'+k')^2/4}{(\ri\omega/2 + \ri k/2 + \sfg D\Sigma_yD)^3} \delta\sfQ_{-p'}\right)\nn \\
    =& -\int\rd^2 x \, \frac{\pi}{32\sqrt{2} \sfg^3}\left(\frac{1}{(1+\epsilon)^6}\tr [ \Sigma_{y,1}, \partial_+\delta \sfQ_{11}]^2 + \tr [ \Sigma_{y,2}, \partial_+\delta \sfQ_{22}]^2\right) \nn \\
    & -\int\rd^2 x \, \frac{\pi}{\sqrt{2}(1+(1+\epsilon)^2)^3 \sfg^3}\left( -\tr (\partial_+\delta \sfQ_{12}(x) \partial_+\delta \sfQ_{21}(x)) + \tr (\Sigma_{y,1} \partial_+\delta \sfQ_{12}(x) \Sigma_{y,2} \partial_+\delta \sfQ_{21}(x))\right) \,. %
\end{align}

We now write down the action around the saddle point in terms of the rescaled matrix field $Q = D^{-1}\sfQ D^{-1}/\sfg$.
We focus on the transversal fluctuation of the rescaled field such that $Q(x)$ is anti-symmetric and orthogonal,
\begin{gather}
    \delta Q_\perp = \frac{1}{2}[\ri\Sigma_y, \delta Q], \quad \delta Q_{\perp,11} = \frac{1}{2}[\ri\Sigma_{y,1}, \delta Q_{11}],  \quad \delta Q_{\perp,22} = \frac{1}{2}[\ri\Sigma_{y,2}, \delta Q_{22}],\quad
    \delta Q_{\perp,\text{off-diag}} = \frac{1}{2}\left[\ri\Sigma_y, \begin{pmatrix}
        0 & \delta Q_{12}  \\
        \delta Q_{21} & 0
    \end{pmatrix} \right],
\end{gather}
where $\delta Q = D^{-1}\delta \sfQ D^{-1}/\sfg$, $\Sigma_{y,1} = \sigma^y$, and $\Sigma_{y,2} = \sigma^y \otimes \mathds{1}_n$. 
These were the gapless fluctuations that kept $Q$ within the target space $\Gamma_{n+1}$ when $\epsilon = 0$; the longitudinal fluctuations are gapped with a $\calO(1)$ mass and will be ignored.
The effective action takes the form
\begin{equation}
\calS_{\epsilon, \eff}[Q] = \int \rd x\rd t \left[-\frac{1}{16g} \tr (\nabla \delta Q_{\perp,22})^2 -\frac{(1+\epsilon)^2}{2(1+(1+\epsilon)^2)^3 g}\tr (\nabla \delta Q_{\perp,\text{off-diag}})^2 - \frac{\epsilon^2(1+\epsilon)\pi^2 g}{2(1+(1+\epsilon)^2)} \tr \delta Q_{\perp,\text{off-diag}}^2 \right] \,,
\end{equation}
where we have ignored the action associated with $\delta Q_{\perp, 11}$ as $\Gamma_1$ is trivial.
The potential is then $m^2 = \calO(\epsilon^2 g^2)$.
The potential term $m^2$ has scaling dimension two, namely relevant under coarse-graining; it becomes $\calO(1)$ at the scale $L = \calO(1/\abs{\epsilon})$.
The action can be expressed in a simpler form by keeping the leading order in $\epsilon$ in the stiffness term
\begin{align}
    \calS_{\epsilon, \eff}[Q] = -\frac{1}{16g} \int \rd x\rd t\tr (\nabla Q)^2 + 4\epsilon^2 \pi^2 g^2 \tr Q_{\text{off-diag}}^2,
\end{align}
where $Q \in \Gamma_{n+1}$, and $Q_{\text{off-diag}}$ is the matrix field out of the reduced target space $\Gamma_n$ and is assumed to be small.

We now comment on the twist expectation value in the field theory for the suboptimal decoder.
In Sec.~\ref{sec:fidelity_subopt}, we derive the scaling of the decoding fidelity for the suboptimal decoder in the dual picture as in Eq.~\eqref{eqn:dual_replicated_subopt_fidelity}.
There, we compute the twist expectation value associated with twisting an odd number of replicas in the constrained subspace $\Gamma_{2n}$.
The action at the microscopic level takes the form
\begin{align}
    \calS^\Lambda_{\epsilon,\eff}[ Q] = -\frac{1}{16g}\int \rd x \rd t  \left[\tr(\nabla  Q)^2 + 4\epsilon^2 \pi^2 g^2 \tr\delta Q^2_{\perp,\text{off-diag}}\right] - \frac{1}{16g} \int \rd x \tr(Q(0^+) - \Lambda Q(0^-) \Lambda )^2.
\end{align}
Here, we only keep the leading order in $\epsilon$ in the first term.
We perform coarse-graining up to scale $L$, leading to an effective 1D action
\begin{align}
     \calS^\Lambda_{\epsilon,\eff}[ Q] = \int_0^\kappa \rd t \left[ -\frac{1}{16 g_R(L)}\tr(\nabla  Q)^2 - \frac{\epsilon^2 L^2 g\pi^2}{4} \tr\delta Q^2_{\perp,\text{off-diag}}\right] - \frac{L}{16g} \tr(Q(0^+) - \Lambda Q(0^-) \Lambda )^2\Big|_{t = 0}.
\end{align}
Note that the reduced target space $\Gamma_n$ has two disconnected components specified by the Pfaffian of $Q$.
Twisting an odd number of replicas changes the Pfaffian of $Q$, which varies the field configuration in the massive direction.
The associated excess free energy is $\calO(m(L)/g_R(L))$, where the renormalized potential at scale $L$ is given by $m(L) = \calO(L\abs{\epsilon}\sqrt{g g_R(L)})$.
We thus obtain the decoding fidelity 
\begin{align}
    \calF^{(n)}_\subopt = \frac{1}{2} + e^{-\calO(L \abs{\epsilon} \sqrt{g/g_R(L)})} = \frac{1}{2} + e^{-\calO( L \abs{\epsilon} \sqrt{\ln L})}.
\end{align}

\end{widetext}

\section{Twist expectation value}\label{app:twist_field}
In this section, we compute the twist expectation value in the non-linear sigma model with target space $\Gamma_n = \SO(2n)/\U(n)$ associated with the optimal decoder.
We consider two-dimensional systems of height $T$ and circumference $L$.
We start with the twist expectation values in the RBIM picture; the analyses for the twist in the dual picture is similar.

\subsection{Twist in the RBIM picture}\label{app:twist_RBIM}
In the RBIM picture, the twist is inserted in the temporal direction.
We analyze the twist expectation values in four different regimes:
\begin{enumerate}[label=(\arabic*)]
    \item $\kappa \gg 1$, $\kappa \gg 1/g_R(L)$;
    \item $\kappa \gg 1$, $\kappa \ll 1/g_R(L)$;
    \item $\kappa \ll 1$, $1/\kappa \gg 1/g_R(T)$;
    \item $\kappa \ll 1$, $1/\kappa \ll 1/g_R(T)$.
\end{enumerate}

\subsubsection{\texorpdfstring{$\kappa \gg 1$}{κ >> 1}}
In the case that the aspect ratio $\kappa \gg 1$, we first coarse-grain up to scale $L$ and obtain an effective sigma model in one dimension,
\begin{align}
\calS^\Lambda_{\eff} = -\int_0^{\kappa} \rd t \, \left[ \frac{1}{2g_R} \tr (\partial_t Q)^2 + \frac{L}{2g_0 a} \tr (Q - \Lambda Q \Lambda)^2 \right],
\end{align}
where $g_R(L)$ is the renormalized coupling at scale and $g$ the bare coupling.
The twist defect becomes the local potential term $\tr (Q - \Lambda Q \Lambda)^2$.

The local potential is relevant under coarse-graining; at a large scale $L/a \gg 1$, it imposes the constraint $Q = \Lambda Q \Lambda$.
As $\Lambda$ is invariant under conjugation by $O \in \SO(2k, 2n-2k) \cap \SO(2n) \cong (\SO(2k) \times \SO(2n-2k)) \times \mathbb{Z}_2$, 
the constraint subspace can be identified with two copies of $\Gamma_k \times \Gamma_{n-k}$ related through conjugation by $\sigma^z \oplus \mathds{1}_{2k-2} \oplus \sigma^z \oplus \mathds{1}_{2n-2k-2}$.

We can thus express the twist field correlation as
\begin{align}
\Phi_{2k} = \frac{\bfZ_0^{(2k)} \bfZ_0^{(2n-2k)}}{\bfZ_0^{(2n)}},
\end{align}
where we add the superscript to label the target space, i.e., $\bfZ_0^{(2k)}$ is the partition function of the sigma model with target space $\Gamma_k = \SO(2k)/\U(k)$ in a 1D system of length $\kappa$.

The one-dimensional sigma model is disordered in the thermodynamic limit and has a correlation length $1/g_R$.
In what follows, we analyze the twist field correlation in two limits $\kappa g_R \gg 1$ and $\kappa g_R \ll 1$, in which the partition function of 1D sigma model can be computed.

\textit{$\kappa \gg 1/g_R$.---}
We first consider the regime $\kappa g_R \gg 1$, in which $\kappa$ is much greater than the correlation length $1/g_R$.
The partition function of the NLsM can be approximated by that of $\calO(\kappa g_R)$ decoupled matrix fields in the coset space $\Gamma_k = \SO(2n) / \U(n)$.  
We have
\begin{align}
    \bfZ_0 = \text{Vol}\left(\Gamma_n\right)^{\calO(\kappa g_R)} \,.
\end{align}
Thus, the twist expectation value is given by
\begin{align}
    \Phi_{2k} = \left(\frac{2\text{Vol}\left(\Gamma_k \times \Gamma_{n-k}\right)}{\text{Vol}\left(\Gamma_n\right)} \right)^{\calO(\kappa g_R)}
    \,.
\end{align}

\textit{$\kappa \ll 1/g_R$.---}
In the opposite limit $\kappa \ll 1/g_R$, the length of the 1D system is much smaller than the correlation length.
The partition function is determined by the quadratic part of the sigma model action.

We perform a rescaling in the temporal direction and obtain the action,
\begin{align}
\calS_{\eff} = -\int_0^1 \rd \tau \, \frac{1}{2g_R\kappa}\tr (\partial_\tau Q)^2
\end{align}
To obtain dependence on $g_R\kappa$, we approximate the action as that of an effectively two-spin problem with the partition function given by
\begin{align}
\bfZ_0 &= \int \rd Q_0 \rd Q_1 e^{\frac{1}{g_R\kappa}\tr(Q_0 - Q_1)^2} \nn\\
&= \text{Vol}\left(\Gamma_n\right)\int_{\Gamma_n} \rd O  e^{\frac{2}{g_R \kappa}(-2n - \tr \ri\Sigma^y O \ri\Sigma^y O^T)},
\end{align}
where we take the periodic boundary condition in the temporal direction.
Since the action is $\SO(2n)$ invariant, we set $Q_0 = \ri\Sigma^y = [0, -\mathds{1}_n; \mathds{1}_n, 0]$ and parameterize $Q_1$ as $Q_1 = O \ri \Sigma^y O^T$, where $O \in \SO(2n)$ is an orthogonal rotation.

In the limit $1/(g_R \kappa) \gg 1$, the partition function is dominated by $O$ that is close to identity.
We express $O$ as $O = e^{X}$ with $X \in \mathfrak{so}(2n) \setminus \mathfrak{u}(n)$, where $X^T = -X$, $\{X, Q_0\} = 0$.
We further express $X$ as $X = \sum_{a = 1}^{n(n-1)} \phi_a T^a$, with $T^a$ being the basis of the Lie algebra, which satisfies $\tr T^a T^b = -2 \delta_{ab}$.
In the case that $g_R \kappa \ll 1$, we take the approximation
\begin{align}
    2n+\tr \ri\Sigma^y O \ri\Sigma^y O^T = \frac{1}{2} \tr \ri \Sigma^y [X, [X, \ri \Sigma^y]] = -2 \tr X^2.
\end{align}
The partition function is then given by the Gaussian integral over free fields $\phi_a$,
\begin{align}
\bfZ_0 \sim \left(g_R \kappa\right)^{n(n-1)/2}.
\end{align}

This leads to the twist expectation value
\begin{align}
    \Phi_{2k} \sim \left(\frac{1}{g_R \kappa}\right)^{k(n-k)}.
\end{align}
We note that the twist field correlation decreases when either the aspect ratio $\kappa$ or the coupling $g_R$ increases, resulting in an increasing decoding fidelity.

\subsubsection{\texorpdfstring{$\kappa \ll 1$}{κ << 1}}
In the case that the aspect ratio $\kappa \ll 1$, we instead coarse-grain up to scale $T$, and obtain an effective 1D model in the spatial direction,
\begin{align}
    &\calS^\Lambda_{\eff} = \nn \\
    &-\int_0^{1/\kappa} \rd x \frac{1}{2g_R}\tr(\partial_x Q)^2 + \frac{T}{2g_0 a}\tr (Q(0) - \Lambda Q(a/T) \Lambda)^2.\label{eq:1d_action_RBIM_twist_short}
\end{align}
The insertion of a twist field at $x = 0$ becomes a local potential, which is relevant and imposes the constraint $Q(0^-) = \Lambda Q(0^+) \Lambda$.

\emph{$1/\kappa \gg 1/g_R$.}---In this regime, the sigma model consists of effectively decoupled spins at the scale of $1/g_R$.
The partition function is invariant under the twist insertion, up to an exponentially small correction $e^{-\calO(g_R/\kappa)}$.
Therefore,
\begin{align}
    \Phi_{2k} = 1 - e^{-\calO(\frac{g_R}{\kappa})}.
\end{align}
This indicates that the twist expectation value is close to unity, giving rise to an almost perfect decoding fidelity.

\emph{$1/\kappa \ll 1/g_R$.}---In this regime, the partition function is governed by the action in Eq.~\eqref{eq:1d_action_RBIM_twist_short}.
The rescaling in the spatial direction allows one to express the partition function as
\begin{align}
\bfZ_{2k} = \int_{Q(1) = \Lambda Q(0) \Lambda} \calD Q \, e^{\int_0^1 \rd x \frac{\kappa}{2g_R}\tr(\partial_x Q)^2}
\end{align}

In the case of large stiffness $\kappa/g_R \gg 1$, the partition function is governed by the saddle point solution.
To obtain the saddle point, we first derive the equation of motion.
First, the variation of the matrix field $Q$ in the sigma model is given by $\delta Q = [\epsilon, Q]$, where $\epsilon$ is an anti-symmetric matrix; $\delta Q$ is anti-symmetric and anti-commutes with $Q$ such that $Q + \delta Q$ is anti-symmetric and orthogonal up to the first order in $\delta Q$.
This leads to the equation of motion of the matrix field
\begin{align}
    [Q, \partial_x^2 Q] = 0.
\end{align}
The equation of motion has the solution $Q(x) = e^{Mx} Q(0) e^{-Mx}$, where $M$ is an anti-symmetric matrix that anti-commutes with $Q(0)$, i.e. $\{M, Q(0)\} = 0$.
Thus, $M$ is determined by $e^{2M} = -Q(1)Q(0) = -\Lambda Q(0) \Lambda Q(0)$.
In the saddle point approximation, the twist expectation value is given by
\begin{align}
    \Phi_{2k} = \int \calD Q\, e^{\int_0^1 \rd x \frac{2\kappa}{g_R}\tr M^2}.
\end{align}

\begin{figure}
    \centering
    \includegraphics[width=\linewidth]{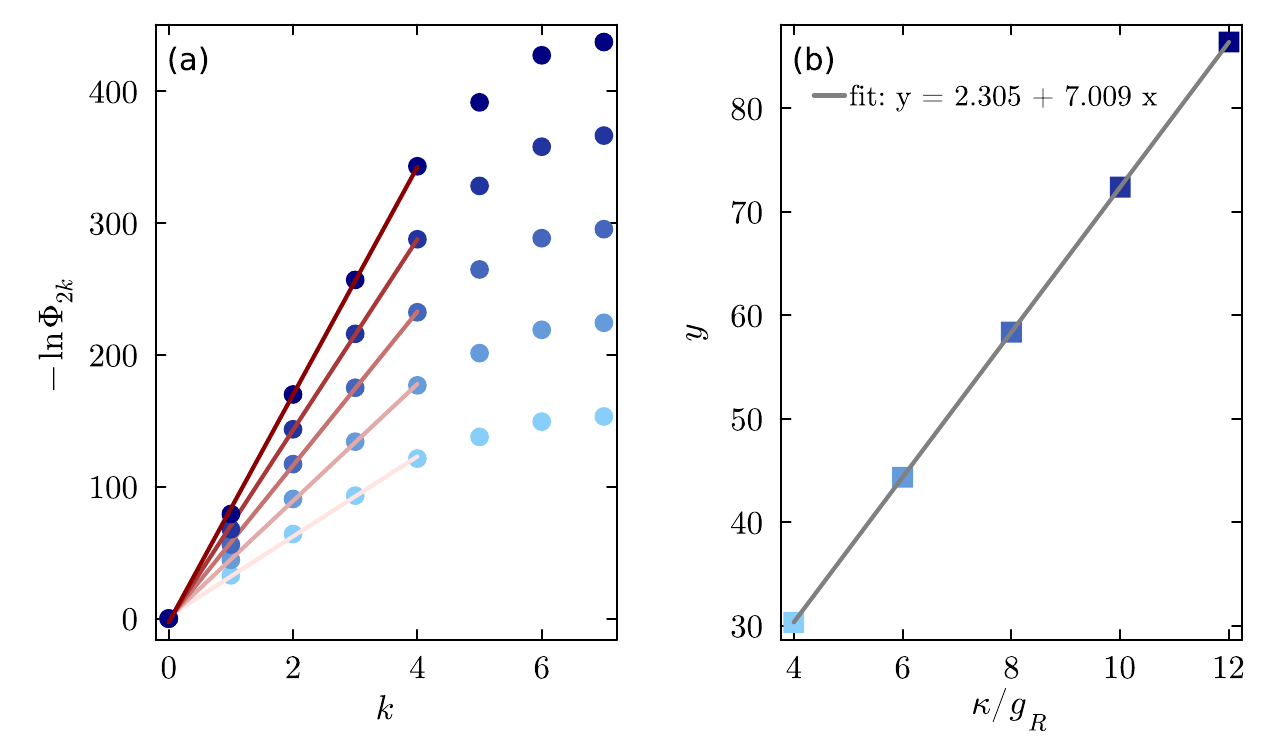}
    \caption{Twist expectation value $\Phi_{2k}$ in the RBIM picture when $1/\kappa \ll 1/g_R$. (a) Twist expectation value as a function of $k$. The exponent $-\ln \Phi_{2k} = y k$ is linear in $k$ for small $k$. The markers with an increasing opacity represent the results for an increasing stiffness $\kappa/g_R$. (b) The linear coefficient $y$ scales linearly with the stiffness $\kappa/g_R$. The replica number is set to be $n = 16$. The results are averaged over $10^5$ samples of the matrix field $Q$ drawn from the Haar measure over $\Gamma_n$.}
    \label{fig:twist_expval}
\end{figure}

Exactly computing the twist expectation value here is challenging.
Our numerical simulation suggests that the twist expectation value decays exponentially in $k$ for $k \ll n$ in the case of large stiffness  (as shown in Fig~\ref{fig:twist_expval}),
\begin{align}
    \Phi_{2k} = \Phi_{2n-2k} = e^{-(\alpha_n + \beta_n \frac{\kappa}{g_R}) k},\label{eq:twist_kappa_less_one_large_stiffness}
\end{align}
where $\alpha_n$, $\beta_n$ are $\calO(1)$ numbers that depend on $n$.
We note that the twist expectation value is the same when twisting $2k$ or $2n -2k$ copies.
We use this form of twist expectation value when analyzing the decoding fidelity in the replica limit.

We also note in passing that the quenched average sets a lower bound to the annealed average due to the convexity of the exponential function
\begin{align}
    \Phi_{2k} &\geq e^{\frac{1}{\Vol(\Gamma_k)} \int_{\Gamma_n} \rd Q \frac{\kappa}{2g_R}(-4n - 2\tr Q \Lambda Q \Lambda)} \nonumber \\
    &= e^{-\frac{\kappa}{g_R}\frac{16k(n-k)}{2n-1}}.
\end{align}

\subsection{Twist in the dual picture}
The analysis of the twist field correlation in the dual picture is similar to that in the RBIM picture.
The only difference is that the twist is inserted in the spatial direction.
In what follows, we present the results in four regimes.

\subsubsection{\texorpdfstring{$\kappa \gg 1$}{κ >> 1}}
We again coarse-grain up to scale $L$. In this case, the twist is inserted in the spatial direction and becomes a local term in the effective 1D theory,
\begin{align}
    &\calS^\Lambda_{\eff} = \nn \\
    &\quad -\int_0^{\kappa} \rd t \frac{1}{2g_R}\tr(\partial_t Q)^2 + \frac{L}{2g_0 a}\tr (Q(0) - \Lambda Q(a/L) \Lambda)^2.\label{eq:1d_action_dual_twist_short}
\end{align}
The twist expectation value exhibits distinct scalings in two regimes
\begin{itemize}
\item \emph{$\kappa \gg 1/g_R$.}---
The 1D sigma model is disordered, and the twist expectation value is 
\begin{align}
    \tilde\Phi_{2k} = 1 - e^{-\calO(\kappa g_R)}.
\end{align}

\item \emph{$\kappa \ll 1/g_R$.}--- The twist expectation value in this regime is governed by the quadratic part of the sigma model action and takes the form
\begin{align}
    \tilde\Phi_{2k} = \tilde \Phi_{2n-2k} = e^{-\left(\alpha_n + \beta_n \frac{1}{\kappa g_R}\right) k}.\label{eq:twist_dual_kappa_greater_1}
\end{align}
\end{itemize}

\subsubsection{\texorpdfstring{$\kappa \ll 1$}{κ << 1}}
Since the twist is instead inserted in the spatial direction, the analysis is simple in the case that $\kappa \ll 1$.
Specifically, we coarse-grain up to scale $T$ and obtain the effective action
\begin{align}
    \calS_{\eff}^{\Lambda} = -\int_0^{1/\kappa} \rd x \, \left[ \frac{1}{2g_R} \tr (\partial_x Q)^2 + \frac{T}{2g_0 a} \tr (Q - \Lambda Q \Lambda)^2 
    \right],
\end{align}
The twist expectation value can be computed in two regimes $g_R/\kappa \gg 1$ and $g_R/\kappa \ll 1$.
\begin{itemize}
\item \emph{$1/\kappa \gg 1/g_R$.}---
The effective 1D sigma model is disordered with correlation length $\calO(1/g_R) \ll 1/\kappa$.
The twist expectation value in the dual picture takes the form
\begin{align}
    \tilde{\Phi}_{2k} = \left(\frac{2\text{Vol}\left(\Gamma_k \times \Gamma_{n-k}\right)}{\text{Vol}\left(\Gamma_n\right)} \right)^{\calO(g_R/\kappa)}
    \,.
\end{align}
In the limit $\kappa \to 0$, the twist field expectation value vanishes. This leads to the decoding fidelity $1/2$.

\item \emph{$1/\kappa \ll 1/g_R$.}--- 
In this case, the length of the 1D sigma model is less than the correlation length.
A similar analysis as in Appendix~\ref{app:twist_RBIM} leads to
\begin{align}
\tilde{\Phi}_{2k} \sim \left(\frac{\kappa}{g_R}\right)^{k(n-k)}.
\end{align}

\end{itemize}

\section{Fidelity of the optimal decoder}\label{app:fidelity_replica_limit}
In this appendix, we use the twist expectation value to determine the fidelity of the optimal decoder.
We first discuss the qualitative scaling of the fidelity as a function of aspect ratio and the overall scale.
We then analyze the replica limit of the fidelity in various regimes.

\subsection{Scaling of the replicated fidelity}
We remark on how the twist expectation value determines the replicated fidelity of the optimal decoder.
First, we consider the surface code with a fixed aspect ratio. The twist expectation values predict how the decoding fidelity changes while increasing the overall scale.
\begin{itemize}
\item \emph{$\kappa \gg 1$}.--- The twist expectation value in this regime is a function of $\kappa g_R$.
The coupling $g_R$ increases with the overall scale for $n \geq 1$, giving rise to a decreasing $\Phi_{2k}$ in the RBIM picture and an increasing $\tilde\Phi_{2k}$ in the dual picture.
This suggests an increasing decoding fidelity with the overall scale.

\item \emph{$\kappa \ll 1$}.--- The twist expectation value in this regime is a function of $g_R/\kappa$.
With an increasing scale, we have an increasing $\Phi_{2k}$ in the RBIM picture and a decreasing $\tilde\Phi_{2k}$ in the dual picture.
This result predicts a decreasing decoding fidelity with the overall scale.

We note that this is a prediction of the fidelity for $\theta$ close to but not exactly at $\pi/4$. The result is valid at an intermediate scale above the mean-free path, $L \gg 1/g_0$, when the sigma model is a valid effective description. 
Right at $\theta = \pi/4$, the network model does not have backscattering, and the fidelity is not governed by the effective sigma model at any scale.
\end{itemize}

The twist expectation value also predicts the scalings of decoding fidelity when the length along one direction is fixed and the length of the other direction varies.
For example, in the case that $L$ is fixed and $T$ increases, the twist expectation value $\Phi_{2k}$ in the RBIM picture decreases and $ \tilde \Phi_{2k}$ in the dual picture increases, leading to an increasing decoding fidelity.
In particular, in the quasi-one-dimensional limit, i.e. the limit of large ($\kappa \gg 1/g_R$) and small ($1/\kappa \gg 1/g_R$) aspect ratios, we obtain decoding fidelities that are consistent with intuition from quantum error correction.
\begin{itemize}
\item \emph{Fixed $L$, $\kappa \gg 1/g_R$.---} In the case that one fixes $L$ and increases the aspect ratio $\kappa \gg 1/g_R$, the twist expectation values are $\Phi_{2k} = e^{-\calO(\kappa g_R)}$ in the RBIM picture and $\tilde \Phi_{2k} = 1 - e^{-\calO(\kappa g_R)}$ in the dual picture.
This leads to a decoding fidelity $1 - e^{-\calO(\kappa g_R)}$.
The result is consistent with the intuition from the quantum error correction; with a fixed $L$, the fidelity is nearly perfect up to a correction that is exponentially decaying in the code distance $T$.
The non-trivial prediction from the sigma model is that the decay coefficient $g_R(L)$ has a scale dependence governed by Eq.~\eqref{eq:n=1_RG_flow}.

\item \emph{Fixed $T$, $1/\kappa \gg 1/g_R$.---} In the case with a fixed $T$ and an increasing $L$, the twist expectation values are $\Phi_{2k} = 1 - e^{-\calO(g_R/\kappa)}$ in the RBIM picture, and $\tilde \Phi_{2k} = e^{-\calO(g_R/\kappa)}$ in the dual picture.
This indicates that the fidelity is $1/2 + e^{-\calO(g_R/\kappa)}$ with a subleading term that is exponentially decaying in $L$.
The result is consistent with our intuition that in the surface code on a quasi-one-dimensional geometry, with a fixed code distance $T$, the fidelity decays exponentially with $L$ when $L \gg T$.
The decay coefficient $g_R(T)$ is again governed by Eq.~\eqref{eq:n=1_RG_flow}.
\end{itemize}

\subsection{Fidelity in the replica limit}
We now analyze the fidelity in the replica limit $n \to 1$ in various regimes.

\subsubsection{\texorpdfstring{$\kappa \gg 1$, $\kappa \ll 1/g_R$}{κ >> 1, κ << 1/gR}}
In this regime, the replica limit of the decoding fidelity can be obtained in the dual picture.

The twist expectation values in the dual picture~\eqref{eq:twist_dual_kappa_greater_1} has an upper and a lower bound,\footnote{Strictly speaking, the bound holds only for $k$ close to $0$ and $n$, in the empirical scaling of the twist expectation values. However, the twist expectation values for intermediate $k$ are subleading in the limit of large stiffness $\kappa \ll 1/g_R$. We thus use these bounds to obtain the fidelity in the replica limit.}
\begin{align}
    e^{-\tilde{c}k}\leq \tilde \Phi_{2k}
    \leq e^{-\tilde{c}k}+e^{-\tilde{c}(n-k)},
\end{align}
where $\tilde{c} = \alpha_n + \beta_n/(\kappa g_R)$.
This leads to the bounds on the replicated fidelity
\begin{align}
    \frac{1}{2} + \frac{e^{-\tilde{c}}}{\left(1+e^{-\tilde{c}}\right)^2} \leq \calF_\opt^{(n)} \leq \frac{1}{2} + \frac{4e^{-\tilde{c}}}{\left(1+e^{-\tilde{c}}\right)^2}.
\end{align}
In the replica limit $n \to 1$, both the upper and the lower bound are $1/2$ with corrections governed by the small parameter $e^{-\beta_1/(\kappa g_R)}$. 
Hence, the fidelity takes the form
\begin{align}
    \calF_\opt = \frac{1}{2} + A e^{-\frac{\beta_1}{\kappa g_R}} + \calO\left(e^{-\frac{2\beta_1}{\kappa g_R}}\right),
\end{align}
where $\beta_1 = \lim_{n \to 1} \beta_n$ is an $\calO(1)$ number.

\subsubsection{\texorpdfstring{$\kappa \gg 1$, $\kappa \gg 1/g_R$}{κ >> 1, κ >> 1/gR}}
The fidelity in this regime can be obtained using the twist expectation value in the RBIM picture, which decays exponentially in the aspect ratio, i.e. $\Phi_{2k} = e^{-\calO(\kappa g_R)}$, for $k \neq 0, n$.
This leads the replicated fidelity $\calF_\opt^{(n)} = 1 - e^{-\calO(\kappa g_R)}$ and the same scaling for the fidelity in the replica limit.

\subsubsection{\texorpdfstring{$\kappa \ll 1$, $1/\kappa \ll 1/g_R$}{κ << 1, 1/κ << 1/gR}}
In this regime, the replica limit of the decoding fidelity can be obtained in the RBIM picture.

We again start with the bounds on the twist expectation value
\begin{align}
    e^{-ck}\leq& \Phi_{2k} \leq e^{-ck}+e^{-c(n-k)}.
\end{align}
where $c = \alpha_n + \beta_n \kappa/g_R$.
This yields the bounds on the replicated fidelity in Eq.~\eqref{eqn:rbim_replicated_opt_fidelity},
\begin{align}
    &\frac{2\sum_{k = 0}^{n-2}\binom{n-2}{k}e^{-ck}}{2 + \sum_{k = 1}^{n-1} \binom{n}{k} [e^{-ck}+e^{-c(n-k)}]} \leq \calF_\opt^{(n)} \nn \\
    &\quad\quad\quad\quad \leq \frac{2 + 2\sum_{k = 1}^{n-2}\binom{n-2}{k}[e^{-ck}+e^{-c(n-k)}]}{2 + \sum_{k = 1}^{n-1} \binom{n}{k} e^{-ck}}.
\end{align}
In the replica limit, we have
\begin{align}
    \calF_\opt = 1 - A e^{-\frac{\beta_1\kappa}{g_R}} + \calO\left(e^{-\frac{2\beta_1\kappa}{g_R}}\right).
\end{align}

\subsubsection{\texorpdfstring{$\kappa \ll 1$, $1/\kappa \gg 1/g_R$}{κ << 1, 1/κ >> 1/gR}}
The fidelity in this regime can be obtained using the twist expectation value in the dual picture, which decays exponentially in the inverse aspect ratio, i.e. $\tilde\Phi_{2k} = e^{-\calO(g_R/\kappa)}$, for $k \neq 0, n$.
The replicated fidelity is then given by $\calF_\opt^{(n)} = 1/2 + e^{-\calO(g_R/\kappa)}$ and so is the fidelity $\calF_\opt$ in the replica limit.

\section{Volume of the target space}\label{app:target_space_volume}
Here, we compute the volume of the target space $\Gamma_n = \SO(2n)/\U(n)$ following~\cite{macdonald_volume, abe_yokota_symmetric_space_volume}.
The $\SO(2n)$-invariant metric on $\Gamma_n$ is induced via the exponential map from the Cartan-Killing form on the Lie Algebra, which is unique up to a normalization depending on the representation.
In this paper, the NLsM field $Q$ is a $2n$-by-$2n$ anti-symmetric orthogonal matrix and can be parameterized by orthogonal matrices $O$ in the vector representation of $\SO(2n)$ while the $\U(n)$ invariant subgroup should be thought of as a direct sum of two copies of the fundamental representation of $\U(n)$.
In both cases, the generators $T^a$ satisfy $\tr T^a T^b = - 2 \delta_{ab}$.  As such, our results differ from~\cite{abe_yokota_symmetric_space_volume} which uses the adjoint representation common in the mathematical literature.

We begin by computing $\Vol(\SO(2n))$.  In the vector representation, $\SO(m) / \SO(m-1) \cong S^{m-1}$ implies
\begin{equation}
    \Vol(\SO(m)) = \Vol(S^{m-1}) \times \Vol(\SO(m-1)) \,,
\end{equation}
where $\mathrm{Vol}(S^{m-1}) = \frac{2 \pi^{m/2}}{\Gamma(m/2)}$.  Using $\Vol(\SO(2)) = 2 \pi$, 
\begin{equation}
    \Vol(\SO(2n)) =  \frac{
    2^{2n-1} \pi^{\frac{(2n-1)(n+1)}{2}}
    G(\frac{3}{2})}{G(n+1) G(n + \frac{1}{2})} \,,
\end{equation}
where $G(z)$ is the ``Barnes G-function'' defined by $G(z) = z \Gamma(z)$ for $z \in \mathbb{C}$ and $G(1) = 1$.

The calculation of $\Vol_{\SO(2n)}(\U(n))$, the volume of the embedding $\U(n) \hookrightarrow \SO(2n)$, is more complicated.
We begin by computing the volume of $\SU(n)$ in the metric induced by the fundamental representation.
Now, $\SU(n)/\SU(n-1) \cong S^{2n-1}$ implies
\begin{align}
    \mathrm{Vol}(\SU(n)) = \sqrt{\frac{n}{2(n-1)}} \times \mathrm{Vol}(&S^{2n-1}) 
    \\
    &\times \mathrm{Vol}(\SU(n-1)) 
    \nonumber
    \,.
\end{align}
Here $\sqrt{n / 2(n-1)}$ is the Jacobian factor that arises from the embedding of $S^{2n-1}$ into $\SU(n)$ in the fundamental representation~\cite{claude_qcd_calculations}. Using $\Vol(\SU(1)) = 1$, we have
\begin{equation}
    \Vol(\SU(n)) = \frac{
    \sqrt{n}\; 2^{\frac{n-1}{2}} \pi^{\frac{(n-1)(n+2)}{2}} 
    }{G(n+1)} \,.
\end{equation}
Next, note that $(\SU(n) \times \U(1)) / \mathbb{Z}_n \cong \U(n)$ under $\phi_1 : (z, U) \mapsto zU$ where $U \in \SU(n)$, $z \in \mathbb{C}$.  On the other hand $\U(n)$ is itself embedded into $\SO(2n)$ under the map $\phi_2 : u \mapsto \begin{pmatrix}
\Re u & \Im u \\
    - \Im u & \Re u
\end{pmatrix}$, where $u \in \mathfrak{u(n)}$ is anti-Hermitian.
This implies that
\begin{equation}
    \Vol_{\SO(2n)}(\U(n)) = 2^{n^2/2} \frac{1}{n} \sqrt{\frac{n}{2}} \times 2 \pi \times \Vol(\SU(n)) \,.
\end{equation}
Here the factor of $2^{n^2/2}$ comes from $\tr \phi_2(u)^2 = -4$ while $\sqrt{\frac{n}{2}}$ is due to $\tr (\ri \mathds{1}_n)^2 = -n$ and the map $\phi_1$.  Thus,
\begin{equation}
    \Vol_{\SO(2n)}(\U(n)) = \frac{(2\pi)^{\frac{n(n+1)}{2}} }{G(n+1)} \,.
\end{equation}
And we conclude that
\begin{equation}
    \Vol\left(\frac{\SO(2n)}{\U(n)}\right) = 2^{\frac{(n-2)(1-n)}{2}}\pi^{\frac{n^2}{2} - \frac{1}{2}}\frac{G(\frac{3}{2})}{G(n+\frac{1}{2})}
    \,.
\end{equation}
This allows writing the invariant volume in the form
\begin{equation}
    \frac{2 \text{Vol}\left(\Gamma_k \times \Gamma_{n-k}\right)}{\text{Vol}\left(\Gamma_n\right)}
    =
    \frac{(2 / \pi)^{k(n-k)} \;G(\frac{3}{2})G(n+\frac{1}{2})}{\sqrt{\pi} \; G(k+\frac{1}{2})G(n-k+\frac{1}{2})}
    \,.
\end{equation}
Note that with this normalization, the above invariant volume is $1$ whenever $k=0$, implying the twist expectation value $1$ in the absence of a twist.

\section{Conductance in the network model}\label{app:network_model_conductance}
Here, we detail the conductance calculation in the network model.
The procedure follows standard techniques~\cite{merz_chalker,cho_fisher,chalker1988percolation,kramer_otsuki_review,venn2022coherent}, which we review here for completeness of presentation.

We begin by detailing the explicit form of the single-particle transfer matrices used in our network model simulation.  
We work with the node transfer matrices $\sfh_{\bfr,\, \bfr + \hat{e}_x}$, $\sfv_{\bfr,\, \bfr + \hat{e}_t}$ corresponding to the $\U(1)$ network model defined in~\eqref{eqn:u1_network_node_transfer_matrices}.  The transfer matrix of the $j$-th row is defined by the $2 L$-by-$2 L$ matrix $\mathsf{T}_{j} = \mathsf{V}_j \mathsf{H}_{j}$ where $\mathsf{V}_{j} = \bigoplus_{i=1}^{L} \sfv_{\bfr,\, \bfr+\hat{e}_t}$ and $\mathsf{H}_{j} = \bigoplus_{i=1}^{L} \sfh_{\bfr,\, \bfr + \hat{e}_x}$.

Next, the orientations of the arrows in the network model (Fig.~$1$ of~\cite{venn2022coherent}) define a $\mathbb{Z}_2$ grading of the vector space of single-particle modes.
For each row $j$, we associate a diagonal matrix $\mathsf{Z}_{j}$ with entries $+1$ ($-1$) corresponding to upward (downward) arrows.  The single particle transfer matrices satisfy the pseudo-unitarity constraint $\mathsf{T}_j^\dagger \mathsf{Z}_{j + 1} \mathsf{T}_{j} = \mathsf{Z}_{j}$.  It is convenient to work in a basis where the $\mathbb{Z}_2$ grading takes a canonical form
\begin{equation}
    \mathsf{Z}_{j} \mapsto \Pi_{j}^{-1} \mathsf{Z}_{j} \Pi_{j}  = 
    \begin{pmatrix}
        \mathds{1}_{L} & 
        \\ & - \mathds{1}_{L}
    \end{pmatrix}\,,
\end{equation}
where $\Pi_{j} \in S_{2L}$ is a permutation matrix.  This induces a transformation on the transfer matrices
\begin{equation}
    \mathsf{T}_{j} \mapsto \tilde{\mathsf{T}}_{j} = \Pi_{j+1}^{-1} \mathsf{T}_{j} \Pi_{j}
    \,.
\end{equation}
The transformation is a gauge transformation as physical observables are related to traces of the combination $\mathsf{T} \mathsf{T}^\dagger$, which is invariant.

As the transfer matrices describe a network model in class D, there exists a gauge with all matrices real~\cite{AZ_10fold,cho_fisher,merz_chalker}.
Because it is more numerically efficient to work with real matrices, we will fix such a gauge.
We begin by defining some notation.  We identify sublattices of the network model (see Fig.~$1$ of~\cite{venn2022coherent}) such that $\sfv^{(A)}$ and $\sfh^{(A)}$ correspond to nodes with upward and rightward pointing arrows, respectively, while $\sfv^{(B)}$ and $\sfh^{(B)}$ correspond to nodes with downward and leftward pointing arrows.

Observe that after the transformation
\begin{widetext}
\begin{align}
    \sfv^{(A)} \mapsto 
    \begin{pmatrix}
    1 & \\
    & \ri
    \end{pmatrix}
    \sfv^{(A)}
    \begin{pmatrix}
    1 & \\
    & -\ri
    \end{pmatrix}
    \,,\;\;\;
    &\sfv^{(B)} \mapsto 
    \begin{pmatrix}
    -\ri & \\
    & 1
    \end{pmatrix}
    \sfv^{(B)}
    \begin{pmatrix}
    -\ri & \\
    & -1
    \end{pmatrix}
    \,,
    \nonumber\\
    \sfh^{(A)} \mapsto 
    \begin{pmatrix}
    -1 & \\
    & 1
    \end{pmatrix}
    \sfh^{(A)}
    \begin{pmatrix}
     -\ri & \\
     & \ri
    \end{pmatrix}
    \,,\;\;\;
    &\sfh^{(B)} \mapsto 
    \begin{pmatrix}
    \ri & \\
    & \ri
    \end{pmatrix}
    \sfh^{(B)}
    \begin{pmatrix}
    1 & \\
    & 1
    \end{pmatrix}
    \,,
\end{align}
\end{widetext}
the node transfer matrices are purely real, with the form
\begin{align}
    \sfh_{\bfr,\, \bfr+\hat{e}_x} &= 
    \begin{pmatrix}
        \cot 2 \theta & \pm \eta_{\bfr,\, \bfr+\hat{e}_x} \csc 2 \theta \\
        \pm\eta_{\bfr,\, \bfr+\hat{e}_x} \csc 2 \theta & \cot 2 \theta
    \end{pmatrix}
    \,, \\
    \sfv_{\bfr,\, \bfr + \hat{e}_t} &= 
    \pm \eta_{\bfr,\, \bfr+\hat{e}_t}
    \begin{pmatrix}
        \cos 2 \theta &  \sin 2 \theta \\
    -    \sin 2 \theta & \cos 2 \theta
    \end{pmatrix}
    \nonumber
    \,.
\end{align}
Here, $\pm$ is $+1$ ($-1$) for the $A$ ($B$) sublattice. 
This is a gauge transformation taking $\mathsf{T} \mapsto \Theta \mathsf{T} \Theta'$ where the $\Theta, \Theta'$ are diagonal phase matrices.

When the transfer matrix $\mathsf{T} = \prod_{j=1}^T \mathsf{T}_j$ after $T$ time steps is written in the canonical form $\tilde{\mathsf{T}}$, it is related to the scattering matrix $\mathsf{S}$ of the 2+1D disordered superconductor~\cite{kramer_otsuki_review}
\begin{equation}
    \tilde{\mathsf{T}} = 
        \begin{pmatrix}
        \mathsf{t} - \mathsf{r}^{'} \mathsf{t}^{'-1}\mathsf{r} & \mathsf{r}^{'} \mathsf{t}^{'-1}
        \\
        - \mathsf{t}^{'-1}\mathsf{r} & \mathsf{t}^{'-1}
    \end{pmatrix}
    \,,
\end{equation}
with $\mathsf{r}, \mathsf{r}^{'}$ and $\mathsf{t}, \mathsf{t}^{'}$ the reflectance and transmission blocks, respectively, of the scattering matrix $\mathsf{S} = \begin{pmatrix}
    \mathsf{t} & \mathsf{r} \\
    \mathsf{r}^{'} & \mathsf{t}^{'}
\end{pmatrix}
$.  The Landauer formula~\cite{d_fisher_landauer_formula,landauer_formula} relates the conductance to the transmission block of the scattering matrix
\begin{equation}
    G = \tr \mathsf{t}^\dagger \mathsf{t} = \tr \mathsf{t}^{' \dagger} \mathsf{t}^{'} \,.
\end{equation}

The transmission block $\mathsf{t}^{'} = \mathsf{B}^{-1}$ is computed numerically by evaluating
\begin{equation}\label{eqn:computing_transmission_part}
    \begin{pmatrix}
        \mathsf{A} \\
        \mathsf{B}
    \end{pmatrix}
    =
    \tilde{\mathsf{T}}
    \begin{pmatrix}
    0_{L} \\ \mathds{1}_{L}
    \end{pmatrix}
    \,.
\end{equation}
The eigenvalues of $\tilde{\mathsf{T}}$ approach $0$ as $T \to \infty$, making an explicit calculation numerically unstable~\cite{kramer_otsuki_review}.  
Instead, we sequentially apply $\tilde{\mathsf{T}}_j$ to the initial state $\mathsf{Q}_0 \equiv \begin{pmatrix}
    0_{L} \\ 
    \mathds{1}_{L}
\end{pmatrix}
$ and apply a QR decomposition every $m$ steps, namely $\mathsf{Q}_{k+1} \mathsf{R}_{k+1} = \tilde{\mathsf{T}}_{(k+1)m} \tilde{\mathsf{T}}_{(k+1)m - 1}\cdots \tilde{\mathsf{T}}_{km+1} \mathsf{Q}_{k}$ with $\mathsf{Q}_k$ having orthonormal columns and $\mathsf{R}_k$ upper triangular.  Thus,~\eqref{eqn:computing_transmission_part} is given by
\begin{equation}
    \begin{pmatrix}
        \mathsf{A} \\
        \mathsf{B}
    \end{pmatrix}
    =
    \mathsf{Q}_{T / m} 
    \mathsf{R}_{T / m} \mathsf{R}_{T / m - 1}\cdots \mathsf{R}_1
    \,.
\end{equation}
In this way, the conductance can be expressed as
\begin{equation}
    G = \tr\left[ \left(\mathsf{B}^{-1}\right)^{\dagger} \mathsf{B}^{-1} \right] \,.
\end{equation}

\section{Syndrome sampling algorithm}\label{app:syndrome_sampling}
In this appendix, we briefly review the syndrome sampling algorithm described in Ref.~\cite{bravyi_coherent}.
Then, we show how the particular contraction order considered by the above authors allows us to view the algorithm as the 1+1D free Majorana dynamics of a fixed number of modes.
Finally, we show how the decoding fidelity can be determined directly from this algorithm.

\subsection{Syndrome sampling algorithm on cylinder}\label{app:syndrome_sampling_cylinder}
In this section, we review the syndrome sampling algorithm developed by~\cite{bravyi_coherent}.
At a high level, the algorithm allows us to sample from the syndrome distribution $\calQ_{\alpha, s}$ indirectly by sampling the joint distribution $\calQ_{\vec{m}}$ of single-qubit measurements in the $X$-basis.
This procedure works because $\calQ_{\vec{m}} \propto \calQ_{s = \partial \vec{m}}$ with a constant proportionality factor independent of $s$.

On the cylinder, the algorithm is performed on a $2$D Majorana state defined on the lattice pictured in Fig.~S3 of Ref.~\cite{venn2022coherent}.
Each qubit on links is represented by $4$ Majorana modes $\gamma_a, \gamma_b, \gamma_c , \gamma_d$ with the constraint $\gamma_a \gamma_b\gamma_c\gamma_d = -1$, known as the C4 encoding.
Single-qubit operators are defined by $X = \ri \gamma_a \gamma_b = \ri \gamma_c \gamma_d$ and $Z = \ri \gamma_b \gamma_c = \ri \gamma_a \gamma_d$.
We refer the readers to~\cite{bravyi_coherent} for further details.

At the start of the algorithm, the Majoranas are prepared in the Gaussian initial state $\ket{\phi_\mathrm{link}}$ defined as a product of paired states $\ri \gamma_a \gamma_b = \pm 1$, where the sign is determined by the orientation of the links in Fig. S3 of~\cite{venn2022coherent}.
This choice of $\ket{\phi_\mathrm{link}}$ corresponds to the decoding problem for the surface code initialized in the $X_L = +1$ state which is relevant for correcting coherent-$Z$ errors.
The joint distribution $\calQ_{\vec{m}}$ of measurement outcomes is then sampled sequentially from the conditional distribution for the $t$-th measurement
\begin{equation}
    \calQ(m_t | m_{t-1} , \dots , m_1) = 
    \frac{
    \calQ(m_t, \dots, m_1)
    }
    {
    \calQ(m_{t-1}, \dots, m_1)
    }
    \,.
\end{equation}
Here, it is crucial that the order of measurements be made such that the set of unmeasured qubits remains connected~\cite{bravyi_coherent}.
Thus, it is convenient to choose a ``spiral'' order where qubits are measured row-by-row along the circumference of the cylinder and starting from one end of the cylinder to the other.
In this case, the conditional probability takes a simple form
\begin{equation}
    \calQ(m_t | m_{t-1} , \dots , m_1) = 
    \kappa_{t}
    \frac{
    \bra{\phi_{t-1}}
    \mathcal{O}_{t}
    \ket{\phi_{t-1}}
    }
    {
    \langle{\phi_{t-1}}|{\phi_{t-1}}\rangle
    }
    \,,
\end{equation}
where $\ket{\phi_{t-1}}$ is the free fermion state after the first $t-1$ rounds of unitaries and measurements.
On the other hand, $\mathcal{O}_t \equiv U^\dagger_t \Pi_t \Pi_t U_t$ represents the combined action of unitaries $U_t$ and measurements $\Pi_t$ on the next measured qubit.
Specifically, $\calO_t$ contains the coherent-$Z$ rotation $U_t = e^{\ri \theta Z} = e^{\ri \theta (\ri \gamma_b \gamma_c)}$ as well as the measurement part
\begin{equation}
    \Pi_t = \frac{1 + m_t \ri \gamma_a\gamma_b}{2} \frac{1 + m_t \ri \gamma_c \gamma_d}{2}
    \,,
\end{equation}
which projects onto measurement outcome $m_t$ while simultaneously enforcing the constraint $\gamma_a\gamma_b\gamma_c\gamma_d = -1$.
In this way, $\calO_t$ is a Gaussian operator and the Gaussianity of $\ket{\phi_{t}}$ can be maintained.
Finally, the prefactor $\kappa_t = 2$ except when making the final measurement, where $\kappa_t = 1$.

\subsection{Algorithm as 1+1D Majorana dynamics}\label{app:syndrome_sampling_dynamics}
\begin{figure*}[t]
\centering
\includegraphics[width=.74 \textwidth  ]{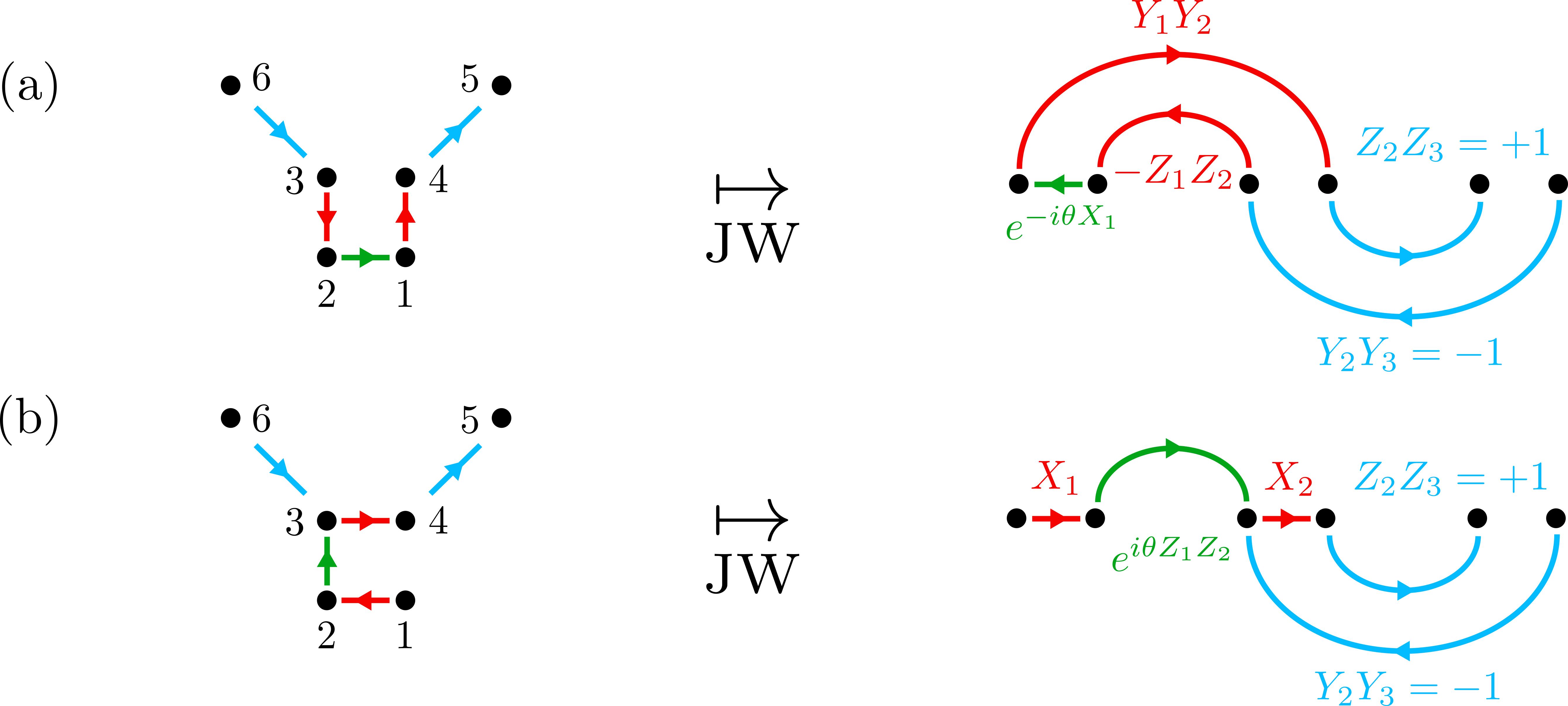}
    \caption{
    In (a) and (b), we present the key features of the 1+1D dynamics defined by the syndrome sampling algorithm for odd and even time steps, respectively, as defined in~\cite{venn2022coherent}.
    The choice of Majorana ordering maps to spin operators through the Jordan-Wigner transformation.
    In blue, we denote stabilizers of the initial state for qubits $2$ and $3$.
    The red (green) represents the projection (unitary) part of the dynamics.
    }
    \label{fig:teleportation}
\end{figure*}
Here, we show that the particular ``spiral'' measurement order described here defines a 1+1D dynamics in symmetry class D.
In fact, the dynamics are equivalent to the contraction of the complex-coupling RBIM in the main text~\eqref{eqn:complex_coupling_RBIM_transfer_matrix_spin_representation}.

In what follows, we will focus on the bulk dynamics.
The basic ingredients of the dynamics are highlighted in Fig.~\ref{fig:teleportation}; at each node, four ancillary Majoranas are introduced in the paired state $\ri \gamma_6 \gamma_3 = \ri \gamma_4 \gamma_5 = +1$, a unitary is applied before the first $4$ Majoranas are disentangled by a pair of projective measurements and discarded.
As a result of the $4$-Majorana stabilizer on each qubit, the projectors are forced to have the same outcome, which we represent by $\eta = \pm 1$.
Finally, we relabel $\gamma_5, \gamma_6 \mapsto \gamma_1, \gamma_2$.
Viewed in this way, the decoding algorithm defines an effective 1+1D dynamics of a system with a fixed number of Majorana modes;
at each time step, we take our system to consist of the $2 L$ Majorana modes living along a given row (circular slice) of the cylinder.
The dynamics of an entire timestep is defined through the combined evolution of all nodes along a given row.
We will now characterize the dynamics at each node.

It is convenient to work in a spin representation defined by the Jordan-Wigner transformation.
These spins (acted upon by $X, Y, Z$) are distinct from the spins appearing in the complex-coupling RBIM~\eqref{eqn:complex_coupling_RBIM_transfer_matrix_spin_representation}, which are acted upon by $\sigma^{x, y, z}$.
We are free to reorder the Majoranas as in Fig.~\ref{fig:teleportation} such that the relevant operators take a simple form after the transformation.

We begin with the odd time step associated with vertical links of the surface code.  
The mapping to the spin representation is given in Fig.~\ref{fig:teleportation}a, with the unitary part of the dynamics performed before the measurements.  
We show that the measurement part of the protocol performs state teleportation from qubit $1 \mapsto 3$ so the effective dynamics is unitary $e^{-\ri \theta X_1}$ after relabeling $3 \mapsto 1$.
Starting from an arbitrary initial state $\ket{\psi_0} = \alpha \ket{0} + \beta \ket{1}$ for qubit $1$, we have
\begin{align}
    \frac{1+Y_1 Y_2}{2} &\frac{1-Z_1 Z_2}{2} \Big[
    \ket{\psi_0}
    \otimes 
    \left( \ket{00} + \ket{11} 
    \right)
    \Big]
    \nonumber \\
    &= 
    \frac{1 + Y_1 Y_2}{2} \left(\alpha \ket{011} + \beta \ket{100}\right)
    \nonumber \\
    &= \left(\ket{10} + \ket{01} \right) \otimes \left(\beta \ket{0} + \alpha \ket{1} \right) \,,
\end{align}
ignoring normalization.
Thus, the measurements perform teleportation into qubit $3$ followed by an $X$ gate.
A similar calculation with $\eta = -1$ exchanges the role of $\alpha$ and $\beta$.
Running the mapping backwards $X_1 \mapsto \ri \gamma_{2j} \gamma_{2j-1} \mapsto - \sigma^x_j$, we conclude that the dynamics is
\begin{equation}
    e^{\ri \theta \sigma^x_j + \ri \frac{\pi}{4} (1 + \eta) (\sigma^x_j  - 1)}
    \,,
\end{equation}
equivalent to $\hat{\mathsf{v}}$ in Eq.~\eqref{eqn:complex_coupling_RBIM_transfer_matrix_spin_representation}.

We now consider the even time step corresponding to horizontal surface code links.
In the spin representation, this is described by Fig.~\ref{fig:teleportation}b, with the unitary applied first.
Starting from an initial state $\ket{\psi_0} = \alpha \ket{+} + \beta \ket{-}$, we have
\begin{widetext}
    \begin{align}
        \frac{1 + X_1}{2} \frac{1+ X_1 X_2}{2}
       & e^{\ri \theta Z_1 Z_2} \left(
        \alpha \ket{+} + \beta \ket{-}
        \right)
        \otimes 
        \left(\ket{++} + \ket{--} \right)
        =
        \frac{1+X_1}{2} e^{\ri \theta Z_1 Z_2}
        \left(\alpha \ket{+++} + \beta \ket{---}\right)
        \nonumber \\
        &=
        \frac{1 + X_1}{2}
        \left(\alpha \cos \theta \ket{+++} + \alpha \ri \sin \theta \ket{--+} + \beta \cos \theta \ket{---} + \beta \ri \sin \theta \ket{++-}\right)
        \nonumber \\
        &= \ket{++} \otimes \left(\alpha \cos \theta \ket{+} + \beta \ri \sin \theta \ket{-}\right) \,.
    \end{align}
\end{widetext}
This corresponds to imaginary time evolution $e^{\beta X_1}$ with $\beta =-\frac{1}{2} \ln \tan \theta$ followed by $e^{\ri \frac{\pi}{4} (1 - X_1)}$ and teleportation of qubit $1\mapsto3$.
A similar calculation with $\eta = -1$ flips the sign of $\beta$.
Mapping backwards $X_1 \mapsto \ri \gamma_{2j} \gamma_{2j+1} \mapsto \sigma^z_j \sigma^z_{j+1}$, we conclude that the dynamics is given by
\begin{equation}
    e^{\left(\beta + \ri \frac{\pi}{4} \right)\eta \sigma^z_j \sigma^z_{j+1} + \ri \frac{\pi}{4}}
    \,,
\end{equation}
equivalent to $\hat{\mathsf{h}}$ in Eq.~\eqref{eqn:complex_coupling_RBIM_transfer_matrix_spin_representation}.

We note that symmetry class D can also be identified by viewing Fig. S3 of~\cite{venn2022coherent} as defining a contraction of a $2$-dimensional Gaussian tensor network.  The tensor network has a known correspondence with a Chalker-Coddington network model and 1+1D free fermion dynamics~\cite{jian2022criticality,jian2023measurement}, with the specific form implying symmetry class D.

\subsection{Decoding fidelity}
In this section, we describe how we obtain the decoding fidelity $\calF$ directly from the syndrome sampling algorithm by extracting the probabilities 
$\calQ_{\alpha | s}$.
This procedure also allows us to determine the corresponding defect free energy $\Delta F$.

We use the syndrome sampling algorithm of~\cite{venn2022coherent,venn_coherent_planar_graph} as reviewed in Appendix~\ref{app:syndrome_sampling_cylinder}.
At each time $t$, we save the conditional probability $\calQ(m_t | m_{t-1}, \dots, m_1)$ as 
\begin{align}
    F_t &= \ln \calQ(m_t|m_{t-1} , \dots , m_1) 
    \nonumber\\
    &= \ln \calQ(m_t, \dots m_1) - \ln \calQ(m_{t-1}, \dots, m_1) \,.
\end{align}
We continue until we have measured every qubit at time $t_f$, after which we compute the telescoping series
\begin{equation}
    F = \sum_{t=1}^{t_f} F_t = \ln \calQ(\vec{m}) - \ln \calQ(m_1) \,.
\end{equation}
Observe that $\calQ(m_1) = 1/2$ independently of $\theta$.

We now run the syndrome sampling algorithm again, this time with ``post-selected''  measurement outcomes $\vec{m} + \vec{\zeta}$, where $\vec{\zeta}$ is a non-contractible loop corresponding to $Z_L$ operator.
We again save the conditional probabilities $\tilde{F}_t$ and compute
\begin{equation}
    \tilde{F} = \sum_{t=1}^{t_f} \tilde{F}_t = \ln \calQ(\vec{m}+\vec{\zeta}) - \ln \calQ(m_1) \,.
\end{equation}
Thus,
\begin{align}
    \Delta F_{s = \partial \vec{m}} := |F - \tilde{F}| 
    &= \abs{\ln\frac{\calQ(\vec{m})}{\calQ(\vec{m}+\vec{\zeta})}}
    \nonumber\\
    &= \abs{\ln\frac{\calQ_{s=\partial\vec{m},\, \alpha=\alpha[\vec{m}]}}{\calQ_{s=\partial\vec{m},\, \alpha = 1 \oplus\alpha[\vec{m}]}}}
    \,.
\end{align}
Here, we use that we may view $\vec{m}$ as an error string defining a given syndrome measurement $s = \partial \vec{m}$ and homology class $\alpha[\vec{m}]$.
As the outcomes $\vec{m}$ from the first run of the algorithm are sampled according to $\calQ_{s}$, we are able to compute the decoding fidelity or defect free energy through a simple average $\langle \cdot \rangle_{N_\mathrm{runs}}$ over $N_\mathrm{runs} \to \infty$ runs of the algorithm.
For example, the optimal probabilistic decoding fidelity is given by
\begin{align}
\Bigg\langle
\frac{1 + \exp\left(2\Delta F_{s = \partial \vec{m}}\right)}{\left(1 + \exp\left(\Delta F_{s = \partial \vec{m}}\right)\right)^2}
\Bigg\rangle_{N_\mathrm{runs}}
&\to \sum_s \calQ_s \sum_\alpha \calQ^2_{\alpha | s}
\nonumber
\\
&= \calF_\mathrm{opt}
\,.
\end{align}
Similarly, one can compute the defect free energy $\Delta F_\mathrm{opt}$ in a similar manner, by through a simple average of $\Delta F_{s = \partial \vec{m}}$ over runs.

For the fidelity and defect free energy corresponding to the suboptimal decoder, the above syndrome sampling algorithm must be run a total of four times.
In the first run, we sample $\vec{m}$ according to $\calQ_s$ with coherent rotation angle $\theta$.
The syndrome sampling algorithm is then run three more times with post-selected measurement outcomes $\vec{m}$ and $\vec{m} + \vec{\zeta}$ and coherent rotation angles $\theta$ and $\theta'$ allowing us to determine $\calQ_{0, s} / \calQ_{1, s}$ and $\calP_{0, s} / \calP_{1, s}$.
Both the decoding fidelity and defect free energy are functions of these ratios.
Finally, we may again take a simple average over runs of the algorithm, producing the average weighted by $\calQ_s$ as $N_\mathrm{runs} \to \infty$.

\section{Fermion description of ballistic metal}\label{app:pi_four_point}
In this Appendix, we determine the decoding fidelity $\calF$ at $\theta = \pi/4$, where the network model describes the ballistic metal.
We assume a cylindrical geometry of circumference $L$ and height $T$.
With a similar calculation, one can obtain the fidelity for a rectangular or toroidal geometry, which we omit here.

To begin, the RBIM partition function $\calZ_{\alpha,\, s}$ is related to a probability amplitude of the transfer matrix~\eqref{eqn:single_copy_rbim_prob_amplitude}
\begin{equation}
    \calZ_{\alpha,\, s} = (\psi_0| \hat{\mathsf H}_{T} \hat{\mathsf T}_{T-1} \cdots \hat{\mathsf T}_1 |\psi_0) \,.
\end{equation}
After a Jordan-Wigner transformation, this can be expressed as $\hat{\mathsf T}_t = \hat{\mathsf V}_{t+1/2} \hat{\mathsf H}_t$ where
\begin{align}
    \hat{\mathsf H}_t &= \prod_{i} \exp\left( \frac{\pi}{4}\eta_{i,\, t} \gamma_{i}^R \gamma_{i+1}^L\right) \,, \\
    \hat{\mathsf V}_{t + 1/2} &= \prod_i \exp\left(- \frac{\pi}{4}\eta_{i,\, t + 1/2}  \gamma_{i}^L \gamma_{i}^R\right) \,,
\end{align}
in terms of Majoranas $\gamma^{R/L}_{i}$ at each spin site $i$, 
with $|\psi_0)$ corresponds to the spin state $|+)^{\otimes L}$ and is stabilized by $\ri \gamma_{i}^L \gamma_{i}^R = +1$ $\forall i$. 
Note that defining $\hat{\mathsf H}$ ($\hat{\mathsf V}$) at (half)-integer times, and the corresponding indexing of $\eta_{i,\, t}$ and $\gamma^{R/L}_{i}$, are conventions adopted for convenience in this appendix, and should not be confused with those used elsewhere in the paper.
We can now write~\cite{bravyi_gosset_pfaffian_overlap}
\begin{equation}\label{eqn:overlap_value_gaussian}
    \calQ_{\alpha,\, s} = |(\psi_0|\psi_{\alpha,\, s})|^2 \propto |\Pf(\Omega + \tilde\Omega)| \,,
\end{equation}
where $|\psi_{\alpha,\, s}) =\hat{\mathsf H}_{T} \hat{\mathsf T}_{T-1} \cdots \hat{\mathsf T}_1  |\psi_0)$.
The covariance matrices $\tilde\Omega$ and $\Omega$ completely characterize the states 
$|\psi_{\alpha,\, s})$ and $|\psi_0)$, respectively, since they are 
fermion Gaussian states.
Specifically,
\begin{equation}
    \Omega_{i,\, j}^{L,\, R} = -\Omega_{i,\, j}^{R,\, L} = \delta_{i,\, j} \,,
\end{equation}
while one can show
\begin{equation}
    \tilde{\Omega}^{R,\, L}_{i-T,\, i+T} = - W_{i-T,\, i+T} \Omega_{i,\,i}^{L,\, R} \,,
\end{equation}
where henceforth, the addition in the spatial index is taken mod $L$ due to the periodic boundary conditions.
We also define the ``string operator'' $W_{i,\, j}$
\begin{align}
    W_{i-T,\, i+T} = \prod_{r=1}^{T-1} \eta_{i-r,\, r+\frac{1}{2}} & \eta_{i+r,\, r+\frac{1}{2}}
    \nonumber\\
    \times \prod_{r=1}^{T}& \eta_{i-r,\, r} \eta_{i-1+r,\, r} \,,
\end{align}
representing the overall sign accrued by the Majorana modes $\gamma^{R/L}_{i}$ as they pass through the bonds $\eta$.

The Pfaffian in~\eqref{eqn:overlap_value_gaussian} can be simplified by noticing that $\Omega + \tilde{\Omega}$ has a certain block structure.
In particular, $\gamma_{i}^L$ is paired with $\gamma_{i}^R$ in $|\psi_0)$ while $\gamma_{i}^R$ is paired with $\gamma_{i+2T}^L$ in $|\psi_{\alpha,\, s})$.
Thus, each block describes the correlations between the $\frac{2 L}{\gcd(2T,\, L)}$ Majorana modes of the form $\gamma^{R/L}_{i_k + 2mT}$ with $m = 0 , 1 , \dots \frac{L}{\gcd(2T,\, L)} - 1$ and $i_k$ a representative site index for the $k$-th block.
The Pfaffian then factorizes over blocks
\begin{equation}
    |\Pf(\Omega + \tilde{\Omega})| = \prod_{k=1}^{\gcd(2T,\, L)} |\Pf(\Omega_k + \tilde{\Omega}_k)| \,.
\end{equation}

We now focus on the calculation of the $k$-th block.
Observe that $\Omega_k$ and $\tilde\Omega_k$ both describe states which are products of paired Majorana states and can each be associated with a partition of the $\frac{2 L}{\gcd(2 T,\, L)}$ indices into pairs.
When $T$ is a multiple of $L$, the block is $2$-by-$2$ and $\Omega_k \propto \tilde\Omega_k$, while in the general case the partitions will be disjoint.
In both cases, it holds that
\begin{equation}
    |\Pf(\Omega_k + \tilde{\Omega}_k)| = | \Pf \Omega_k + \Pf \tilde\Omega_k| = 1 + \Pf\tilde\Omega_k \,,
\end{equation}
where we fixed $\Pf \Omega = +1$.
With this convention,
\begin{align}
    \Pf \tilde\Omega_k 
    &= - \prod_{m=1}^{\frac{L}{\gcd(2 T,\, L)}} \tilde\Omega^{R,\, L}_{i_k + (2m-1)T,\, i_k +(2m+1)T}
    \\
    &= -W^{(k)}_{\alpha,\, s} \times (-1)^{\frac{L}{\gcd(2 T,\, L)}} \,,
\end{align}
where
\begin{equation}
    W^{(k)}_{\alpha,\, s} = \prod_{m=1}^{\frac{L}{\gcd(2 T,\, L)}} W_{i_k + (2m-1)T,\, i_k +(2m+1)T} \,.
\end{equation}
First, consider the case of trivial syndrome measurements $\calQ_{\alpha,\, s=0}$.
In this case, $\Omega_k + \tilde{\Omega}_k$ has the same Pfaffian for every $k$.
Furthermore, because the total fermion parity is even, $\calZ_{\alpha=1,\, s=0}$ corresponds to the case where all $\eta = +1$ are such that $W_{\alpha=1} = +1$.
In the case of $\calZ_{\alpha=0,\, s=0}$, we insert $\eta = -1$ at even times along a defect in the time direction; $W_{\alpha = 0}$ now measures the parity of the winding number of the combined string operator $W_{\alpha=0} = (-1)^{\frac{2 T}{\gcd(2T,\, L)}}$.
This winding number will depend on the parity of $L / \gcd(T,\, L)$.
\begin{itemize}
    \item $L /\gcd(T,\, L)$ is odd.
    In this case, $T = 2^\mu R$ where $2^\mu$ is the largest power of two which divides $L$ and $R \in \mathbb{N}$.
    This implies $\gcd(2 T,\, L) = \gcd(T, L)$ such that $W_{\alpha=0} = W_{\alpha=1} = +1$.
    Thus, $\Pf\tilde\Omega_k = +1$ for all $k$ and $\calQ_{\alpha| s=0} = 1/2$.
    \item $L /\gcd(T,\, L)$ is even.  In this case, we factor $L = 2^\mu O_L$ and $T = 2^{\mu-\delta} O_T$ where $\delta \geq 1$ is integer and $O_L$, $O_T$ are odd.
    Now $\gcd(2 T,\, L) = 2 \gcd(T,\, L)$ such that $W_{\alpha=0} W_{\alpha=1} = (-1)^{\frac{O_T}{\gcd(O_L, O_T)}} = -1$.
    Thus, when $\frac{L}{2 \gcd(T,\, L)}$ is also even, we have $\calQ_{\alpha=1 | s=0} = 1$, otherwise $\calQ_{\alpha = 0 | s=0} = 1$.
\end{itemize}

Finally, in case of a general syndrome measurement outcome, the majority of the above argument still applies, except it is now possible that $\calQ_s= \calQ_{0,\, s} + \calQ_{1,\, s} = 0$, depending on the signs of $W^{(k)}_{\alpha,\, s}$.
However, whenever $\calQ_s > 0$, it must be that $| \Pf(\Omega_k + \tilde\Omega_k)|$ are either all simultaneously zero or simultaneously non-zero.
Furthermore, the product $W_{\alpha=0,\, s} W_{\alpha=1,\, s}$ is independent of $s$, so the sign of $\Pf \Omega_k$ relative to $\tilde{\Omega}_k$ is as well.
Since the decoding fidelity can be equivalently written
\begin{equation}
    \calF = \sum_{s | \calQ_s > 0} \calQ_s \max_\alpha \calQ_{\alpha | s} \,,
\end{equation}
we conclude that $\calF = 1/2$ whenever $T$ is an integer multiple of $2^\mu$, the largest power of two which divides $L$.  
Otherwise, $\calF = 1$.

\section{Additional numerics}
\subsection{Numerics for the error model with uniform rotation angle}
In this appendix, we provide numerical results for the bipartite entanglement entropy and defect free energy distribution, which are additional observables that distinguish the two replica limits associated with optimal and suboptimal decoders.
Our results are for the error model with spatially uniform rotation angle $\theta$, as in Sec.~\ref{sec:other_nlsm_predictions} of the main text.

\begin{figure}[h]
\centering
\includegraphics[width=.95 \linewidth  ]{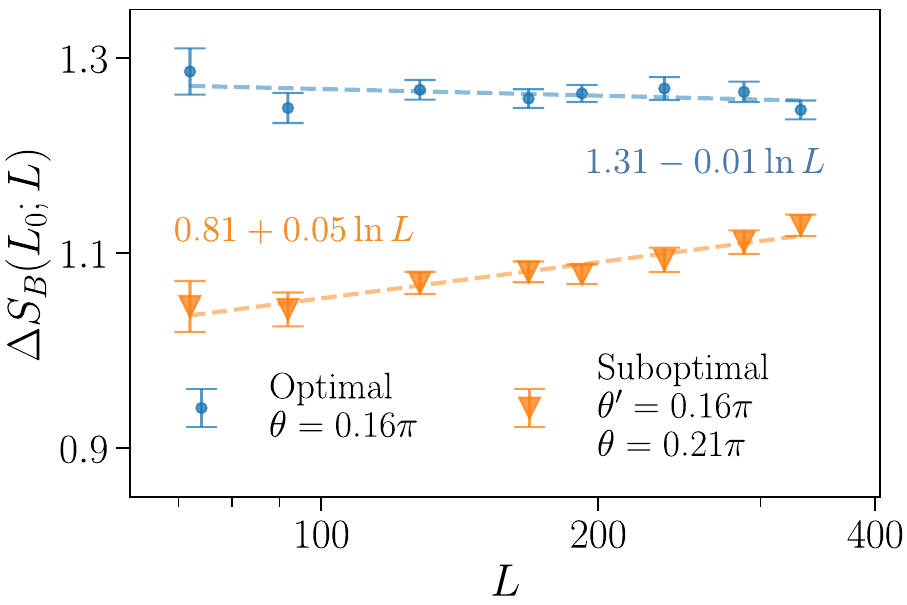}
    \caption{
    The half-system bipartite von-Neumann entropy $S_B$ after time $T=2L$.
    For the optimal decoder $\theta = 0.16\pi$ while for the suboptimal $\theta' = 0.16\pi$ with $\theta = 0.21\pi$.
    At the numerically accessible scale, the NLsM predicts $S_B \sim a \ln L + b \ln^2 L$.
    On the $y$-axis we plot $\Delta  S_B(L_0;\, L) := (S_B(L) - S_B(L_0))/\ln L/L_0$ with $L_0 = 52$.
    The data is consistent with fits to $a + b \ln L$ with $b >0$ for the suboptimal and $b < 0$ for the optimal decoder.
    For the optimal decoder, the fit to the $a + b\ln L$, as well as the small value of $b$, is consistent with $g_0^2 \ln L \ll 1$ near the metallic fixed point. 
    Data generated with $450$ to $5000$ samples.
    }
    \label{fig:bipartite_entropy}
\end{figure}
\subsubsection{Bipartite entanglement entropy}
The distinct RG flows associated with the different decoders can also be distinguished through the bipartite entanglement entropy in the steady state of the free fermion dynamics defined by the syndrome sampling algorithm.
As analyzed in Ref.~\cite{fava2023nonlinear}, the bipartite entanglement entropy is given by
\begin{align}\label{eqn:bipartite_scaling_prediction}
S_B \propto \int_{\ln a}^{\ln L} \frac{\rd s}{g_R(e^s)}.
\end{align}
We thus obtained the distinct predictions in two replica limits:
\begin{itemize}
\item In the limit $n \to 0$, the bipartite entropy is $S_B \propto g_0^{-1} \ln L + \ln^2 L$.
\item In the limit $n \to 1$, we have $S_B \propto g_0^{-1} \ln L -  2 g_0\ln^2 L$. Again, this result is valid when $g_0^2 \ln L \ll 1$.
\end{itemize}

\begin{figure*}[t]
\centering
\includegraphics[width=.95 \textwidth  ]{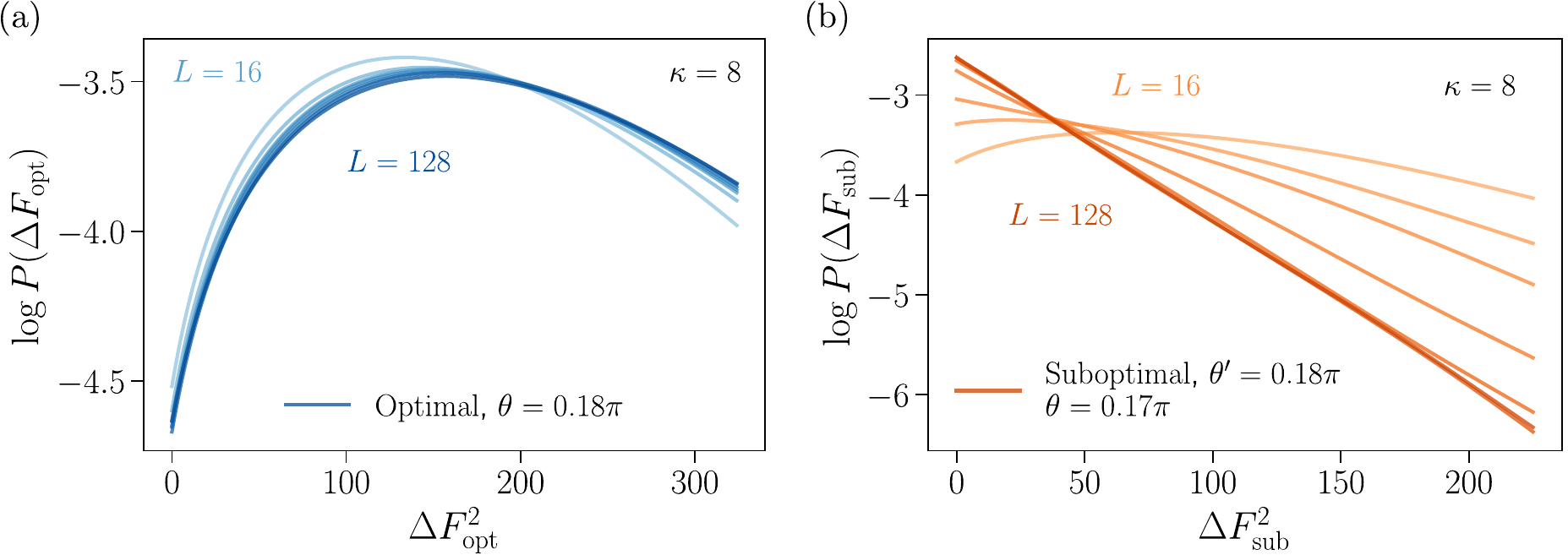}
    \caption{
    Distribution $P(\Delta F)$ of the defect free energy associated with the optimal (a) and suboptimal (b) decoders at fixed $\kappa = 8$.
    The distribution is bimodal for the optimal decoder (a), while it is well fit to a Gaussian distribution with zero mean for the suboptimal decoder (b).
    The distribution is determined using a Gaussian kernel with variance set by Scott's rule~\cite{scotts_rule}.
    In both figures, curves for various system sizes $L = 16, 24, 32, 48, 64, 92, 128$ are presented with increasing opacity.
    For the optimal decoder $\theta = 0.18\pi$, while for the suboptimal $\theta'=0.18\pi$ and $\theta = 0.17\pi$.
    Data points are averaged over $9000$ to $15000$ samples.
    }
    \label{fig:dfe_dist}
\end{figure*}

The two distinct replica limits can be distinguished by the different signs of the $\ln^2 L$ term, which in numerically accessible system sizes appears as a subleading correction to the $\ln L$ scaling from the bare coupling.
To this end, we consider the quantity 
\begin{equation}\label{eqn:discrete_log_derivative}
    \Delta S_B(L_0; L) = \frac{S_B(L) - S_B(L_0)}{\ln L - \ln L_0}
    \,,
\end{equation}
where the numerator removes a constant and the denominator removes a factor of $\ln L$.  Here, $L_0 = 52$ is the smallest system size we consider.

We simulate the bipartite entropy after time $T = 2L$ starting from the initial product state and under the syndrome sampling dynamics described in Appendix~\ref{app:syndrome_sampling_dynamics}.
In Fig.~\ref{fig:bipartite_entropy}, we find that $\Delta S_B$ is consistent with a fit to $a + b \ln L$, with $b > 0$ for the suboptimal decoder whereas $b < 0$ and small for the optimal decoder, in agreement with the non-linear sigma model prediction.

\subsubsection{Distribution of defect free energy}
In the main text, we obtain analytic predictions for the defect free energy $\Delta F$ associated with both decoders through the effective NLsM.

In addition to this, we have also observed empirically that the two decoders can be distinguished through the distribution of the defect free energy $P(\Delta F)$ over the syndrome configurations, as shown in Fig.~\ref{fig:dfe_dist}.
Here, $\Delta F_\opt := \ln\abs{\calQ_{0,s}/\calQ_{1,s}}$ and $\Delta F_\subopt := \ln\abs{\calP_{0,s}/\calP_{1,s}}$, and the syndrome $s$ is drawn from the true Born distribution $\calQ_s$.
For the optimal decoder, we observe that $P(\Delta F_\mathrm{opt})$ is bimodal, with the centers of the two peaks separating as $L$ is increased at fixed aspect ratio $\kappa$.
Having a bimodal distribution is compatible with the optimal decoder being in the decodable phase for $\theta < \pi / 4$.

On the other hand, the distribution $F(\Delta F_\mathrm{sub})$ for the suboptimal decoder is qualitatively different.
Specifically, the distribution is well fit to a Gaussian distribution with zero mean and decreasing variance as $L$ increases.
This is consistent with the suboptimal decoder being in the non-decodable thermal metal phase at this value of $(\theta$, $\theta')$.

\begin{figure*}[t]
\centering
\includegraphics[width=.95 \textwidth  ]{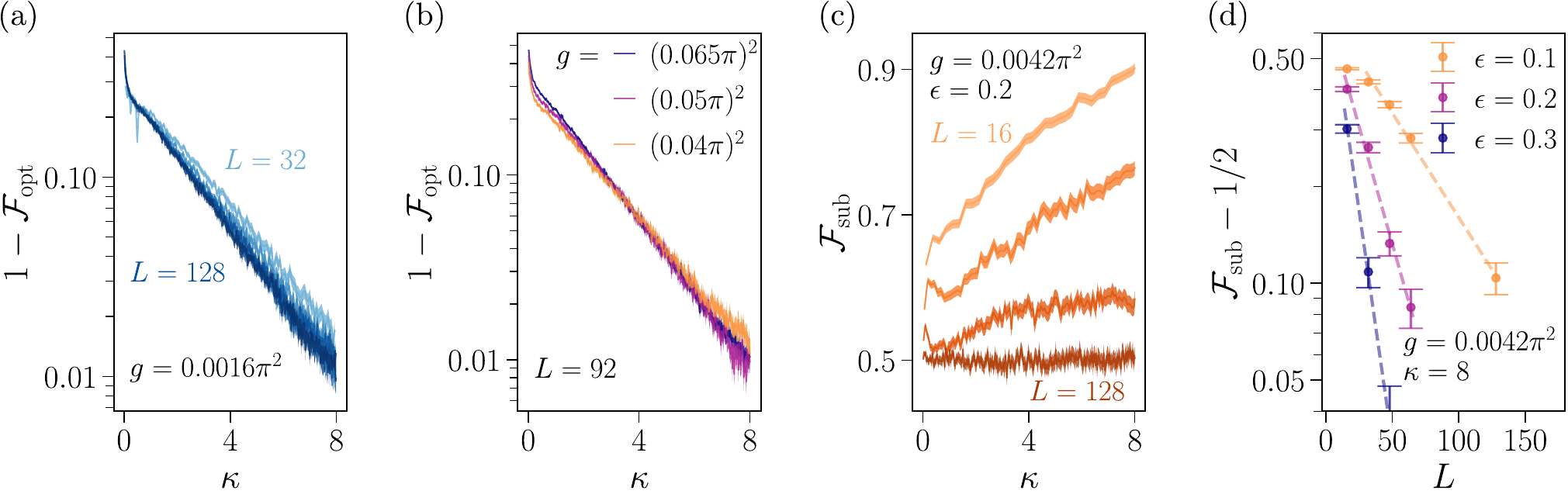}
    \caption{
    Decoding fidelity for the error model with rotation angles $\theta_\ell$ drawn from a Gaussian distribution with variance $g$.
    (a) Fidelity $\calF_\mathrm{opt}$ of the optimal decoder as a function of $\kappa$ for various system sizes $L$. When $\kappa \gg 1$, $\calF_\mathrm{opt}$ increases with scale $L$.
    (b) $\calF_\mathrm{opt}$ as a function of $\kappa$ for different variances $g$.
    As the variance $g$ is increased, $\calF_\mathrm{opt}$ decreases for $\kappa \ll 1$ and increases for $\kappa \gg 1$.
    (c) Fidelity $\calF_\mathrm{sub}$ of the suboptimal decoder as a function of $\kappa$. The estimated and the true rotation angle are related by $\left(\frac{\pi}{4} - \theta'_\ell\right)(1+ \epsilon) = \frac{\pi}{4} - \theta_\ell$. At large system sizes, $\calF_\mathrm{sub}$ decays to $1/2$.
    (d) The fidelity $\calF_\mathrm{sub}$ decays to $1/2$ exponentially in the system size $L$, with a rate that increases with $\epsilon$.
    Numerical results are averaged over $1200$ to $3000$ samples.
    Error bars are indicated by the linewidth.
    }
    \label{fig:random_theta_fid}
\end{figure*}
\begin{figure}[h]
\centering
\includegraphics[width=.95 \linewidth  ]{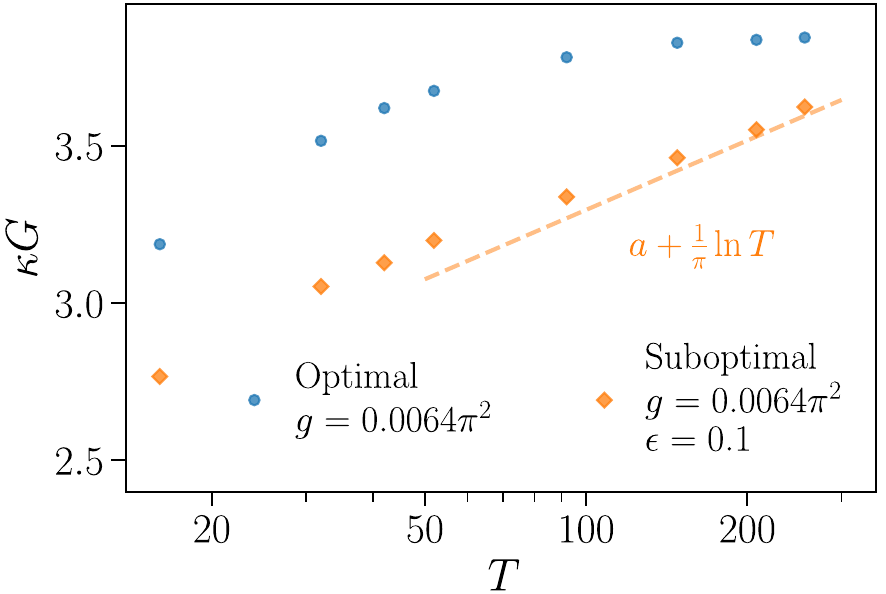}
    \caption{The conductivity $\kappa G$ in the network model 
    associated with the optimal (blue) and suboptimal (orange) decoder.
    The rotation angle $\theta_\ell$ is chosen from a Gaussian distribution with variance $g$ and mean $\pi / 4$.
    For the suboptimal decoder, $\left(\frac{\pi}{4} - \theta'_\ell\right)(1+ \epsilon) = \frac{\pi}{4} - \theta_\ell$.
    The conductivity for the suboptimal decoder fits to the scaling $\pi^{-1} \ln T$ and is distinct from that for the optimal decoder.
    Numerical results are averaged over $250$ to $550$ samples and are generated in the system with $T = L/4$.
    All errorbars are within the marker size.}
    \label{fig:compare_random_decoder_conductance}
\end{figure}

\subsection{Numerics for the error model with random rotation angles}\label{app:additional_numerics_non-uniform}
In this appendix, we examine the predictions of NLsM in the surface code with coherent errors of random rotation angles.
We consider the rotation angle $\theta_\ell$ on each edge drawn independently from a Gaussian distribution in Eq.~\eqref{eq:angle_distribution}.
The NLsM is derived microscopically as an effective theory for decoding with this error model. 
In the main text, we have presented the numerical results for the other error model with uniform rotation angles.
The results in this appendix and in the main text suggest that the predictions of NLsM hold generally for the decoding problems governed by the network model in Class D.

We begin by verifying the behavior of the decoding fidelity in Fig.~\ref{fig:random_theta_fid}.
For the optimal decoder, we find that $\calF_\mathrm{opt}$ increases with $L$ for fixed $\kappa \gg 1$ and is an increasing function of $\kappa$ for fixed $L$.
Moreover, we observe the ``trend reversal'' in the fidelity as a function of the renormalized coupling $g_R(L)$.
We tune $g_R(L)$ by increasing the variance $g$ and observe that $\calF_\mathrm{opt}$ increases with $g$ for $\kappa \gg 1$ and decreases with $g$ for $\kappa \ll 1$.
On the other hand, $\calF_\mathrm{sub}$ decays exponentially in the system size $L$.
Furthermore, the rate of the exponential decay is an increasing function of $\epsilon$, which parameterizes the extent to which the decoder's estimate $\theta'_\ell$ differs from the true rotation angle $\theta_\ell$.
This behavior of both fidelities is qualitatively the same as that of the uniform $\theta$ model and agrees with the NLsM prediction.

Next, we examine the RG flow of the coupling $g_R(L)$ by computing the conductivity $\kappa G$ of the corresponding network model (presented in Fig.~\ref{fig:compare_random_decoder_conductance}).
We observe $\pi^{-1} \ln T$ scaling for the suboptimal decoder, while such a scaling is absent for the optimal decoder, which agrees with the expectation from the NLsM in distinct replica limits.

\bibliographystyle{ieeetr}
\bibliography{refs.bib}
\end{document}